\documentclass[
reprint,
amsmath,amssymb,
aps,
]{revtex4-2}
\usepackage{amsthm}
\usepackage{graphicx}
\usepackage{dcolumn}
\usepackage{bm}
\usepackage{hyperref}

\usepackage{algorithm}
\usepackage{algpseudocode}
\usepackage{verbatim}
\usepackage{xcolor}

\usepackage{subfigure} 
\usepackage{float}

\newtheorem{theorem}{Theorem}
\newtheorem{corollary}{Corollary}
\newtheorem{lemma}{Lemma}

\def\>{\rangle}
\def\<{\langle}
\def\polylog{{\rm polylog}}
\def\poly{{\rm poly}}

\def\R{\mathbb{R}}

\def\D{\Delta}
\def\y{\boldsymbol{y}}
\def\z{\boldsymbol{z}}

\allowdisplaybreaks
\allowbreak
\begin{document}


\title{Quantum Gram-Schmidt Processes and Their Application to Efficient State Read-out for Quantum Algorithms}

\author{Kaining Zhang$^1$}
\author{Min-Hsiu Hsieh$^2$}
\author{Liu Liu$^1$}
\author{Dacheng Tao$^1$}
\affiliation{$^1$School of Computer Science, Faculty of Engineering, University of Sydney, Australia}
\affiliation{$^2$Hon Hai Quantum Computing Research Center, Taipei, Taiwan}

\date{\today}

\begin{abstract}

Many quantum algorithms that claim speed-up over their classical counterparts only generate quantum states as solutions instead of their final classical description. The additional step to decode quantum states into classical vectors normally will destroy the quantum advantage in most scenarios because all existing tomographic methods require runtime that is polynomial with respect to the state dimension. 
In this work, we present an efficient read-out protocol that yields the classical vector form of the generated state, so it will achieve the end-to-end advantage for those quantum algorithms.
Our protocol suits the case that the output state lies in the row space of the input matrix, of rank $r$, that is stored in the quantum random access memory. 
{The quantum resources for decoding the state in $\ell^2$ norm with $\epsilon$ error  require $\poly(r,1/\epsilon)$ copies of the output state and $\poly(r, \kappa^r,1/\epsilon)$ queries to the input oracles, where $\kappa$ is the condition number of the input matrix.} With our read-out protocol, we completely characterise the end-to-end resources for quantum linear equation solvers and quantum singular value decomposition. One of our technical tools is an efficient quantum algorithm for performing the Gram-Schmidt orthonormal procedure, which we believe, will be of independent interest. 
\end{abstract}


\maketitle


\section{Introduction} 

{Quantum algorithms have been popular for decades, due to the potential advantage in varying fields including physical simulations~\cite{PhysRevX.8.031022, RevModPhys.92.015003, PhysRevLett.126.062001}, combinatorial optimization~\cite{Guerreschi_2019, PRXQuantum.1.020312}, and linear algebra~\cite{prakash2014quantum}.
Notably, the latter has induced an independent subfield known as the quantum machine learning (QML)~\cite{biamonte2017quantum, havlivcek2019supervised}, which involves quantum linear algebra~\cite{harrow2009quantum, PhysRevA.97.012327, PRXQuantum.2.010315}, quantum learning protocols~\cite{PhysRevLett.113.130503, lloyd2013quantum, 10.1145/3106700.3106710, kerenidis2019q}, and quantum neural networks \cite{kapoor2016quantum, bausch2020recurrent}.}
These quantum algorithms have shown to achieve speed-ups over their classical counterparts.

Despite the claimed quantum speed-up, most QML algorithms suffered from both the input and the read-out problems. Specifically,  the input problem tackles the issue of efficient state preparation, namely, encoding the classical data, potentially of tantamount size, into quantum states. 
A few techniques \cite{harrow2009quantum, PhysRevLett.100.160501, kapoor2016quantum, gilyen2019distributional} have been proposed to address this problem, and among them, the quantum random access memory (QRAM) oracle model \cite{PhysRevLett.100.160501} has become, arguably, the most popular method in the domain of machine learning applications. 
{It has induced interesting outcomes in quantum algorithms for tasks such as the linear system solver \cite{harrow2009quantum, wossnig2018quantum, PhysRevA.101.022316}, the singular value decomposition \cite{PhysRevA.97.012327}, support-vector machines \cite{PhysRevLett.113.130503, li2015experimental, kerenidis2019quantum}, supervised and unsupervised learning \cite{lloyd2013quantum, kerenidis2019q}, neural networks \cite{allcock2018quantum, kerenidis2019quantum_dcnn}, and other machine learning tasks \cite{kerenidis2019quantum_gauss, kerenidis2016quantum, kerenidis2018quantum}.}
Generally, for a data matrix $A \in \R^{m \times d}$, the corresponding QRAM oracle could be {prepared} by using $O(\polylog(md))$ quantum operations with $O(md)$ physical resources \cite{PhysRevLett.100.160501} stored in a binary tree data structure~\footnote{Another implementation of the QRAM oracle is proposed in Ref. \cite{kerenidis2016quantum}, which requires $O(k \polylog(md))$ quantum operations and physical resources for $m \times d$ matrix with $k$ non-zero elements}. Although the QRAM oracle is criticized for the requirement of large physical resources, recent works \cite{PhysRevLett.123.250501,PRXQuantum.2.020311} have proven possible the practical implementation of the QRAM oracle.

On the other hand, the read-out problem addresses recovery of classical description from the output quantum state that contains the classical solutions. In order to preserve the quantum advantage of the underlining quantum algorithm, the output state needs to be decoded efficiently.  
For some quantum algorithms, such as the quantum recommendation system \cite{kerenidis2016quantum}, the read-out issue is relatively mild because the classical solution can be obtained by only a few measurements on the output state. {In general, most machine learning problems demand classical solutions in vector form, for example, finding solutions to linear systems. Hence, the read-out problem of these quantum algorithms could be critical. }
However, protocols for efficiently decoding the output quantum states into classical vectors remain little explored~\cite{aaronson2015read}. 

The task of recovering the unknown quantum state from measurements, which is also known as Quantum State Tomography (QST), is one of the fundamental problems in quantum information science. QST has attracted significant interest from both theoretical \cite{PhysRevLett.105.150401, kyrillidis2018provable, haah2017sample, o2016efficient, cramer2010efficient, xin2017quantum} and experimental \cite{haffner2005scalable, riebe2006process, RevModPhys.81.299, PRXQuantum.2.010318, PhysRevX.10.031064, PhysRevX.5.041006, PRXQuantum.2.010307} perspectives in recent years. The best general tomography method \cite{o2016efficient} could reconstruct a $d\times d$ density matrix $\rho$ for the unknown state with rank $r$ by using  $n=O(rd\epsilon^{-2})$ copies to the state, which implies $O(d\epsilon^{-2})$ copy complexity for the pure state case $\rho=|\bm{v}\>\<\bm{v}|$. 
We remark that most of QML algorithms that output a $d$-dimensional state as the solution claim the time complexity polylogarithmical to $d$. Thus, directly using state tomography methods for state read-out in QML is computationally expensive and would offset the gained quantum speedup.
Since the required number $n$ is proven optimal for both cases \cite{o2016efficient},  any further improvement on $n$ could be achieved only by assuming special prior knowledge on state $\rho$. 
For example, QST via local measurements provides efficient estimation for states which can be determined by local reduced density matrices \cite{xin2017quantum} or states with a low-rank tensor decomposition \cite{cramer2010efficient}. However, the output states generated by QML algorithms normally do not have these structures. 

In contrast with the assumptions in the QST scenarios, the output states generated by most QML algorithms \emph{do} have inherent relationship between the solution vector and the input data, commonly represented as a matrix. Specifically, the solution vector normally lies in the row space of the input data matrix. Notable examples that satisfy the aforementioned condition include: (1) the quantum SVD algorithm where the singular value $\sigma_i$ and corresponding singular vectors $|\bm{u}_i\>$ and $|\bm{v}_i\>$ for matrix $A=\sum_{i} \sigma_i \bm{u}_i \bm{v}_i^T$; and (2)
the quantum linear system solver for linear system $A\bm{x}=\bm{b}$ whose solution state $|\bm{x}\> \propto A^{-1} \bm{b}$ lies in the row space of $A$. Most machine learning problems can be reduced to these two categories  \cite{aaronson2015read}. Hence, finding efficient read-out protocols for them that go beyond the standard QST limit will be extremely desirable in the field of QML.  

In this work, we design an efficient state read-out protocol that works for QML algorithms which involve a $r$-rank input matrix $A\in\mathbb{R}^{m\times d}$ stored in the quantum random access memory (QRAM), and the output state $|\bm{v}\>$ lies in the row space of $A$. 
Instead of obtaining coefficients $\{v_i\}$ by measuring the state $|\bm{v}\>=\sum_{i=1}^{n} v_i |i\>$ in the standard orthonormal basis $\{|i\>\}$, our key technical contribution is an efficient method to obtain the classical description $x_i$ in the complete basis spanned by the rows $\{{A}_{g(i)}\}_{i=1}^{r}$ of $A$, so that $|\bm{v}\>=\sum_{i=1}^{r} x_i |{A}_{g(i)}\>$, {where the mapping $g(i):[r] \rightarrow [m]$ denotes the indices of rows selected as the basis.}
Our state read-out protocol requires $\tilde{O}(\poly(r))$ copies of the output states and $\tilde{O}(\poly(r, \kappa^r))$ queries to input oracles, where $r$  is the rank of the input matrix and $\kappa=\sigma_{\max}(A)/\sigma_{\min}(A)$ is the condition number of the input matrix. 
{We remark that the low-rank matrix assumption is common in machine learning models \cite{kulis2006learning, yao2018large, udell2019big}.}
Compared to previous QST methods which require at least ${O}(d\epsilon^{-2})$ copies of pure states, our protocol is much more efficient given $r\ll n$ with small condition numbers, and more importantly, the complexity does not depend on the system dimension. Finally, combining our read-out protocol with quantum SVD or quantum linear system solver yields an end-to-end complexity that takes $\tilde{O}(\poly(r,\kappa^r,\log(md)))$ queries to input oracles.

During the whole read-out protocol, we develop a quantum generalization of the Gram-Schmidt Orthonormalization process.
Our quantum Gram-Schmidt Process (QGSP) algorithm can construct a complete basis, by sampling a set of rows $\{{A}_{g(i)}\}_{i=1}^{r}$ of the input $A$, with $\tilde{O}(\poly(r, \kappa^r))$ queries to QRAM oracles. 
Since the vector orthonormalization is a crucial procedure in linear algebra as well as machine learning \cite{wang2016unsupervised, zhao2017gram, bansal2018can}, an efficient quantum algorithm will be of independent interest. Notice that there are some related works for the construction of orthogonal states \citep{vanner2013quantum, jevzek2014orthogonalization, Coelho_2016, Havlicek_2018}. However, these results deviate from standard Gram-Schmidt process and their applications are also limited. Ref.~\cite{vanner2013quantum} is only applicable to the single-qubit system, while Refs.~\cite{jevzek2014orthogonalization, Coelho_2016} only generate a state that is orthogonal to the input state and their complexity depends on the system dimension. Ref. \cite{Havlicek_2018} constructs orthogonal states from original states by lifting the dimension of the original Hilbert space, and cannot select a complete basis as standard Gram-Schmidt process does. Consequently, our proposed QGSP algorithm avoids all these restrictions and can be proven to be efficient. 

Specifically, we have the following result for QGSP. 
{\begin{theorem}[Informal]
By using ${O}(r^{27} \kappa^{14r})$ queries to QRAM  oracles of the matrix $A$, we could find a group of linearly independent rows $\{A_{g(i)}\}_{i=1}^r$, where $r$ and $\kappa$ is the rank and the condition number of $A$, respectively.
\end{theorem}}

\medskip
\textit{Main Result.} The main result for our state read-out protocol is as follows. 

\begin{theorem}\label{whole_theorem_1}
For the $d$-dimensional state $|\bm{v}\>$ lies in the row space of a matrix $A \in \R^{m \times d}$ with rank $r$ and the condition number $\kappa$, the classical form of $|\bm{v}\>$ could be obtained by using $O(r^4 \epsilon^{-2} )$ queries to the state $|\bm{v}\>$ and $O(r^{27} \kappa^{14r} + r^{18} \kappa^{8r} \epsilon^{-2} )$ queries to QRAM oracles of $A$, such that the $\ell^2$ norm error is bounded in $\epsilon$.
\end{theorem}

Further discussion about the applications of our main result will be delayed in Section~\ref{sec_app}. Instead, we will move on to formally define the framework of the state read-out protocol.

\section{State Read-Out Framework}

{In this section, we explain our protocol in detail.} Since $A\in\mathbb{R}^{m\times d}$ is of rank $r$, we can identify a set of $r$ linearly independent vectors $\{|{A}_{g(i) }\>\}_{i=1}^{r}$ selected from all rows of $A$ so that the output state can be rewritten as $|\bm{v}\rangle=\sum_{i=1}^r x_i|{A}_{g(i)}\rangle.$ Our goal is accomplished if we can determine $\{x_i\}_{i=1}^{r}$ efficiently. 
Following this, our algorithm consists of two major parts, a subroutine to sample a set of $r$ linearly independent rows $\{|{A}_{g(i) }\>\}_{i=1}^{r}$ from all rows of $A$ and a subroutine to calculate $\{x_i\}$, {which will be introduced in following subsections, respectively.}

\subsection{Complete Basis Sampling}

We begin with the first subroutine. The {Quantum Gram-Schmidt Process (QGSP)} in Algorithm~\ref{Basis} is developed to generate a complete row basis, by performing a quantum version of the adaptive sampling. 
The advantage of our adaptive sampling is that those rows, which have larger orthogonal part to the row space of previous sampled row submatrix, will be sampled with a larger probability. 
This ensures that the complete basis is nonsingular, and will improve the accuracy of the estimation of the coefficients in the second subroutine. 

\begin{algorithm}[H]
\caption{Quantum~Gram-Schmidt~Process~(QGSP)}
\label{Basis}
\begin{algorithmic}[1]
\Require
QRAM oracles $V_A$ and $U_A$ in Eqs.~(\ref{eq_QRAM_2}) and (\ref{eq_QRAM_1}). 
\Ensure
A group of orthonormal states $\{|\bm{t}_i\>\}_{i=1}^{r}$. An index set of the complete basis: $S_I = \{g(i)\}_{i=1}^{r}$.
\State Initialize the index set $S_I= \emptyset $.\label{Basis_one}
\For{$\ell=1$ to $r$}
\State Run the quantum circuit in Fig~\ref{QGSP_circuit}. Measure the third register and post-select on result $0$. Measure the first register to obtain an index $g(\ell)$. Update the index set $S_I = S_I \cup \{g(\ell)\}$.
\EndFor
\end{algorithmic}
\end{algorithm}

\begin{figure}[t]
\includegraphics[width=0.48\textwidth]{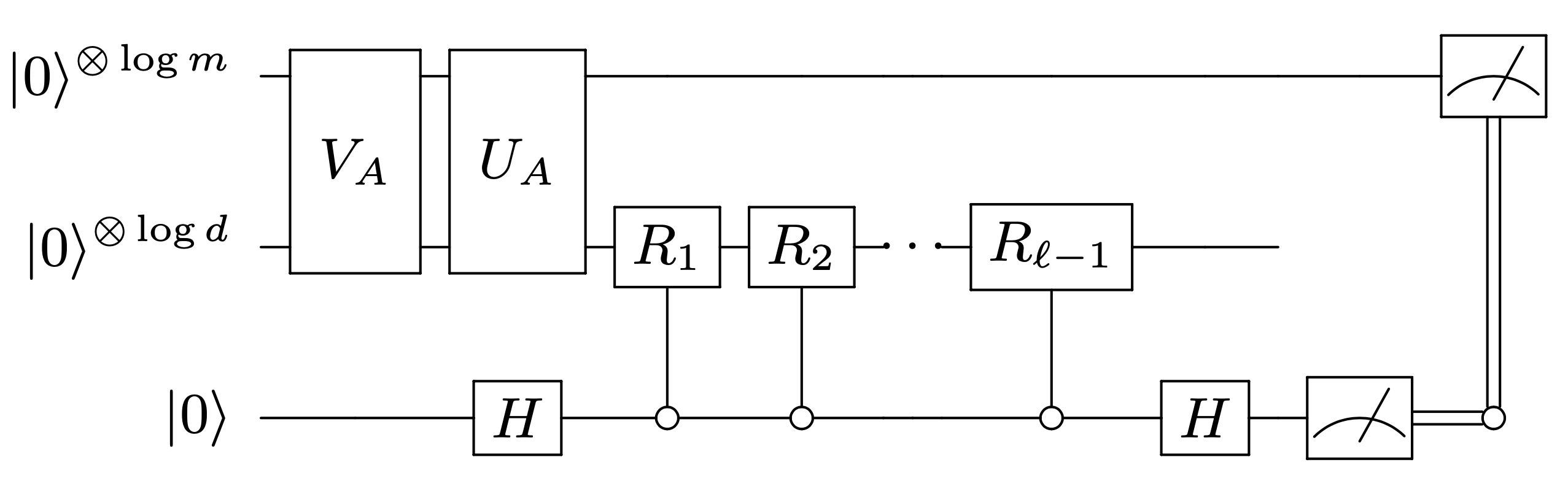}
\caption{Quantum circuit for the $\ell$-th iteration in the QGSP. {Oracles $V_A$ (\ref{eq_QRAM_1}) and $U_A$ (\ref{eq_QRAM_2}) are employed for encoding rows of the matrix $A$. Gates $C(R_i)$ (\ref{main_control_R_i}) are used for extracting the orthogonal part of rows from existing basis rows. After unitary operations, we measure the third register and post-select on the result $0$, to generate the state $|\phi_2^{(\ell)}\>$~(\ref{eq_post_phi}) from the state $|\phi_1^{(\ell)}\>$~(\ref{eq_before_phi}). Finally, the new index is sampled by measuring the first register of the state $|\phi_2^{(\ell)}\>$.}}
\label{QGSP_circuit}
\end{figure}

Now we analyze the QGSP in detail. We utilize QRAM oracles $V_A$ and $U_A$ to encode the matrix $A$ in the amplitude of quantum states:
\begin{align} 
|0\> \xrightarrow{V_A}  &\ \sum_{i=1}^{m} {\|{A}_{i}\|}/{\|{A}\|_F}  |i\> ,\label{eq_QRAM_2} \\
|i\> |0\> \xrightarrow{U_A} &\ |i\> |{A}_{i}\> \equiv \sum_{j=1}^{d}  {A_{ij}}/{\|{A}_{i}\|} |i\> |j\>, \forall i \in [m], \label{eq_QRAM_1}
\end{align}
where $A_{ij}$, $A_i$, and $\|A\|_F$ denote the $(i,j)$-th element, the $i$-th row, and the Frobenius norm of $A$, respectively.
In the first iteration of the QGSP, an {index} $g(1)$ is sampled from the set $[m]:=\{1,2,\cdots,m\}$ with the probability $\text{Pr}^{(1)}(i)=\|A_{i}\|^2/\|A\|_F^2$, where $i\in[m]$. {Let $|\bm{t}_1\>:=|A_{g(1)}\>$ be the first basis vector.}
{The remaining basis vectors are generated inductively.} Assume a set of orthogonal states $\{|\bm{t}_i\>\}_{i=1}^{\ell-1}$ has been generated in the previous $\ell-1$ iterations. 
To proceed to the $\ell$-th iteration, we perform the quantum circuit illustrated in Fig.~\ref{QGSP_circuit}, which first creates the state
\begin{equation}
\sum_{j=1}^{m} \frac{\|{A}_j\|}{\|{A}\|_F}  |j\>|{A}_j\> |0\>,
\end{equation}
with the help of input oracles $U_A$ and $V_A$. Then a Hadamard gate is applied to the third register, followed by a sequence of controlled $R_i$ gates 
\begin{equation}\label{main_control_R_i}
C(R_i) = R_i \otimes |0\>\<0|+ I \otimes |1\>\<1|,
\end{equation}
where the unitary $R_i = I-2|\bm{t}_i\>\<\bm{t}_i|$. {Next,} another Hadamard gate is applied to the third register, {and the quantum state evolves into:}
\begin{align}\label{eq_before_phi}
|\phi_1^{(\ell)}\> {}&= \frac{1}{\|{A}\|_F} 
\sum_{j=1}^m \|{A}_j\| |j\> \otimes \nonumber \\
& \bigg[ \Big( |{A}_j\>-\sum_{i=1}^{\ell-1}|\bm{t}_i\>\<\bm{t}_i|{A}_j\> \Big) |0\>-\sum_{i=1}^{\ell-1}|\bm{t}_i\>\<\bm{t}_i|{A}_j\> |1\>\bigg].
\end{align}
After all unitary operations, we measure the third register and post-select on result $0$ {with the success probability:}
\begin{equation}\label{definition_P_l}
P_{\ell} = \frac{1}{\|A\|_F^2}  \sum_{j=1}^{m} \|{A}_j\|^2 \Big\| |{A}_j\>-\sum_{i=1}^{\ell-1}|\bm{t}_i\>\<\bm{t}_i|{A}_j\> \Big\|^2,
\end{equation}
and the post-selected state (without the third register) is
\begin{equation}\label{eq_post_phi}
|\phi_2^{(\ell)}\> = \frac{1}{\sqrt{P_\ell}\|{A}\|_F} 
\sum_{j=1}^m \|{A}_j\| |j\> 
 \Big[ |{A}_j\>-\sum_{i=1}^{\ell-1}|\bm{t}_i\>\<\bm{t}_m|{A}_j\> \Big].
\end{equation}
We need roughly $1/P_\ell$ copies of $|\phi_1^{(\ell)}\>$ to generate the state $|\phi_2^{(\ell)}\>$. Finally, we measure the first register for a new basis index $g(\ell)$ and a new orthogonal state $|\bm{t}_{\ell}\>$:
\begin{equation}
|\bm{t}_{\ell}\> = \frac{1}{Z_{\ell}}
\bigg[|{A}_{g(\ell)}\>-\sum_{i=1}^{\ell-1} |\bm{t}_i\>\<\bm{t}_i|{A}_{g(\ell)}\> \bigg],\label{4.1_t_l+1}
\end{equation}
where $Z_{\ell}$ is the normalizing constant.
{Specifically, denote the probability of the outcome $g(\ell)$ being $j\in[m]$ by $\text{Pr}^{(\ell)} (j)$, and let $S_I=\{g(i)\}_{i=1}^{\ell-1}$.} We have 
\begin{align}
\text{Pr}^{(\ell)} (j) &= \frac{ \| A_j\|^2  \left\| ( \prod_{k=1}^{\ell-1} R_k + I)  |A_{j}\> \right\|^2 }{ \sum_{i=1}^m \| A_j\|^2  \left\| ( \prod_{k=1}^{\ell-1} R_k + I) |A_{j}\> \right\|^2 },\label{sample_prop_t} \\
&\equiv \frac{\| A_{j} - \pi_{S_I} (A_{j})\|^2}{ \sum_{i=1}^m \| A_{i} - \pi_{S_I} (A_{i}) \|^2 },\label{sample_prop}
\end{align}
where $\pi_{S_I}(A_j)$ denotes the projection of the row $A_j$ on the row space of the submatrix $A(S_I,\cdot) \in \R^{(\ell-1) \times n}$. In other words, the new index is sampled with the probability proportional to the norm of orthogonal part of the row $A_{g(\ell)}$ to the current basis set $S_I$.
After $r$ iterations, we could obtain the index set $S_I=\{g(i)\}_{i=1}^r$ such that $\{A_{g(i)}\}_{i=1}^{r}$ forms a linearly independent basis.
We remark that orthonormal states $\{|\bm{t}_i\>\}_{i=1}^{r}$ are generated from $\{|A_{g(i)}\>\}_{i=1}^{r}$ by performing Gram-Schmidt orthogonalization. {Thus, an orthonormal basis could be also generated after the implementation of Algorithm~\ref{Basis}.}

The technical difficulty of constructing the circuit in Fig.~\ref{QGSP_circuit} comes from  efficient implementation of the controlled version of reflection $R_\ell = I-2|\bm{t}_\ell\>\<\bm{t}_\ell|$, since we do not have additional quantum memory to store $\{|\bm{t}_\ell\>\}$ generated during the algorithm. To overcome this {problem}, we note that the state $|\bm{t}_\ell\>$ lies in $\text{span}\{|A_{g(i)}\>\}_{i=1}^{\ell}$, so that $|\bm{t}_\ell\> = \sum_{i=1}^{\ell} z_{i\ell} |A_{g(i)}\>$ for some coefficients $\{z_{i\ell}\}_{i=1}^{\ell}$. Instead, we could generate $|\bm{t}_i\>$ by the linear combination of unitary (LCU) method \cite{childs2012hamiltonian} with post-selections. Let $C_\ell$ be the Gram matrix of $\{|A_{g(i)}\>\}_{i=1}^{\ell}$, and let $C_{\ell-1}$ be the submatrix of $C_{\ell}$ by deleting the last row and column. The following lemma shows that the coefficient vector $\bm{z}_\ell=(z_{1\ell},\cdots,z_{\ell\ell})^T$ has a compact expression that only depends on the Gram matrices.  The proof is provided in Appendix~\ref{proof_cal_z_j}.

\begin{lemma}\label{4_1_lemma_calculate_z_j}
The coefficients in $|\bm{t}_\ell\> = \sum_{i=1}^{\ell} z_{i \ell} |{A}_{g(j)}\> $ could be written in the vector form $\bm{z}_\ell = \sqrt{\frac{|C_{\ell}|}{|C_{\ell-1}|}}C_\ell^{-1} \bm{e}_\ell$, where $\bm{e}_\ell=(0,0,\cdots,0,1)^T \in \R^{\ell}$ and $|X|$ denotes the determinant of a matrix $X$.
\end{lemma}

We remark that each element in the matrix $C_\ell$, i.e., the inner product between quantum states $\{|A_{g(i)}\>\}_{i=1}^{\ell}$, is unknown and needs to be estimated in practice. The error on elements in $C_\ell$ would influence the accuracy of coefficients $\bm{z}_\ell$, and consequently, impacts the whole complexity of the state read-out protocol. 
Let $\tilde{\bm{t}}_{\ell} = \sum_{i=1}^{\ell} \tilde{z}_{i\ell} A_{g(i)}/\|A_{g(i)}\|$ be the perturbed vector of $\bm{t}_\ell$, where $\{\tilde{z}_{i\ell}\}_{i=1}^{\ell}$ are the coefficients calculated following Lemma~\ref{4_1_lemma_calculate_z_j} with noisy Gram matrices $\tilde{C}_\ell$.  Denote $\sigma_{\min}(C_\ell)$ as the least singular value of $C_\ell$. 
We have the following Lemma~\ref{4_1_lemma_error_t_l} to bound $\| \tilde{\bm{t}}_\ell - \bm{t}_\ell \|$, whose proof is given in Appendix~\ref{proof_norm_error_t_l}.

\begin{lemma}\label{4_1_lemma_error_t_l}
If each element in $\tilde{C}_\ell$ deviates from that in $C_\ell$ by at most {$\epsilon_C \leq \frac{\sigma_{\min}^2 (C_\ell) }{80 \ell^{5/2}} \epsilon_R$}, then for any $\epsilon_R \in (0,1)$,
the $\ell^2$ norm of the error between $\bm{t}_\ell$ and $\tilde{\bm{t}}_\ell $ is bounded as 
\begin{equation}\label{eq_norm_error_t_l}
\| \tilde{\bm{t}}_\ell - \bm{t}_\ell \| \leq   \frac{\epsilon_R}{10},	
\end{equation}
where $\tilde{\bm{t}}_{\ell} = \sum_{i=1}^{\ell} \tilde{z}_{i\ell} A_{g(i)}/\|A_{g(i)}\|$.
\end{lemma}

{Lemma~\ref{4_1_lemma_calculate_z_j} and \ref{4_1_lemma_error_t_l} complete preconditions to generate the state $|\bm{t}_\ell\>$ through the LCU method. }
Then, given copies of $|\bm{t}_\ell\>\<\bm{t}_\ell|$, we can implement {the controlled version of the gate} ${R_\ell} = I-2|\bm{t}_\ell\>\<\bm{t}_\ell| = e^{-i\pi |\bm{t}_\ell\>\<\bm{t}_\ell|} $ with the help of the Hamiltonian simulation developed in Quantum PCA~\cite{Lloyd_2014}, as explained in Lemma~\ref{4.1_R_l_main}. 

\begin{lemma}\label{4.1_R_l_main}
Given Eq.~(\ref{eq_norm_error_t_l}) in Lemma~\ref{4_1_lemma_error_t_l}, 
the state $|\bm{t}_\ell\>$ could be prepared using $O(\ell \sigma_{\min}^{-1/2}(C_\ell))$ queries to the oracle $U_A$ with the $\ell^2$ norm error bounded by $\epsilon_R/5$.
The operation $C(R_\ell)$ could be prepared using $O(\ell \sigma_{\min}^{-1/2}(C_\ell)\epsilon_R^{-1})$ queries to the oracle $U_A$ with the spectral norm error of $R_\ell$ bounded by $\epsilon_R$.
\end{lemma}

The proof is provided in Appendix~\ref{proof_4.1_R_l_main}. As a natural corollary, the Gram-Schmidt orthonormal basis $\{|\bm{t}_\ell\>\}_{\ell=1}^{r}$ could be provided using $O(r^2 \sigma_{\min}^{-1/2}(C_r))$ queries to the oracle $U_A$.

Notice that the complexity of implementing $C(R_\ell)$ depends on the least singular value of the Gram matrix $C_\ell$, which is largely affected by the choice of the sampled basis $\{|A_{g(i)}\>\}_{i=1}^{\ell}$. A too small $\sigma_{\min}(C_\ell)$ will significantly increase the number of queries to the oracles. 
Notice that a group of basis with a small least singular value tends to have less probability being sampled, e.g., the probability of sampling a linearly dependent basis is $0$ by Eq.~(\ref{sample_prop}). 
Through further analysis, we prove that the expectation of $\sigma_{\min}(C_\ell)$ with the distribution formed by Eq.~(\ref{sample_prop_t}) is lower bounded as: 
\begin{equation}
\mathop{\mathbb{E}}\limits_{{\text{Pr}}^{(1)}} \cdots \mathop{\mathbb{E}}\limits_{{\text{Pr}}^{(\ell)}} [\sigma_{\min}(C_\ell)] \geq	\frac{r-\ell+1}{\ell r} \kappa^{2-2\ell}.
\end{equation}
This statement also holds \emph{approximately} if we 
take into account the error of implementing each $R_i$ for $i \in [\ell-1]$, {as provided in Lemma~\ref{bound_sigma_noisy}.} 

\begin{lemma}\label{bound_sigma_noisy}
Given that each gate $R_i$ in Algorithm~\ref{Basis} is implemented with error bounded by {$\epsilon_R = \frac{1}{3r^5 \kappa^{2r}}$}, where $r$ and $\kappa$ is the rank and the condition number of $A$, respectively, we have
\begin{align*}
\mathbb{E}_{\tilde{P}} [\sigma_{\min}({C}_\ell)] &\geq \frac{2}{3} \mathop{\mathbb{E}}\limits_{{\rm Pr}^{(1)}} \cdots \mathop{\mathbb{E}}\limits_{{\rm Pr}^{(\ell)}}  [\sigma_{\min} (C_\ell)], 
\end{align*}
where the distribution
\begin{equation}\label{main_distribution_P_noisy}
\tilde{P}(s_1,\cdots,s_\ell) = \tilde{{\rm Pr}}^{(1)}(s_1) \cdots \tilde{{\rm Pr}}^{(\ell)}(s_\ell)
\end{equation}
follows from Eq.~(\ref{sample_prop_t}) using noisy gates $\tilde{R}_i$.
\end{lemma}
The proof is very technical with lengthy steps.  Hence we delay their introduction to Appendix~\ref{app_min_singular}.

{As a result, we could perform Algorithm~\ref{Basis} for a few times to generate a basis with bounded least singular value.
The conclusion is summarized in Theorem~\ref{Basis_theorem_16} whose proof is given in Appendix~\ref{4.1_error_runtime}.}

\begin{theorem}\label{Basis_theorem_16}
By using {$O(r^{27} \kappa^{14r})$} queries to input oracles {$V_A$~(\ref{eq_QRAM_2}) and $U_A$~(\ref{eq_QRAM_1}), we} could find a group of {linearly independent} states $\{|A_{g(i)}\>\}_{i=1}^r$, such that the least singular value of the Gram matrix $C_r$ formed by $\{|A_{g(i)}\>\}_{i=1}^r$ is greater than $\frac{1}{2r^2 \cdot \kappa^{2r-2}}$, where $r$ and $\kappa$ is the rank and the condition number of $A$, respectively.
\end{theorem}

\subsection{Coefficient Calculation}

Next we focus on the second subroutine. 
Once the row basis has been selected, which now we denote as $\{\bm{s}_i\}_{i=1}^{r}$ for simplicity, the read-out problem reduces to obtaining coordinates $\{x_i\}_{i=1}^r$ in the description \(|\bm{v}\> = \sum_{i=1}^r x_i |\bm{s}_i\>\). 
The steps are outlined in Algorithm~\ref{coordinate_main}.

\begin{algorithm}[H]
\caption{State Read-out}
\label{coordinate_main}
\begin{algorithmic}[1]
\Require
QRAM oracle $U_A$. Copies of state $|\bm{v}\>$. Orthonormal basis $\{|\bm{t}_i\>\}_{i=1}^r$. The precision parameter $\epsilon$. 
\Ensure
Coordinates $\{x_i\}_{i=1}^{r}$ in \(|\bm{v}\> = \sum_{i=1}^r x_i |\bm{s}_i\>\) that guarantees a $\epsilon$ accuracy under $\ell^2$ norm.
\State Estimate the value $a_i^2=|\<\bm{v}|\bm{t}_i\>|^2$, for $i \in [r]$ by SWAP Test. Mark $k := \text{argmax}_{i\in [r]}	{a}_i^2$.
\State Run the circuit in Fig.~\ref{sign_circuit_main} to estimate ${a}'_i = \<\bm{t}_k|\bm{v}\>\<\bm{v}|\bm{t}_i\>$ for $i \in [r]$. Normalize the vector $\bm{a} = {\bm{a}'}/\|{\bm{a}'}\|$.
\State Output the solution as $\bm{x} = Z \bm{a}$, where $Z$ is given in Eq.~(\ref{ortho_basis_Z}).
\end{algorithmic}
\end{algorithm} 

The idea of Algorithm~\ref{coordinate_main} is fairly natural. Since the QGSP algorithm generates orthonormal states $\{|\bm{t}_i\>\}_{i=1}^{r}$, we could first calculate the coordinate of state $|\bm{v}\>$ under the basis $\{|\bm{t}_i\>\}_{i=1}^{r}$: $|\bm{v}\> = \sum_{i=1}^{r} a_i |\bm{t}_i\>,$ and then transfer the orthonormal basis to the row basis $\{\bm{s}_i\}_{i=1}^{r}$: 
\begin{equation}\label{ortho_basis_Z}
(\bm{t}_1,\cdots,\bm{t}_r) = (\frac{\bm{s}_1}{\|\bm{s}_1\|},\cdots, \frac{\bm{s}_r}{\|\bm{s}_r\|}) Z,
\end{equation}
where $Z=[z_{ij}]_{r \times r}$ is the transformation matrix.
The coordinates $\{x_i\}_{i=1}^r$ is given as: $\bm{x} = Z \bm{a}$.

The crucial part of Algorithm~\ref{coordinate_main} is to calculate the coefficient  $a_i=\<\bm{v}|\bm{t}_i\>, \forall i \in [r]$. However, the overlap estimation techniques based on {the Hadamard Test \cite{aharonov2009polynomial}} could not be directly employed for estimating the state overlap, since the unitaries for generating the states are required. This drawback limits most quantum algorithms, e.g., the quantum linear system solver, that require post-selection to yield the {solution state} easily.
Another choice is the SWAP test \cite{PhysRevLett.87.167902} that only requires copies of states. However, directly using the quantum SWAP test could only obtain the estimation to the value $|\<\bm{v}|\bm{t}_i\>|^2$, while $\text{sign}(a_i)$ remains unknown.
To overcome this difficulty, we could  assume that the state $|\bm{v}\>$ has the positive overlap with one of the basis, say $|\bm{t}_k\>$, and take the value  
\begin{equation}\label{definition_a_i}
a_i=\text{sign} \big( \<\bm{t}_k|\bm{v}\>\<\bm{v}|\bm{t}_i\> \big) |\<\bm{v}|\bm{t}_i\>|=\frac{\<\bm{t}_k|\bm{v}\>\<\bm{v}|\bm{t}_i\>}{|\<\bm{t}_k|\bm{v}\>|}
\end{equation}
as the state overlap. This assumption is equivalent to adding a global phase $0$ or $e^{i\pi}=-1$ on $|\bm{v}\>$, and will not affect the extraction of the classical description.

\begin{figure}[H]
\includegraphics[width=0.48\textwidth]{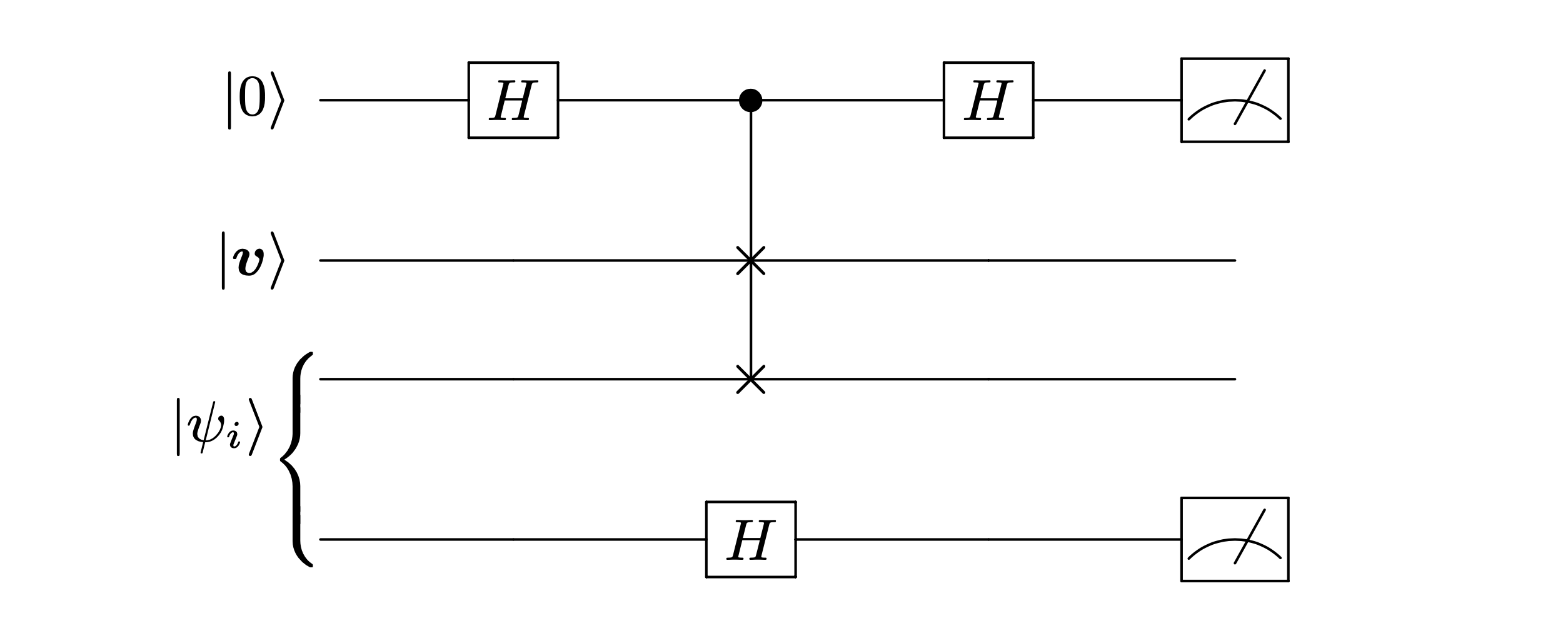}
  \caption{Quantum circuit for estimating $\<\bm{t}_k|\bm{v}\>\<\bm{v}|\bm{t}_i\>$, where $|\psi_i\> = \frac{1}{\sqrt{2}} (|\bm{t}_k\>|0\> + |\bm{t}_i\>|1\>)$.} 
  \label{sign_circuit_main}
\end{figure}

We construct a variant of the SWAP Test, illustrated in Fig.~\ref{sign_circuit_main} for estimating ${a}'_i = \<\bm{t}_k|\bm{v}\>\<\bm{v}|\bm{t}_i\>$. It is easy to see that the probability of the measurement outcomes `$00$' and `$11$' yields the value ${a}'_i$:
\begin{equation}\label{probability_a_i_a_k}
P_{\textnormal{same}} = P_{00} + P_{11} = \frac{1+ \<\bm{t}_k|\bm{v}\>\<\bm{v}|\bm{t}_i\>}{2}	= \frac{1+{a}'_i}{2}.
\end{equation}
Similar to the SWAP Test, the proposed quantum circuit provides a $\epsilon$-error estimation to the value $\<\bm{t}_k|\bm{v}\>\<\bm{v}|\bm{t}_i\>$ with $\tilde{O}(\epsilon^{-2})$ measurements.
Notice that a larger $|\<\bm{t}_k|\bm{v}\>|$ is preferred to obtain more accurate estimations of $a_i$ in Eq.~(\ref{definition_a_i}) through the estimations of $a_i'$ in Eq.~(\ref{probability_a_i_a_k}).
Thus, we mark $k := \text{argmax}_{i\in [r]} |\<\bm{t}_i|\bm{v}\>|^2$ by using the SWAP Test, before the estimations of $\{a_i'\}_{i=1}^r$ by running the circuit in Fig.~\ref{sign_circuit_main}.

The difficulty of implementing the quantum circuit in Fig.~\ref{sign_circuit_main} is to efficiently prepare the state $(|\bm{t}_k\>|0\> + |\bm{t}_i\>|1\>)/\sqrt{2}$. We apply the linear combination of unitaries (LCU) method again such that $(|\bm{t}_k\>|0\> + |\bm{t}_i\>|1\>)/\sqrt{2}$ could be prepared with query complexity $O(r \sigma_{\min}^{-1/2}(C_r))$. See Appendix~\ref{app_coordinate} for detail. 
By using this circuit {along with the SWAP Test}, we could approximately calculate the coordinates $\{x_i\}_{i=1}^r$.
The error and time complexity of Algorithm~\ref{coordinate_main} is provided in Theorem~\ref{solution_123}, with proof given in Appendix~\ref{app_coordinate}.

\begin{theorem}
\label{solution_123}
Algorithm~\ref{coordinate_main} provides a classical description $\bm{v} = \sum_{i=1}^r x_i {A}_{g(i)}/\|{A}_{g(i)}\|$ with $\ell^2$ norm error bounded in $\epsilon$, by using \(O({r^4} \epsilon^{-2})\) copies of state $|\bm{v}\>$ and $O({r^{10}} \sigma_{\min}^{-4}(C_r) \epsilon^{-2} )$ queries to input oracles.
\end{theorem}
Thus, our state read-out protocol only requires $\tilde{O}(\poly(r) \epsilon^{-2})$ copies of the unknown quantum state. 
The required state copy complexity is independent from the dimension of the state, which makes our algorithm more efficient than previous QST methods \cite{o2016efficient} in the low-rank case, since the latter needs at least $O(d \epsilon^{-2})$ copies.
We remark that the combination of Theorem~\ref{Basis_theorem_16} and Theorem~\ref{solution_123} yields the main result in Theorem~\ref{whole_theorem_1}.

\section{Applications} \label{sec_app}
As introduced in previous text, our read-out protocol suits the case that the output state of the quantum algorithm lies in the row space of the input matrix. We remark that this assumption is naturally satisfied by  many proposed quantum algorithms in the field of machine learning and linear algebra. 
In this section, we discuss the end-to-end versions of two existing quantum algorithms: the quantum singular value decomposition (SVD) algorithm and the quantum linear system solver, when employing our state read-out protocol for generating classical solutions.

\subsection{Quantum singular value decomposition}
We begin with the {quantum singular value decomposition} protocol.
For a given $r$-rank input matrix $A = \sum_{i=1}^r \sigma_i \bm{u}_i \bm{v}_i^T \in \R^{m \times d}$, there is:
\begin{equation}
\bm{v}_i = \frac{1}{\sigma_i}( \bm{u}_i^T A)^T	= \frac{1}{\sigma_i} \sum_{j=1}^m u_i^{(j)} A_{j}, \forall j \in [m],
\end{equation}
so any singular vector $\bm{v}_i$ lies in the row space $\text{span}\{A_{i}\}_{i=1}^{m}$.
Given QRAM oracles of the matrix $A$, quantum SVD allows to perform the operation \(\sum_j \beta_j |\bm{v}_j\> \rightarrow \sum_j \beta_j |\bm{v}_j\> |\hat{\sigma}_j\>\) with complexity \(O(\polylog(md)\|A\|_F \epsilon^{-1})\) such that \(\hat{\sigma}_j \in \sigma_j \pm \epsilon\) with high probability. Consider the state
\begin{equation*}
|0\>|0\> \stackrel{V_A U_A}{\longrightarrow}	\frac{\sum_{i=1}^m \sum_{j=1}^d A_{ij} |i\>|j\>}{\|A\|_F} = \frac{\sum_{i=1}^r \sigma_i |\bm{u}_i\> |\bm{v}_i\>}{\|A\|_F} 
\end{equation*}
as the input to the quantum SVD algorithm to generate the state $\frac{1}{\|A\|_F} \sum_{i=1}^r \sigma_i |\bm{u}_i\> |\bm{v}_i\> |\hat{\sigma}_i\>$. Then the measurement on the eigenvalue register could collapse the state to different eigenstates $|\bm{u}_i\>|\bm{v}_i\>$ with probability $\frac{\sigma_i^2}{\|A\|_F^2}$. Thus, any target state $|\bm{v}_i\>$ could be prepared with complexity $O(\polylog(md)\|A\|_F^3 \Delta_\sigma^{-1} \sigma_i^{-2})$, where $\Delta_\sigma$ is the eigen gap of the matrix $A$. Using this result along with {Theorem~\ref{whole_theorem_1},} we could derive the end-to-end complexity for SVD as follows.

\begin{corollary}
\label{readout_svd}
The classical form of any eigenstate $|\bm{v}_i\>$ of $A$ could be obtained by using $O(\kappa^{14r} \poly (r,\log (md)) \frac{\|A\|_F}{\Delta_\sigma \epsilon^2})$ queries to the input oracle of $A$, such that the $\ell^2$ norm error is bounded in $\epsilon$.
\end{corollary}

\subsection{Quantum linear system solver}

There has been an increasing interest in quantum machine learning \cite{PhysRevLett.113.130503, lloyd2013quantum, PhysRevLett.109.050505} and linear algebra \cite{kerenidis2018quantum, kerenidis2019quantum} algorithms following the quantum linear system solver proposed by Harrow, et al. \cite{harrow2009quantum}. The first quantum linear system solver was proposed especially for the sparse case by Hamiltonian simulation, and several other different linear system solvers \cite{childs2017quantum, wossnig2018quantum} have been proposed subsequently for the general case. Here we consider the quantum solver \cite{wossnig2018quantum} which encodes the input matrix $A \in \R^{d \times d}$ into the QRAM model. 

For matrix $A = \sum_{i=1}^r \sigma_i \bm{u}_i \bm{v}_i^T$, the solution could be written as:
\begin{equation}\label{example_ax_b}
\bm{x} = A^+ \bm{b},	
\end{equation}
where $A^+ = \sum_{i=1}^r \frac{1}{\sigma_i} \bm{v}_i \bm{u}_i^T$ is the pseudo-inverse matrix of $A$. Equation~(\ref{example_ax_b}) gives $\bm{x} = \sum_{i=1}^r \frac{1}{\sigma_i} \bm{u}_i^T \bm{b} \bm{v}_i \in \text{span}\{\bm{v}_i\}_{i=1}^{r}$, which means $\bm{x}$ also lies in the row space $\text{span}\{A_{i}\}_{i=1}^{n}$ by using the previous conclusion about eigenvectors.

For the linear system $A\bm{x} = \bm{b}$, the solution state $|\bm{x}\> = |A^{+}\bm{b}\>$ could be prepared in time $O(\kappa^2 \polylog(d) \|A\|_F \epsilon^{-1})$ with $\ell^2$ norm error bounded in $\epsilon$, where $\kappa$ is the condition number of $A$.
Then we could derive the end-to-end complexity for the quantum linear system solver as follows.

\begin{corollary}
\label{readout_hhl}
The classical form of the solution state $|A^+ \bm{b}\>$ for the linear system $A\bm{x} = \bm{b}$ could be obtained by using $O(\kappa^{14r} \poly(r, \log d ) \frac{\|A\|_F}{\epsilon^{3}} )$ queries to input oracles of $A$, such that the $\ell^2$ norm error is bounded in $\epsilon$.
\end{corollary}

\section{Conclusion and discussion}

In this work, we developed an efficient state read-out framework for quantum algorithms which involve a low-rank input matrix and the output state $|\bm{v}\>$ lies in the row space of the input matrix. 
The proposed framework takes $\tilde{O}(\poly(r) \epsilon^{-2})$ copies of the output state and $\tilde{O}(\poly(r, \kappa^r) \epsilon^{-2})$ queries to input oracles for providing $\epsilon$ error bounded classical description. Thus, our protocol preserves the quantum speed-up at the state read-out step of these quantum algorithms for the case that the rank $r$ and the condition number $\kappa$ are small, relative to the system dimension $d$. We analyzed the feasibility of our framework for quantum algorithms including the quantum SVD and the QRAM-based linear system solver in the low-rank case.

{Recently, several quantum-inspired classical algorithms \cite{3313276.3316310, chia2018quantuminspired, jethwani2019quantuminspired, du2019quantuminspired} have been developed as challenges to quantum advantage on machine learning tasks. 
Since QRAM oracles are employed in this work, we would like to emphasize the difference between these classical algorithms and the proposed read-out protocol. Note that the state read-out is a ``pure quantum" task which aims to generate the classical form of the unknown quantum state. However,  the quantum-inspired algorithms are developed for solving certain linear algebra problems if certain data structure and query access are allowed.

{Finally, we believe that the proposed results about decoding the pure state could be extended into the mixed-state case.} A quick outline of the procedure is as follows. We could first employ the quantum PCA \cite{Lloyd_2014} to perform the eigen-decompositions, and then to decode the eigenstates using our protocol. Another future direction is to improve our read-out framework such that the complexity is polynomial in both the rank and the condition number.

\appendix

\section{Proof of Lemma~\ref{4_1_lemma_calculate_z_j}}
\label{proof_cal_z_j}

\begin{proof}

Denote $\bm{s}_i:=A_{g(i)}$ for the simplicity of notation.
Consider the state:
\begin{equation}\label{4_1_lemma_calculate_z_j_t_l_eq}
|\bm{t}_{\ell}\>= \sum_{i=1}^{\ell} z_{i \ell} |\bm{s}_i\> = \frac{1}{Z_{\ell}}(|\bm{s}_{\ell}\>-\sum_{i=1}^{\ell-1} |\bm{t}_i\>\<\bm{t}_i|\bm{s}_{\ell}\>), 
\end{equation}
where $Z_\ell$ {has another formulation} obtained by multiplying {$\<\bm{t}_\ell|$} on both sides
\begin{equation}\label{4.1_z_l_1}
1 = \<\bm{t}_\ell|\bm{t}_\ell\> = \frac{1}{Z_\ell} \<\bm{t}_\ell|\bm{s}_\ell\> = \frac{1}{Z_\ell} \sum_{i=1}^{\ell} z_{i\ell} \<\bm{s}_i|\bm{s}_\ell\>.
\end{equation}
The restriction that $|\bm{t}_{\ell}\>$ is normalized and is orthogonal to states $|\bm{s}_1\>,|\bm{s}_2\>,\cdots|\bm{s}_{\ell-1}\>$ could yield:
\begin{align}
&\<\bm{s}_j| \bm{t}_\ell\> = \sum_{i=1}^{\ell} z_{i \ell} \<\bm{s}_j|\bm{s}_i\> = 0,\ \forall j \in [\ell-1], \label{4.1_x_1_1}\\	
& \<\bm{t}_\ell|\bm{t}_\ell\> = \sum_{j=1}^{\ell} \sum_{i=1}^{\ell} z_{j \ell} z_{i \ell} \<\bm{s}_j|\bm{s}_i\> = 1. \label{4.1_x_1_2}
\end{align}
Rewrite Equation~(\ref{4.1_z_l_1}) and (\ref{4.1_x_1_1}) in the vector form:
\begin{equation}\label{4.1_x_2_1}
{C}_{\ell} \bm{z}_\ell = Z_\ell	\bm{e}_\ell.
\end{equation}
Equation~(\ref{4.1_x_1_2}) could be written as:
\begin{align*}
1 &= \sum_{i,j=1}^{\ell} z_{i \ell} z_{j \ell} \<\bm{s}_j|\bm{s}_i\> =\bm{z}_{\ell}^T C_\ell \bm{z}_\ell \\
&= Z_\ell^2 \bm{e}_\ell^T C_\ell^{-1} \bm{e}_\ell  = Z_\ell^2 \frac{|C_{\ell-1}|}{|C_\ell|} ,
\end{align*}
where the third equation derives from $\bm{z}_\ell = Z_\ell C_\ell^{-1} \bm{e}_\ell$ by Equation~(\ref{4.1_x_2_1}) and the last equation is derived by noticing that the $(\ell,\ell)$-th element of $C_\ell^{-1}$ is $\frac{|C_{\ell-1}|}{|C_\ell|}$. Thus, we obtain
\begin{equation}\label{value_Z_l}
Z_\ell = \sqrt{ \frac{|C_\ell|}{|C_{\ell-1}|} } .
\end{equation}
Finally, solving (\ref{4.1_x_2_1}) is trivial
\begin{equation}\label{appendix_3}
\bm{z}_\ell = Z_\ell C_{\ell}^{-1} \bm{e}_\ell = \sqrt{ \frac{|C_\ell|}{|C_{\ell-1}|} } C_{\ell}^{-1} \bm{e}_\ell .
\end{equation}
\end{proof}

\section{Proof of Lemma~\ref{4_1_lemma_error_t_l}}
\label{proof_norm_error_t_l}

\begin{proof}
We denote $\|\cdot\|$ as the $\ell^2$ norm and the spectral norm for vectors and matrices.

First notice that 
\begin{align}
\| \bm{t}_\ell - \tilde{\bm{t}}_\ell \|^2 &= \bm{t}_\ell^T \bm{t}_\ell - 2 \bm{t}_\ell^T \tilde{\bm{t}}_\ell + \tilde{\bm{t}}_\ell^T \tilde{\bm{t}}_\ell \\
&= \bm{z}_\ell^T C_\ell \bm{z}_\ell - 2 \bm{z}_\ell^T C_\ell \tilde{\bm{z}}_\ell + \tilde{\bm{z}}_\ell^T C_\ell \tilde{\bm{z}}_\ell  \label{4_1_lemma_calculate_z_j_2_1}\\
&= \D \z_\ell^T C_\ell \D \z_\ell  \label{4_1_lemma_calculate_z_j_2_2} \\
&\leq \|C_\ell\| \|\D\z_\ell\|^2  \label{4_1_lemma_calculate_z_j_2_3},
\end{align}
where $C_\ell$ in Eq.~(\ref{4_1_lemma_calculate_z_j_2_1}) is the Gram matrix of $\{|A_{g(i)}\rangle\}_{i=1}^\ell$, and  $\D\z_\ell = \tilde \z_\ell - \z_\ell$ in Eq.~(\ref{4_1_lemma_calculate_z_j_2_2}). 
Since  $\|C_\ell\| \leq \text{Tr}[C_\ell]=\ell$, we can obtain the desired result; namely,
\begin{align}
\| \bm{t}_\ell - \tilde{\bm{t}}_\ell \|^2 &\leq \frac{\epsilon_R^2}{100},  \label{4_1_lemma_calculate_z_j_2_4}
\end{align}
if the following claim is true:
\begin{align}
 \|\D\z_\ell\| &\leq \frac{\epsilon_R}{10 \ell^{1/2}} .\label{4_1_lemma_calculate_z_j_3_8}
\end{align}

To prove Eq.~(\ref{4_1_lemma_calculate_z_j_3_8}), let us introduce some more notation.  
Denote by $\tilde{C}_\ell$ and $\tilde{C}_{\ell-1}$ the perturbed Gram matrices of ${C}_\ell$ and ${C}_{\ell-1}$, respectively. Let $\tilde{Z}_\ell = \sqrt{ \frac{|\tilde{C}_\ell|}{|\tilde{C}_{\ell-1}|} }$ and 
\begin{equation}\label{4_1_lemma_calculate_noisy_z}
\tilde{\bm{z}}_\ell = \tilde{Z}_\ell \tilde{C}_{\ell}^{-1} \bm{e}_\ell.
\end{equation}
Let $\D C_\ell = \tilde{C}_\ell - C_\ell$,  and $\D Z_\ell = \tilde{Z}_\ell - Z_\ell$. 
Then, 
\begin{align}
&{}\ \ \ \ \|\D\z_\ell\| \nonumber \\
&= \| \tilde{C}_{\ell}^{-1} \tilde{Z}_{\ell} \bm{e}_\ell - C_{\ell}^{-1} Z_{\ell} \bm{e}_\ell \| \label{4_1_lemma_calculate_z_j_3_1} \\
&= \| (C_{\ell} + \D C_{\ell})^{-1} (Z_{\ell} \bm{e}_\ell + \D Z_\ell \bm{e}_\ell) - C_{\ell}^{-1} Z_\ell \bm{e}_\ell \| \label{4_1_lemma_calculate_z_j_3_2} \\
&= \|(C_{\ell} + \D C_{\ell})^{-1} [(Z_{\ell} \bm{e}_\ell + \D Z_\ell \bm{e}_\ell)\nonumber \\ 
&\qquad\qquad\qquad\qquad - (C_{\ell} + \D C_{\ell}) C_{\ell}^{-1} Z_\ell \bm{e}_\ell ] \|  \\
&= \| (C_{\ell} + \D C_{\ell})^{-1} (\D Z_\ell \bm{e}_\ell - \D C_{\ell} C_{\ell}^{-1} Z_\ell \bm{e}_\ell) \| \label{4_1_lemma_calculate_z_j_3_3} \\
& \leq \| (C_{\ell} + \D C_{\ell})^{-1}\| \cdot (|\D Z_\ell| + \| \D C_{\ell}\| \| C_{\ell}^{-1} \| Z_\ell ) \label{4_1_lemma_calculate_z_j_3_4} 
\end{align}
where 
Eq.~(\ref{4_1_lemma_calculate_z_j_3_4}) follows from the triangular inequality.

Since each element in $\tilde{C}_\ell$ diviates from that in $C_\ell$ by at most {$\epsilon_C \leq \frac{\sigma_{\min}^2 (C_\ell) }{80 \ell^{5/2}} \epsilon_R$}, we could obtain
\begin{align}
\|\D C_\ell\| &\leq \|\D C_\ell\|_F \leq \sqrt{\ell^2 \epsilon_C^2} = \ell \epsilon_C ,\label{norm_delta_C_1}
\end{align}
and 
\begin{align}
\|(C_\ell + \D C_\ell)^{-1}\| &= \frac{1}{\sigma_{\min}(C_\ell + \D C_\ell)}  \\
&\leq \frac{1}{\sigma_{\min}(C_\ell) - \|\D C_\ell\|}  \label{norm_delta_C_2_2}\\
&\leq \frac{1}{\sigma_{\min}(C_\ell) - \ell {\epsilon_C}} \label{norm_delta_C_2_3} \\
&\leq \frac{80}{79} \sigma_{\min}^{-1}(C_\ell). \label{norm_delta_C_2}
\end{align}
Eq.~(\ref{norm_delta_C_2_2}) follows from the Weyl's inequality 
$$|\sigma_{\min}(C_\ell+\D C_\ell) -\sigma_{\min}(C_\ell)| \leq \|\D C_\ell\|.$$ 
Eq.~(\ref{norm_delta_C_2_3}) employs Eq.~(\ref{norm_delta_C_1}).
Eq.~(\ref{norm_delta_C_2}) follows because {
{$$\ell \epsilon_C \leq \frac{\sigma_{\min}^2 (C_\ell)}{80\ell^{3/2}} \epsilon_R \leq \frac{1}{80} \sigma_{\min}(C_\ell).$$}}

Together with Eqs.~(\ref{norm_delta_C_1}), (\ref{norm_delta_C_2}) {and $\|C_\ell^{-1}\| = \sigma_{\min}^{-1}(C_\ell)$}, Eq.~(\ref{4_1_lemma_calculate_z_j_3_4}) is upper bounded by
\begin{align}
\|\D\z_\ell\| \leq \frac{80}{79} \sigma_{\min}^{-1}(C_\ell) (|\D Z_\ell| + \ell {\epsilon_C} \sigma_{\min}^{-1}(C_\ell) Z_\ell) \label{4_1_lemma_calculate_z_j_3_5}.
\end{align}
To finish the proof of Eq.~(\ref{4_1_lemma_calculate_z_j_3_8}), we only need to bound 
\begin{align}
 |\D Z_\ell| 
& \leq 4 \ell^2 \sigma_{\min}^{-1}(C_\ell) {\epsilon_C}  Z_\ell \label{4_1_lemma_calculate_z_j_4_7}.
\end{align}
If Eq.~(\ref{4_1_lemma_calculate_z_j_4_7}) were true,  we could further bound $\|\D \z_\ell\|$ from Eq.~(\ref{4_1_lemma_calculate_z_j_3_5}) as follows:
\begin{align}
 \|\D\z_\ell\|  &\leq \frac{80}{79} \sigma_{\min}^{-2}(C_\ell) 5 \ell^2  {\epsilon_C} Z_\ell\label{4_1_lemma_calculate_z_j_3_7} \\
&\leq \frac{{\epsilon_R}}{10 \ell^{1/2}}, 
\end{align}
where $\ell > 1$, ${\epsilon_C} \leq \frac{\sigma_{\min}^2 (C_\ell) }{80 \ell^{5/2}} {\epsilon_R}$ and 
\begin{equation}\label{4_1_lemma_tmp01}
Z_\ell = \left\| |\bm{s}_\ell\> - \sum_{i=1}^{\ell-1} |\bm{t}_i\>\<\bm{t}_i|\bm{s}_\ell\> \right\| \leq 1.
\end{equation}

The last part of this section is to prove Eq.~(\ref{4_1_lemma_calculate_z_j_4_7}). 
To further analyze this term, we utilize the bound on the determinant of the perturbed matrix \cite[page 113]{godunov2013guaranteed}:
\begin{align}
\left| \frac{|C_\ell + \D C_\ell|-|C_\ell|}{|C_\ell|}\right| &\leq \frac{\ell \|C_\ell^{-1}\| \|\D C_\ell\|}{1-\ell \|C_\ell^{-1}\| \|\D C_\ell\|}.
\label{perturbed_determinant}
\end{align}
We can obtain 
\begin{align}
\left| \frac{|C_{\ell-1} + \D C_{\ell-1}|-|C_{\ell-1}|}{|C_{\ell-1}|}\right| &\leq \frac{(\ell-1) \|C_{\ell-1}^{-1}\| \|\D C_{\ell-1}\|}{1-(\ell-1) \|C_{\ell-1}^{-1}\| \|\D C_{\ell-1}\|} \nonumber \\
&\leq \frac{\ell \|C_\ell^{-1}\| \|\D C_\ell\|}{1-\ell \|C_\ell^{-1}\| \|\D C_\ell\|},
\label{perturbed_determinant_l_1}
\end{align}
where the second inequality follows by noticing that the function $f(x)=\frac{x}{1-x}$ is monotonically increasing and the property that the range of singular values of the submatrix is contained in that of the original matrix:
\begin{align*}
(\ell-1) \|C_{\ell-1}^{-1}\| \|\D C_{\ell-1}\| &\leq \ell \|C_{\ell-1}^{-1}\| \|\D C_{\ell-1}\|\\
&= \ell \sigma_{\min}^{-1} (C_{\ell-1}) \sigma_{\max}(\D C_{\ell-1}) \\
&\leq \ell \sigma_{\min}^{-1} (C_{\ell}) \sigma_{\max}(\D C_{\ell}) \\
&= \ell \|C_{\ell}^{-1}\| \|\D C_{\ell}\|. 
\end{align*}
Consequently, we have the bound on the term $|\D Z_\ell|$:
\begin{align}
&{}\ \ |\D Z_\ell| = \left|\frac{\tilde{Z}_\ell}{Z_\ell} -1\right| Z_\ell \label{4_1_lemma_calculate_z_j_4_1}\\
&= \left|\sqrt{\frac{|\tilde{C}_\ell|}{|C_\ell|}} \sqrt{\frac{|C_{\ell-1}|}{|\tilde{C}_{\ell-1}|}} -1\right| Z_\ell \label{4_1_lemma_calculate_z_j_4_2} \\
&= \left|\sqrt{\frac{|{C}_\ell+\D C_\ell|}{|C_\ell|}} \sqrt{\frac{|C_{\ell-1}|}{|{C}_{\ell-1}+\D C_{\ell-1}|}} -1\right| Z_\ell \label{4_1_lemma_calculate_z_j_4_8} \\
&\leq {\max\Bigg(   \sqrt{\frac{1}{1-\ell \|C_\ell^{-1}\|\|\D C_\ell\|} \frac{1-\ell \|C_\ell^{-1}\| \|\D C_\ell\|}{1-2\ell \|C_\ell^{-1}\| \|\D C_\ell\|}} -1, \nonumber } \\
&\ {1- \sqrt{ \frac{1-2\ell \|C_\ell^{-1}\| \|\D C_\ell\|}{1-\ell \|C_\ell^{-1}\| \|\D C_\ell\|}{(1-\ell \|C_\ell^{-1}\|\|\D C_\ell\|)}}  \Bigg) Z_\ell }\label{4_1_lemma_calculate_z_j_4_3} 
\end{align}
where Eq.~(\ref{4_1_lemma_calculate_z_j_4_3}) is derived by employing the following equivalent form of Eqs.~(\ref{perturbed_determinant}) and (\ref{perturbed_determinant_l_1}):
\begin{align*}
\frac{|C_\ell + \D C_\ell|}{|C_\ell|} &\geq \frac{1-2\ell \|C_\ell^{-1}\| \|\D C_\ell\|}{1-\ell \|C_\ell^{-1}\| \|\D C_\ell\|} ,\\
\frac{|C_\ell + \D C_\ell|}{|C_\ell|} &\leq \frac{1}{1-\ell \|C_\ell^{-1}\| \|\D C_\ell\|}, \\
\frac{|C_{\ell-1} + \D C_{\ell-1}|}{|C_{\ell-1}|}  &\geq \frac{1-2\ell \|C_{\ell}^{-1}\| \|\D C_{\ell}\|}{1-\ell \|C_{\ell}^{-1}\| \|\D C_{\ell}\|},\\
\frac{|C_{\ell-1} + \D C_{\ell-1}|}{|C_{\ell-1}|}  &\leq \frac{1}{1-\ell \|C_{\ell}^{-1}\| \|\D C_{\ell}\|}.
\end{align*}
Since $\max(A,B) \leq A+B$ for $A,B\geq 0$, Eq.~(\ref{4_1_lemma_calculate_z_j_4_3}) yields
\begin{align}
& \leq \left( \sqrt{\frac{1}{1-2\ell \|C_\ell^{-1}\| \|\D C_\ell\|}} {-} \sqrt{1-2\ell \|C_\ell^{-1}\| \|\D C_\ell\|} \right) Z_\ell \label{4_1_lemma_calculate_z_j_4_4} \\
& = \frac{2\ell \| C_\ell^{-1}\| \|\D C_\ell\|}{\sqrt{1-2\ell \|C_\ell^{-1}\| \|\D C_\ell\| }} Z_\ell \label{4_1_lemma_calculate_z_j_4_5} \\
&\leq \frac{2\ell^2 \sigma_{\min}^{-1}(C_\ell) {\epsilon_C}}{\sqrt{1-2\ell^2 \sigma_{\min}^{-1}(C_\ell) {\epsilon_C} }} Z_\ell \label{4_1_lemma_calculate_z_j_4_6} \\
& \leq 4 \ell^2 \sigma_{\min}^{-1}(C_\ell) {\epsilon_C}  Z_\ell. 
\end{align}
Eq.~(\ref{4_1_lemma_calculate_z_j_4_6}) is derived by using Eqs.~(\ref{norm_delta_C_1}), (\ref{4_1_lemma_tmp01}) and $\|C_\ell^{-1}\| = \sigma_{\min}^{-1}(C_\ell)$.
The last equation holds because
\begin{align*}
\sqrt{1-2\ell^2 \sigma_{\min}^{-1}(C_\ell) {\epsilon_C} } &\geq \sqrt{1-  2\ell^2 \sigma_{\min}^{-1}(C_\ell) \frac{\sigma_{\min}^2 (C_\ell) }{80 \ell^{5/2}} {\epsilon_R} } \\
&\geq \sqrt{1-\frac{1}{4}} \geq \frac{1}{2},
\end{align*} 
{which is obtained by using the bound of $\epsilon_C$ and $\frac{\sigma_{\min}(C_\ell)}{40\ell^{1/2}} \epsilon_R \leq \frac{1}{4}$}.

\end{proof}

\section{Proof of Lemma~\ref{4.1_R_l_main}}
\label{proof_4.1_R_l_main}

\begin{proof}
The main idea is to firstly derive the error analysis of $|\bm{t}_\ell\>$ and $R_\ell$, followed by the development of the LCU protocol.
Denote $\bm{s}_i:=A_{g(i)}$ for the simplicity of notation.
We begin from the assumption that 
\begin{equation}\label{proof_4.1_R_l_main_0}
\|\tilde{\bm{t}}_\ell - \bm{t}_\ell\| \leq \frac{{\epsilon_R}}{10},	
\end{equation}
where $\tilde{\bm{t}}_\ell = \sum_{i=1}^\ell \tilde{z}_{i\ell} \bm{s}_i/\|\bm{s}_i\|$.
Then the $\ell^2$ norm of the error of the state $|\bm{t}_\ell\>$ is bounded as follows.
\begin{align}
\||\tilde{\bm{t}}_\ell\> - |\bm{t}_\ell\>\| &\leq \| \tilde{\bm{t}}_\ell - |\tilde{\bm{t}}_\ell\> \| + \| \tilde{\bm{t}}_\ell - |{\bm{t}}_\ell\>\| \label{proof_4.1_R_l_main_1_1} \\
&= \left| \| {\tilde{\bm{t}}_\ell \| -1 }\right| {\| | \tilde{\bm{t}}_\ell \> \|} + \| \tilde{\bm{t}}_\ell - {\bm{t}}_\ell \| \label{proof_4.1_R_l_main_1_2} \\
&= \left| \| \tilde{\bm{t}}_\ell \| - \| {\bm{t}}_\ell \| \right| + \| \tilde{\bm{t}}_\ell - {\bm{t}}_\ell \| \label{proof_4.1_R_l_main_1_3} \\
&\leq \| \tilde{\bm{t}}_\ell - {\bm{t}}_\ell \| + \| \tilde{\bm{t}}_\ell - {\bm{t}}_\ell \| \label{proof_4.1_R_l_main_1_4} \\
&\leq \frac{{\epsilon_R}}{5}.\label{proof_4.1_R_l_main_1_5}
\end{align}
Eqs~(\ref{proof_4.1_R_l_main_1_1}-\ref{proof_4.1_R_l_main_1_5}) are derived by using $\|\bm{t}_\ell\|=1$, the triangular inequality, and Eq~(\ref{proof_4.1_R_l_main_0}).
We could further provide the spectral norm of the error of the gate $R_\ell$:
\begin{align}
&{}\ \ \ \ \| \tilde{R}_\ell - {R}_\ell \| \nonumber \\
&=  \left\| (I-2|\tilde{\bm{t}}_\ell\>\<\tilde{\bm{t}}_\ell|) - (I-2 |{\bm{t}}_\ell\>\<{\bm{t}}_\ell|) \right\| \label{proof_4.1_R_l_main_2_2} \\
&= 2 \left\| |\tilde{\bm{t}}_\ell\>\<\tilde{\bm{t}}_\ell| - |{\bm{t}}_\ell\>\<{\bm{t}}_\ell| \right\| \label{proof_4.1_R_l_main_2_3} \\
&\leq 2 \left\| |\tilde{\bm{t}}_\ell\>\<\tilde{\bm{t}}_\ell| - |\tilde{\bm{t}}_\ell\>\<{\bm{t}}_\ell| \right\|_F + 2 \left\| |\tilde{\bm{t}}_\ell\>\<{\bm{t}}_\ell| - |{\bm{t}}_\ell\>\<{\bm{t}}_\ell| \right\| \label{proof_4.1_R_l_main_2_4} \\
&= 2 \left\| |\tilde{\bm{t}}_\ell\> - |{\bm{t}}_\ell\> \right\| + 2 \left\| |\tilde{\bm{t}}_\ell\> - |{\bm{t}}_\ell\> \right\|\label{proof_4.1_R_l_main_2_5} \\
&\leq \frac{4}{5} {\epsilon_R}. \label{proof_4.1_R_l_main_2_6}
\end{align}
Eq.(\ref{proof_4.1_R_l_main_2_2}) is derived due to the definition of $R_\ell$.
Eq.(\ref{proof_4.1_R_l_main_2_4}) is derived by using the triangular inequality. Eq.(\ref{proof_4.1_R_l_main_2_6}) is derived by using Eq.(\ref{proof_4.1_R_l_main_1_5}).

Now we provide a framework to implement operations $C(\tilde{R}_\ell)$ using coefficients $\{\tilde{z}_{j \ell}\}_{j=1}^\ell$.
We could first prepare the pure state $\tilde{\rho}_{\ell} = |\tilde{\bm{t}}_{\ell}\>\<\tilde{\bm{t}}_{\ell}|$ by the linear combination of unitaries method as follows. 
Firstly, initialize the state $|0\>^{\otimes \log m} |0\>^{\otimes \log n} |0\>$. Then, we apply Hadamard operations on the {last $\log \ell$ qubits} in the first register to create the state:
\begin{equation*}
\frac{1}{\sqrt{\ell}} \sum_{i=1}^{\ell} |i\> |0\>|0\>.
\end{equation*}
Next, we employ the operation
\begin{equation}\label{U_index}
U_{\text{index}} = 	\prod_{i=1}^\ell \left(I-|i\>\<i| -|g(i)\>\<g(i)| + |i\>\<g(i)| + |g(i)\>\<i| \right)
\end{equation}
to swap states $|i\>$ and $|g(i)\>, \forall i \in [\ell]$, to yield the state:
\begin{equation*}
\frac{1}{\sqrt{\ell}} \sum_{i=1}^\ell |g(i)\>|0\>|0\>.	
\end{equation*}
The unitary $U_{\text{index}}$ could be implemented by $O(\ell)$ operations. Then we employ the oracle $U_A$ on the first and the second register, followed by the unitary $U_{\text{index}}^\dag$, to yield:
\begin{align*}
\frac{1}{\sqrt{\ell}}\sum_{i=1}^{\ell}  |i\>|A_{g(i)}\> |0\> \equiv \frac{1}{\sqrt{\ell}}\sum_{i=1}^{\ell}  |i\>|\bm{s}_i\> |0\>. 
\end{align*}
Denote $\tilde{z}_{\ell}  \equiv \max_{i}|\tilde{z}_{i \ell}|$. Then we perform the controlled rotation 
$$\sum_{i=1}^\ell |i\>\<i| \otimes e^{-i \sigma_y \arccos(\tilde{z}_{i \ell}/\tilde{z}_{\ell})} + \sum_{i=\ell+1}^{m} |i\>\<i| \otimes I$$ 
on the third register, conditioned on the first register $|i\>$, to obtain:
\begin{align*}
\frac{1}{\sqrt{\ell}}\sum_{i=1}^{\ell}  |i\>|\bm{s}_i\> \left(\frac{\tilde{z}_{i \ell}}{\tilde{z}_{\ell}}|0\> + \sqrt{1-\frac{\tilde{z}_{i \ell}^2}{\tilde{z}_{\ell}^2}}|1\>\right). 
\end{align*}
Finally, we employ Hadamard operations on {last $\log \ell$ qubits in} the first  register, to obtain the state
\begin{align*}
 &  \frac{1}{\ell}\sum_{i=1}^{\ell} |0\> \frac{\tilde{z}_{i \ell}}{\tilde{z}_{\ell}} |\bm{s}_i\> |0\> +\ orthogonal\ garbage\ state\\
=& \frac{\|\tilde{\bm{t}}_\ell\|}{\ell \cdot \tilde{z}_{\ell}} |0\> |\tilde{\bm{t}}_{\ell}\> |0\> +\ orthogonal\ garbage\ state.
\end{align*}
The measurement on the first and the third registers of the final state could yield state $|\tilde{\bm{t}}_{\ell}\>$ with success probability $\|\tilde{\bm{t}}_\ell\|^2/\ell^2 \tilde{z}_{\ell}^2$, so we could prepare the state $|\bm{t}_{\ell}\>$ with ${O}(\ell \tilde{z}_{\ell}/\|\tilde{\bm{t}}_\ell\|)$ queries to $U_A$ by using the amplitude amplification method~\cite{brassard2002quantum}.

Note that operations $\tilde{R}_\ell = I -2|\tilde{\bm{t}}_\ell\>\<\tilde{\bm{t}}_\ell|$ can be viewed as the unitary with Hamiltonian $\tilde{\rho}_\ell = |\tilde{\bm{t}}_\ell\>\<\tilde{\bm{t}}_\ell|$:
\begin{align*}
e^{-i\pi \tilde{\rho}_\ell} &= 1 + (-i\pi \tilde{\rho}_\ell) + \frac{1}{2!}(-i \pi \tilde{\rho}_\ell)^2 + \cdots\\
&= 1 - \tilde{\rho}_\ell + \tilde{\rho}_\ell \left[  1 + (-i\pi) + \frac{1}{2!} (-i\pi)^2 + \cdots \right] \\
&= 1 - \tilde{\rho}_\ell + \tilde{\rho}_\ell e^{-i\pi} \\
&= I -2|\tilde{\bm{t}}_\ell\>\<\tilde{\bm{t}}_\ell|.
\end{align*}
Therefore, by using the Hamiltonian simulation method developed in Quantum PCA~\cite{Lloyd_2014}, the controlled version of $\tilde{R}_{\ell}$ could be performed with error {$\epsilon_R/5$} consuming {$O(5\pi^2/\epsilon_R)=O(1/\epsilon_R)$} copies of $\tilde{\rho}_{\ell} $.
Taking the complexity of generating state $|\tilde{\bm{t}}_{\ell}\>$ into account, we could implement operation $C(\tilde{R}_{\ell})$ with the error of $\tilde{R}(\ell)$ bounded as {$\epsilon_R/5$}, by using $O(\ell \max_i |\tilde{z}_{i \ell}|/(\|\tilde{\bm{t}}_\ell\| {\epsilon_R}))$ queries to $U_A$. 
We remark that the $\ell^2$ norm of vector $\tilde{\bm{z}}_\ell$ is bounded as
\begin{equation*}
1 = \<\tilde{\bm{t}}_\ell|\tilde{\bm{t}}_\ell\> = \frac{\tilde{\bm{z}}_\ell^T C_\ell \tilde{\bm{z}}_\ell}{\| \tilde{\bm{t}}_\ell \|^2} \geq \frac{\|\tilde{\bm{z}}_\ell\|^2}{\| \tilde{\bm{t}}_\ell \|^2} \sigma_{\min}(C_\ell),
\end{equation*}
which yields:
\begin{equation}\label{bound_z_l}
\frac{\max_{i} |\tilde{z}_{i \ell}|}{\| \tilde{\bm{t}}_\ell \|} \leq \frac{\|\tilde{\bm{z}}_\ell\|}{\| \tilde{\bm{t}}_\ell \|} \leq \sigma_{\min}^{-1/2}(C_\ell).
\end{equation}
So the query complexity for implementing $C(\tilde{R}_\ell)$ could be bounded as $O(\ell \sigma_{\min}^{-1/2}(C_\ell) {\epsilon_R^{-1}})$. {By considering the distance between $R_\ell$ and $\tilde{R}_\ell$ in Eq.~(\ref{proof_4.1_R_l_main_2_6}), we could then implement the controlled version of the gate $R_\ell$ with error bounded by {$\epsilon_R$}.}
Now we have proved Lemma~\ref{4.1_R_l_main}.

\end{proof}

\section{ Proof of Lemma~\ref{bound_sigma_noisy} }
\label{app_min_singular}
{In this section, we prove Lemma~\ref{bound_sigma_noisy}.
Before we detail main technical procedures, we first provide some useful theoretical bounds in Lemma~\ref{bound_on_P_l} and Lemma~\ref{bound_sigma}.}

\begin{lemma}\label{bound_on_P_l}
The probability $P_\ell$ defined in Eq.~(\ref{definition_P_l}) is bounded by $$\frac{\sum_{i=\ell}^{r} \sigma_i^2}{\|A\|_F^2} \leq P_\ell \leq \frac{\sum_{i=1}^{r-\ell+1} \sigma_i^2}{\|A\|_F^2}, $$ where $\sigma_1 \geq \sigma_2 \geq \cdots \geq \sigma_r$ are singular values of $A$.
\end{lemma}

\begin{proof}
Denote the singular value decomposition
$$ A = \sum_{i=1}^r \sigma_i \bm{u}_i \bm{v}_i^T. $$
Since the state $|\bm{t}_i\>$ is the linear sum of rows $\{A_{j}\}_{j=1}^{m}$, while each row is the linear sum of singular vectors:
\begin{equation}\label{P_l_Aj}
A_{j} = \sum_{i=1}^{r} \sigma_i u_i^{(j)} \bm{v}_i,
\end{equation}
we can further write:
\begin{equation}\label{P_l_ti}
|\bm{t}_i\> = \sum_{j=1}^r w_{ij}|\bm{v}_j\>.
\end{equation}
Rewrite Eq.~(\ref{definition_P_l}) as:
\begin{align}
P_\ell &= \frac{1}{\|A\|_F^2} \sum_{j=1}^{m} \left[ \|A_{j}\|^2 -\sum_{i=1}^{\ell-1} \|A_{j}\|^2 |\<\bm{t}_i|A_{j}\>|^2 \right]\label{P_l_1_2}\\
&= 1-\frac{1}{\|A\|_F^2} \sum_{j=1}^{m} \sum_{i=1}^{\ell-1} \left[ \sum_{k=1}^{r} w_{ik} \sigma_k u_k^{(j)} \right]^2 \label{P_l_1_3},
\end{align}
where Eq.~(\ref{P_l_1_3}) comes from Eq.~(\ref{P_l_Aj}) and Eq.~(\ref{P_l_ti}). Expand the square term in Eq.~(\ref{P_l_1_3}) yields: 
\begin{align}
P_\ell &= 1-\frac{1}{\|A\|_F^2} \sum_{j=1}^{m} \sum_{i=1}^{\ell-1} \Bigg[ \sum_{k=1}^{r} w_{ik}^2 \sigma_k^2 ({u_k^{(j)}})^2 \nonumber \\
&{}\ \ \ \ \ \ \ \ + \sum_{k \neq k'}^{r} w_{ik}w_{ik'} \sigma_k \sigma_{k'} u_k^{(j)}u_{k'}^{(j)} \Bigg]\\
&= 1-\frac{1}{\|A\|_F^2} \sum_{i=1}^{\ell-1} \sum_{k=1}^r w_{ik}^2 \sigma_k^2 \label{P_l_2_2}\\
&= 1-\frac{1}{\|A\|_F^2} \sum_{k=1}^{r} c_k \sigma_k^2, \label{P_l_2_3}
\end{align}
where Eq.~(\ref{P_l_2_2}) follows because $\sum_{j=1}^m u_k^{(j)} u_{k'}^{(j)} = \bm{u}_k^T \bm{u}_{k'} = \delta_{kk'}$, and  we denote $c_k=\sum_{i=1}^{\ell-1} w_{ik}^2$ in Eq.~(\ref{P_l_2_3}). 

Define the $r$-dimensional vector $\bm{w}_i=\sum_{k=1}^r w_{ik}\bm{e}_k$. Since $\<\bm{t}_i|\bm{t}_j\> = \delta_{ij} = \sum_{k=1}^r w_{ik} w_{jk} = \bm{w}_i^T \bm{w}_j$, vectors in set $\{\bm{w}_i\}_{i=1}^{\ell-1}$ are orthogonal  with each other. We can add $\bm{w}_{\ell},\cdots \bm{w}_{r}$ such that $\{\bm{w}_i\}_{i=1}^{r}$ forms an orthonormal basis in the $r$-dimensional space. Denote the matrix $W=(\bm{w}_1,\bm{w}_2,\cdots,\bm{w}_r)$. Since $W^{T}W=I$, we have:
\begin{equation}\label{P_l_ck1}
0 \leq c_k = \sum_{i=1}^{\ell-1} w_{ik}^2 \leq \sum_{i=1}^r w_{ik}^{2}         = [WW^T]_{kk} = 1, \forall k \in [r].	
\end{equation}
Note that 
\begin{equation}\label{P_l_ck2}
\sum_{k=1}^{r} c_k = \sum_{i=1}^{\ell-1} \sum_{k=1}^r w_{ik}^2 = \sum_{i=1}^{\ell-1} [WW^T]_{ii}=\ell-1.
\end{equation}
Hence by using Eqs.~(\ref{P_l_2_3}-\ref{P_l_ck2}) and $\|A\|_F^2 = \sum_{i=1}^r \sigma_i^2$, we could obtain the lower and upper bounds for $P_\ell$ as follows.
\begin{align}
P_\ell &\geq 1-\frac{1}{\|A\|_F^2}\sum_{i=1}^{\ell-1} \sigma_i^2 = \frac{\sum_{i=\ell}^{r} \sigma_i^2}{\|A\|_F^2}, \label{lower_bound_P}\\
P_\ell	&\leq 1-\frac{1}{\|A\|_F^2} \sum_{i=r-\ell+2}^r \sigma_i^2 = \frac{\sum_{i=1}^{r-\ell+1} \sigma_i^2}{\|A\|_F^2}.\label{upper_bound_P}
\end{align} 

\end{proof}

\begin{lemma}\label{bound_sigma}
Denote $P$ to be the distribution of the adaptive sampling following from the Eq.~(\ref{sample_prop}):
\begin{equation}\label{eq_lemma5_tmp1}
P(s_1,\cdots s_{\ell}) = {\rm{Pr}}^{(1)}(s_1) {\rm{Pr}}^{(2)}(s_2) \cdots {\rm{Pr}}^{(\ell)}(s_\ell) ,
\end{equation}
where $s_{\ell} \in [m]$ denotes the index of the row $\bm{s}_{\ell}$ in the matrix $A\in \R^{m \times d}$. 
Then
$$\mathbb{E}_P\left[ \sigma_{\min} (C_{\ell}) \right] \geq \frac{r-\ell+1}{\ell r} \kappa^{2-2\ell}. $$
\end{lemma}

\begin{proof}
By the Cauchy-Schwarz Inequality, we have:
\begin{equation}
\mathbb{E}_P [\sigma_{\min} (C_{\ell})] \cdot \mathbb{E}_P [\sigma_{\min}^{-1} (C_{\ell})] \geq 1.
\end{equation}
If the following inequality were true,
\begin{equation}\label{bound_sigma_inverse_eq}
\mathbb{E}_P\left[ \sigma_{\min}^{-1} (C_{\ell}) \right] \leq \frac{\ell r}{r-\ell+1} \kappa^{2\ell-2},
\end{equation} 
then we could reach the conclusion of this lemma:
\begin{align}
\mathbb{E}_P [\sigma_{\min} (C_{\ell})] &\geq \frac{1}{\mathbb{E}_P [\sigma_{\min}^{-1} (C_{\ell})]} 	\\ 
&\geq \frac{r-\ell+1}{\ell r} \kappa^{2-2\ell}.
\end{align}

To prove Eq.~(\ref{bound_sigma_inverse_eq}), we first rewrite it as follows:
\begin{align}
&{}\ \ \ \ \mathbb{E}_{P}	[\sigma_{\min}^{-1}(C_{\ell})] \nonumber \\
&= \sum_{s_1=1}^{m}\cdots \sum_{s_{\ell}=1}^{m} P(s_1, \cdots, s_{\ell}) \sigma_{\min}^{-1}(C_{\ell}) \\
&= \sum_{s_1 =1}^m \cdots \sum_{s_{\ell}=1}^m \frac{\|\bm{s}_1\|^2 \cdots \|\bm{s}_{\ell}\|^2 }{\Sigma^{(1)} \cdots \Sigma^{(\ell)}}	|C_{\ell}| \sigma_{\min}^{-1}(C_{\ell}) \label{sigma_inverse_2_1}.
\end{align}
In Eq.~(\ref{sigma_inverse_2_1}), we rewrite $P(s_1, \cdots, s_{\ell})$ with Eq.~(\ref{eq_lemma5_tmp1}) and  
\begin{align}
\text{Pr}^{(\ell)}(s_{\ell}) 
&= \frac{\|\bm{s}_{\ell}\|^2  \| |\bm{s}_{\ell}\> - \sum_{i=1}^{\ell-1} |\bm{t}_i\>\<\bm{t}_i|\bm{s}_{\ell}\> \|^2}{\Sigma^{(\ell)}} \label{sigma_inverse_0_1}\\
&= \frac{\|\bm{s}_{\ell}\|^2 Z_{\ell}^2}{\Sigma^{(\ell)}} \label{sigma_inverse_0_2}\\
&= \frac{\|\bm{s}_{\ell}\|^2}{\Sigma^{(\ell)}} \frac{|C_{\ell}|}{|C_{\ell-1}|}, \label{sigma_inverse_0_3}
\end{align}
where, in Eq.~(\ref{sigma_inverse_0_1}), we denote 
\begin{equation}\label{sigma_inverse_normalization}
\Sigma^{(\ell)} = \sum_{s_{\ell}=1}^m \| \bm{s}_{\ell} - \sum_{i=1}^{\ell-1} \bm{t}_i \bm{t}_i^T \bm{s}_\ell \|^2,
\end{equation}
Eq.~(\ref{sigma_inverse_0_2}) is derived from $Z_\ell^2 = \||\bm{s}_\ell\>-\sum_{i=1}^{\ell-1}|\bm{t}_i\>\<\bm{t}_i|\bm{s}_\ell\>\|^2$, and Eq.~(\ref{sigma_inverse_0_3}) is due to Eq.~(\ref{value_Z_l}).

Continuing from Eq.~(\ref{sigma_inverse_2_1}), it holds 
\begin{align}
&{}\ \ \ \ \mathbb{E}_{P}	[\sigma_{\min}^{-1}(C_{\ell})] \nonumber \\
&\leq \sum_{s_1 =1}^m \cdots \sum_{s_{\ell}=1}^m \frac{\|\bm{s}_1\|^2 \cdots \|\bm{s}_{\ell}\|^2 }{\Sigma^{(1)} \cdots \Sigma^{(\ell)}}	\sum_{i=1}^{\ell} |C_{\ell}^{(i)}| \label{sigma_inverse_2_2} \\
&= \sum_{i=1}^{\ell} \sum_{s_i=1}^m \|\bm{s}_i\|^2 \sum_{s_j=1, j \neq i}^m \frac{\prod_{j=1,j \neq i}^{\ell} \|\bm{s}_j\|^2 }{\Sigma^{(1)} \cdots \Sigma^{(\ell)}} |C_{\ell}^{(i)}|, \label{sigma_inverse_2_3}
\end{align}
 where Eq.~(\ref{sigma_inverse_2_2}) uses 
\begin{equation*}\label{sigma_inverse_1}
\sigma_{\min}^{-1}(C_{\ell})	= \sigma_{\max}(C_{\ell}^{-1}) \leq \text{Tr} (C_{\ell}^{-1}) = \frac{\sum_{i=1}^{\ell} |C_{\ell}^{(i)}|}{|C_{\ell}|},
\end{equation*}
with $C_{\ell}^{(i)} \in \R^{(\ell-1) \times (\ell-1)}$ being the principal submatrix of $C_{\ell}$ by removing the $i$-th row and column, and Eq.~(\ref{sigma_inverse_2_3}) follows by rearranging the sum order.

Next, we will provide a lower bound on the denominator term $\Sigma^{(1)} \cdots \Sigma^{(\ell)}$ in Eq.~(\ref{sigma_inverse_2_3}). Note that for any $1\leq j \leq \ell$, $\Sigma^{(j)}$ only depends on the matrix $A$ and indices $({s}_1, {s}_2,\cdots,{s}_{j-1})$, so it can be viewed as the function of $({s}_1, {s}_2,\cdots,{s}_{j-1})$ when treating $A$ as the constant matrix, namely,
\begin{align}
\Sigma^{(j)} &:= \Sigma^{(j)}({s}_1, {s}_2,\cdots,{s}_{j-1}) \\
&= \|A\|_F^2 P_j, \label{sigma_inverse_Sigma}
\end{align} 
where Eq.~(\ref{sigma_inverse_Sigma}) comes from the definition of $P_j$ in Eq.~(\ref{definition_P_l}). 
By employing the lower and upper bounds of $P_j$ in {Eqs.~(\ref{lower_bound_P}) and (\ref{upper_bound_P})}, we could bound the function $\Sigma^{(j)}$ as
\begin{align}
\sum_{i=j}^{r} \sigma_{i}^2 (A) \leq \Sigma^{(j)} \leq \sum_{i=1}^{r-j+1} \sigma_{i}^2 (A), \label{sigma_inverse_4_0}
\end{align} 
where {$\sigma_1(A) \geq \sigma_2(A) \geq \cdots \geq \sigma_r(A)$ denote singular values of $A$, and } Eq.~(\ref{sigma_inverse_4_0}) holds for any choice of linearly independent row vectors for $\Sigma^{(j)}$. 
{Then Eq.~(\ref{sigma_inverse_4_0}) yields }
\begin{align}
&{}\ \ \ \ \Sigma^{(j)}({s}_1,\cdots,{s}_{j-1}) \nonumber \\
& \geq \sum_{i=j}^{r} \sigma_{i}^2 (A) \label{sigma_inverse_4_1}\\
& \geq \sum_{i=j}^{r} \sigma_{i}^2 (A) \frac{\Sigma^{(j)}({s}_1,\cdots, {s}_{i-1}, {s}_{i+1}, \cdots, {s}_j)}{\sum_{i=1}^{r-j+1} \sigma_{i}^2 (A)}  \label{sigma_inverse_4_2}\\
& \geq \frac{\sigma_{\min}^2(A)}{\sigma_{\max}^2(A)} \Sigma^{(j)}({s}_1, \cdots, {s}_{i-1}, {s}_{i+1}, \cdots, {s}_j) \label{sigma_inverse_4_3}
\end{align}
where Eq.~(\ref{sigma_inverse_4_2}) holds true because of the second inequality in Eq.~(\ref{sigma_inverse_4_0}):
\begin{equation}
\frac{\Sigma^{(j)}({s}_1,\cdots, {s}_{i-1}, {s}_{i+1}, \cdots, {s}_j)}{\sum_{i=1}^{r-j+1} \sigma_{i}^2 (A)}\leq 1.
\end{equation}

Continuing from Eq.~(\ref{sigma_inverse_2_3}) and with the inequality $\Sigma^{(\ell)} \geq \sum_{k=\ell}^r \sigma_k^2(A)$ in Eq.~(\ref{sigma_inverse_4_0}), we obtain the first inequality below:
\begin{align}
&{}\ \ \ \ \mathbb{E}_{P}	[\sigma_{\min}^{-1}(C_{\ell})] \nonumber \\
\leq &{} \sum_{i=1}^{\ell} \sum_{s_i=1}^m {\|\bm{s}_i\|^2} \sum_{s_j=1, j \neq i}^m \frac{\prod_{j=1,j \neq i}^{\ell} \|\bm{s}_j\|^2 }{\Sigma^{(1)} \cdots \Sigma^{(\ell-1)} \sum_{k=\ell}^{r} \sigma_k^2(A)} |C_{\ell}^{(i)}| \label{sigma_inverse_5_4}\\
\leq &{} \left(\frac{\sigma_{\max}^2(A)}{\sigma_{\min}^2(A)} \right)^{\ell-2} \sum_{i=1}^{\ell} \sum_{s_i=1}^m \frac{\|\bm{s}_i\|^2}{\sum_{k=\ell}^{r} \sigma_{k}^2 (A)} \nonumber \\
&{}\ \ \ \ \ \ \ \ \cdot \sum_{s_j=1, j \neq i}^m \frac{\prod_{j=1,j \neq i}^{\ell} \|\bm{s}_j\|^2 }{\Sigma^{'(1)} \cdots \Sigma^{'(\ell-1)}} |C_{\ell}^{(i)}| \label{sigma_inverse_5_3},
\end{align}
where in Eq.~(\ref{sigma_inverse_5_3}) we denote 
\begin{equation}
\Sigma^{'(j)} = \left\{
\begin{aligned}
& \Sigma^{(j)} ({s}_1, {s}_2, \cdots, {s}_{j-1}), \forall j<i+1, \\
&\Sigma^{(j)} ({s}_1, {s}_2, \cdots, {s}_{i-1}, {s}_{i+1}, \cdots, {s}_{j}), \forall j\geq i+1.
\end{aligned}
\right.
\end{equation}
and employ Eq.~(\ref{sigma_inverse_4_3}).
Notice that in Eq.~(\ref{sigma_inverse_5_3}), 
\begin{equation}
\sum_{s_j=1, j \neq i}^m \frac{\prod_{j=1,j \neq i}^{\ell} \|\bm{s}_j\|^2 }{\Sigma^{'(1)} \cdots \Sigma^{'(\ell-1)}} |C_{\ell}^{(i)}| =1
\end{equation}
which can be interpreted as the probability for sampling $({s}_1, {s}_2, \cdots, {s}_{i-1}, {s}_{i+1}, \cdots, {s}_{\ell})$ over all choice of indices. Finally, Eq.~(\ref{sigma_inverse_5_3}) further leads to
\begin{align}
&{}\ \ \ \ \mathbb{E}_{P}	[\sigma_{\min}^{-1}(C_{\ell})] \nonumber \\
&\leq  \left(\frac{\sigma_{\max}^2(A)}{\sigma_{\min}^2(A)} \right)^{\ell-2} \sum_{i=1}^{\ell} \sum_{s_i=1}^m \frac{\|\bm{s}_i\|^2}{\sum_{k=\ell}^{r} \sigma_{k}^2 (A)} \\
&\leq \frac{\ell\|A\|_F^2}{(r-\ell+1)\sigma_{\min}^2(A)} \left(\frac{\sigma_{\max}^2(A)}{\sigma_{\min}^2(A)} \right)^{\ell-2} \label{sigma_inverse_6_2}\\
&\leq \frac{\ell r}{r-\ell+1} \left(\frac{\sigma_{\max}^2(A)}{\sigma_{\min}^2(A)} \right)^{\ell-1}\label{sigma_inverse_6_3}\\
&= \frac{\ell r}{r-\ell+1} \kappa^{2\ell-2},
\end{align}
where $\sum_{s_i=1}^m \|\bm{s}_i\|^2 = \|A\|_F^2$ and $\sum_{k=\ell}^r \sigma_k^2(A) \geq (r-\ell+1) \sigma_{\min}^2(A)$ are used to derive Eq.~(\ref{sigma_inverse_6_2}), and $\|A\|_F^2 = \sum_{i=1}^r \sigma_i^2(A) \leq r \sigma_{\max}^2(A)$ is used to derive Eq.~(\ref{sigma_inverse_6_3}).

\end{proof}

Instead of the distribution $P$ defined in Eq.~(\ref{eq_lemma5_tmp1}), the perturbed distribution $\tilde{P}$ is employed due to noisy gates $\tilde{R}_i, \forall i \in [r]$ in Algorithm~\ref{Basis}.
{For simplicity, we denote $\Pi_\ell=\prod_{i=1}^{\ell}R_i$ and $\tilde{\Pi}_\ell=\prod_{i=1}^{\ell} \tilde{R}_i, \forall \ell \in [r]$, then the sampling distributions could be rewritten as 
\begin{align}
{\text{Pr}}^{(\ell)}(s_{\ell})
&:= {P}(s_{\ell}|s_1, \cdots, s_{\ell-1})  \nonumber \\ 
&= \frac{1}{{\Sigma}^{(\ell)}} \|\bm{s}_{\ell}\|^2 \left\| \frac{{\Pi}_{\ell-1}+I}{2}|\bm{s}_{\ell}\> \right\|^2, \label{eq_P} \\
\tilde{\text{Pr}}^{(\ell)}(s_{\ell}) &:= \tilde{P}(s_{\ell}|s_1, \cdots, s_{\ell-1})\nonumber \\
& = \frac{1}{\tilde{\Sigma}^{(\ell)}} \|\bm{s}_{\ell}\|^2 \left\| \frac{\tilde{\Pi}_{\ell-1}+I}{2}|\bm{s}_{\ell}\> \right\|^2, \label{eq_noisyP}
\end{align}
where 
\begin{align}
{\Sigma}^{(\ell)} &= \sum_{s_{\ell}=1}^{m}\|\bm{s}_{\ell}\|^2 \left\| \frac{{\Pi}_{\ell-1}+I}{2}|\bm{s}_{\ell}\> \right\|^2 \label{eq_sigma}\\
\tilde{\Sigma}^{(\ell)} &= \sum_{s_{\ell}=1}^{m}\|\bm{s}_{\ell}\|^2 \left\| \frac{\tilde{\Pi}_{\ell-1}+I}{2}|\bm{s}_{\ell}\> \right\|^2 \label{eq_noisysigma}
\end{align}
are corresponding normalization factors.}
Now we prove Lemma~\ref{bound_sigma_noisy}.

\begin{proof}

{The main idea is that, if the following statement holds true for any $0\leq j \leq \ell-1$:
\begin{align}
& \mathop{\mathbb{E}}\limits_{\tilde{\text{Pr}}^{(j+1)}} \mathop{\mathbb{E}}\limits_{{\text{Pr}}^{(j+2)}} \cdots \mathop{\mathbb{E}}\limits_{{\text{Pr}}^{(\ell)}} [\sigma_{\min}({C}_{\ell})] \geq \nonumber \\
&  (1-\frac{1}{6\ell}) \mathop{\mathbb{E}}\limits_{{\text{Pr}}^{(j+1)}} \cdots \mathop{\mathbb{E}}\limits_{{\text{Pr}}^{(\ell)}} [\sigma_{\min}({C}_{\ell})] - \frac{1}{6\ell} \mathbb{E}_P [\sigma_{\min}(C_{\ell})],\label{middle_4_5}
\end{align}
then we could provide a lower bound on the expectation of $\sigma_{\min}({C}_{\ell})$ with the distribution $\tilde{P}$ inductively.} Specifically, we could obtain
\begin{align}
&{}\ \ \ \ \ \mathbb{E}_{\tilde{P}}	[\sigma_{\min}({C}_{\ell})]= \mathop{\mathbb{E}}\limits_{\tilde{\text{Pr}}^{(1)}} \mathop{\mathbb{E}}\limits_{\tilde{\text{Pr}}^{(2)}} \cdots \mathop{\mathbb{E}}\limits_{\tilde{\text{Pr}}^{(\ell)}}[ \sigma_{\min}({C}_{\ell}) ]\nonumber \\
&\geq  \mathop{\mathbb{E}}\limits_{\tilde{\text{Pr}}^{(1)}} \cdots \mathop{\mathbb{E}}\limits_{\tilde{\text{Pr}}^{(\ell-1)}} (1-\frac{1}{6\ell}) \mathop{\mathbb{E}}\limits_{\text{Pr}^{(\ell)}} [\sigma_{\min}({C}_{\ell})] \nonumber \\ 
&\qquad\qquad\qquad  - \frac{1}{6\ell} \mathbb{E}_{{P}} [\sigma_{\min} (C_{\ell})]\label{eq_lemma501}  \\
&\geq \mathop{\mathbb{E}}\limits_{\tilde{\text{Pr}}^{(1)}}\cdots \mathop{\mathbb{E}}\limits_{\tilde{\text{Pr}}^{(\ell-2)}} (1-\frac{1}{6\ell})^2 \mathop{\mathbb{E}}\limits_{{\text{Pr}}^{(\ell-1)}}  \mathop{\mathbb{E}}\limits_{\text{Pr}^{(\ell)}} [\sigma_{\min}({C}_{\ell})] \nonumber \\
&{}\qquad\qquad - \frac{1}{6\ell} \mathbb{E}_{{P}} [\sigma_{\min} (C_{\ell})] - \frac{1}{6\ell} \mathbb{E}_{{P}} [\sigma_{\min} (C_{\ell})] \label{eq_lemma502} \\
{}& \vdots \nonumber \\
&\geq (1-\frac{1}{6\ell})^\ell \mathop{\mathbb{E}}\limits_{{\text{Pr}}^{(1)}} \cdots \mathop{\mathbb{E}}\limits_{{\text{Pr}}^{(\ell)}} [\sigma_{\min}({C}_{\ell})] - \frac{\ell}{6\ell} \mathbb{E}_{{P}} [\sigma_{\min}(C_{\ell})] \label{eq_lemma503} \\
&\geq \frac{2}{3} \mathbb{E}_{{P}} [\sigma_{\min}(C_{\ell})], \nonumber
\end{align}
where $\tilde{\text{Pr}}^{(j)}$ is defined in Eq.~(\ref{eq_noisyP}), 
Eqs.~(\ref{eq_lemma501})-(\ref{eq_lemma503}) follow from Eq.~(\ref{middle_4_5}), and we employ 
$$\left(1-\frac{1}{6\ell}\right)^\ell \geq  1-\frac{1}{6} $$ to obtain the last inequality.

To prove Eq.~(\ref{middle_4_5}), we need a lower bound on the distribution $\tilde{\text{Pr}}^{(j+1)}$, which could be derived as follows. 
\begin{align}
&{}\ \ \ \ \tilde{P}^{(j+1)}(s_{j+1}) \nonumber \\
&= \frac{\|\bm{s}_{j+1}\|^2 \<\bm{s}_{j+1}|\frac{2I+\tilde{\Pi}_{j} + \tilde{\Pi}_{j}^\dag}{4} |\bm{s}_{j+1}\> }{ \sum_{s_{j+1}=1}^{m} \|\bm{s}_{j+1}\|^2 \<\bm{s}_{j+1}|\frac{2I+\tilde{\Pi}_{j} + \tilde{\Pi}_{j}^\dag}{4} |\bm{s}_{j+1}\> } \label{noise_distribution_1} \\
&\geq \frac{\|\bm{s}_{j+1}\|^2 \big( \<\bm{s}_{j+1}|\frac{\Pi_j +I}{2} |\bm{s}_{j+1}\> -\|\frac{\tilde{\Pi}_j - \Pi_j}{2}\| \big) }{ \sum_{s_{j+1}=1}^{m} \|\bm{s}_{j+1}\|^2 \big( \<\bm{s}_{j+1}|\frac{\Pi_j +I}{2} |\bm{s}_{j+1}\> + \|\frac{\tilde{\Pi}_j - \Pi_j}{2}\| \big) }  \label{noise_distribution_2} \\
&= \frac{ \Sigma^{(j+1)} \text{Pr}^{(j+1)}(s_{j+1}) - \|\bm{s}_{j+1}\|^2  \|\frac{\tilde{\Pi}_j - \Pi_j}{2}\| }{ \Sigma^{(j+1)} + \|A\|_F^2   \|\frac{\tilde{\Pi}_j - \Pi_j}{2}\|  }  \label{noise_distribution_3} \\
&\geq \frac{ \Sigma^{(j+1)} \text{Pr}^{(j+1)}(s_{j+1}) - \|\bm{s}_{j+1}\|^2 \frac{j {\epsilon_R}}{2} }{ \Sigma^{(j+1)} + \|A\|_F^2   \frac{j {\epsilon_R}}{2} } \label{noise_distribution_4},
\end{align}
where Eq.~(\ref{noise_distribution_1}) is derived by using Eqs.~(\ref{eq_noisyP}) and (\ref{eq_noisysigma}). 
{Eq.~(\ref{noise_distribution_2}) is obtained by noticing 
\begin{equation*}
- \| \frac{\tilde{\Pi}_j - \Pi_j}{2} \| \leq \<\bm{s}_{j+1}| \frac{\tilde{\Pi}_j - \Pi_j}{2} |\bm{s}_{j+1}\> \leq \| \frac{\tilde{\Pi}_j - \Pi_j}{2} \|. 
\end{equation*} }
Eq.~(\ref{noise_distribution_3}) is derived by using Eqs.~(\ref{eq_P}) and (\ref{eq_sigma}). 
Eq.~(\ref{noise_distribution_4}) is derived by noticing 
$$\|\tilde{\Pi}_j-{\Pi}_j\| \leq \sum_{i=1}^{j} \|\tilde{R}_i - R_i\| \leq j {\epsilon_R},$$
{where we denote by {$\epsilon_R = \frac{1}{3r^5\kappa^{2r}}$} the error bound on each $R_i$, as provided in the assumption of this Lemma.
Notice that
\begin{equation}\label{upper_bound_sigma_min}
0\leq \sigma_{\min}(C_\ell) \leq \frac{\text{Tr}[C_\ell]}{\ell} \leq 1
\end{equation}
holds for any choice of row vectors.
}
For simplicity, in Eq.~(\ref{middle_4_5}), we denote 
$$ X := \mathop{\mathbb{E}}\limits_{{\text{Pr}}^{(j+2)}} \cdots \mathop{\mathbb{E}}\limits_{{\text{Pr}}^{(\ell)}} [\sigma_{\min}({C}_{\ell})] \in [0,1], $$
and proceed as follows
\begin{align}
&{}\ \ \ \ \mathop{\mathbb{E}}\limits_{\tilde{\text{Pr}}^{(j+1)}} \mathop{\mathbb{E}}\limits_{{\text{Pr}}^{(j+2)}} \cdots \mathop{\mathbb{E}}\limits_{{\text{Pr}}^{(\ell)}} [\sigma_{\min}({C}_{\ell})] = \mathop{\mathbb{E}}\limits_{\tilde{\text{Pr}}^{(j+1)}}  [X]\nonumber \\
&= \sum_{s_{j+1}=1}^{m} \tilde{P}^{(j+1)}(s_{j+1}) \cdot X \label{expect_C_l_noisy_1} \\
&\geq \sum_{s_{j+1}=1}^{m} \left(\frac{ \Sigma^{(j+1)} \text{Pr}^{(j+1)}(s_{j+1}) - \|\bm{s}_{j+1}\|^2 \frac{j {\epsilon_R}}{2} }{ \Sigma^{(j+1)} + \|A\|_F^2   \frac{j {\epsilon_R}}{2} }\right) X \label{expect_C_l_noisy_1}
\end{align}
where the inequality employs Eq.~(\ref{noise_distribution_4}). Using the identity $\sum_{s_{j+1}=1}^{m}  \|\bm{s}_{j+1}\|^2 = \|A\|_F^2$ and {$X \leq 1$}, Eq.~(\ref{expect_C_l_noisy_1}) further yields 
\begin{align}
&\geq \frac{\Sigma^{(j+1)} \mathop{\mathbb{E}}\limits_{{\text{Pr}}^{(j+1)}}[ X]}{\Sigma^{(j+1)} + \|A\|_F^2 \frac{j {\epsilon_R}}{2}} - \frac{\|A\|_F^2 \frac{j {\epsilon_R}}{2}}{\Sigma^{(j+1)} + \|A\|_F^2 \frac{j {\epsilon_R}}{2}} \label{expect_C_l_noisy_2} \\
&= \frac{P_{j+1} }{P_{j+1} + \frac{j {\epsilon_R}}{2}} \mathop{\mathbb{E}}\limits_{{\text{Pr}}^{(j+1)}} [X] - \frac{  \frac{j {\epsilon_R}}{2}}{P_{j+1} + \frac{j {\epsilon_R}}{2}} \label{expect_C_l_noisy_3} \\
&\geq \frac{P_{j+1} }{P_{j+1} + \frac{j {\epsilon_R}}{2}} \mathop{\mathbb{E}}\limits_{{\text{Pr}}^{(j+1)}} [X] - \frac{1}{6\ell} \mathbb{E}_P [\sigma_{\min}(C_{\ell})] \label{expect_C_l_noisy_4} \\
&\geq (1-\frac{1}{6\ell}) \mathop{\mathbb{E}}\limits_{{\text{Pr}}^{(j+1)}} [X] - \frac{1}{6\ell} \mathbb{E}_P [\sigma_{\min}(C_{\ell})]. \label{expect_C_l_noisy_5}
\end{align}
Eq.~(\ref{expect_C_l_noisy_3}) is obtained by using $\Sigma^{(j+1)} = \|A\|_F^2 P_{j+1}$ in Eq.~(\ref{sigma_inverse_Sigma}). Eq.~(\ref{expect_C_l_noisy_4}) is derived by noticing that 
\begin{align}
\frac{\frac{j {\epsilon_R}}{2}}{P_{j+1} + \frac{j {\epsilon_R}}{2}} &\leq \frac{\frac{j \kappa^{-2r}}{6r^5} }{\frac{r-j}{r}\kappa^{-2}} \\
&= \frac{j}{r(r-j)} \frac{\kappa^{2-2r}}{6r\cdot r^2} \label{expect_C_l_noisy_6} \\
&\leq \frac{1}{6\ell}  \mathbb{E}_P [\sigma_{\min}(C_{\ell})].\label{expect_C_l_noisy_7}
\end{align}
The first inequality follows from  ${\epsilon_R} = \frac{1}{3r^5\kappa^{2r}}$ and 
\begin{equation}\label{lower_bound_P_j}
P_{j+1} \geq \frac{\sum_{i=j+1}^{r} \sigma_i^2(A)}{\|A\|_F^2} \geq \frac{(r-j)\sigma_{\min}^2(A)}{r \sigma_{\max}^2(A)} =  \frac{r-j}{r} \kappa^{-2},
\end{equation}
where the first inequality uses Lemma~\ref{bound_on_P_l}.
Eq.~(\ref{expect_C_l_noisy_7}) holds due to the lower bound in Lemma~\ref{bound_sigma}.

Finally, Eq.~(\ref{expect_C_l_noisy_5}) is derived by using 
\begin{align}
\frac{P_{j+1}}{P_{j+1} + \frac{j {\epsilon_R}}{2}} &\geq 1- \frac{1}{6\ell} \mathbb{E}_P [\sigma_{\min}(C_{\ell})] \geq 1-\frac{1}{6\ell},
\end{align}
which holds due to Eqs.~(\ref{expect_C_l_noisy_7}) and (\ref{upper_bound_sigma_min}).

\end{proof}

\section{Proof of Theorem~\ref{Basis_theorem_16}}
\label{4.1_error_runtime}

\begin{proof}

We sketch the main idea of the proof first. 
We could implement Algorithm~\ref{Basis} for $N$ times to guarantee sampling out one basis which satisfies the conditions
\begin{equation}\label{aim_lower_bound}
{\rm cond}^{(\ell)}: \left\{ \sigma_{\min}(C_\ell) \geq \frac{1}{2 r^2 \kappa^{2r-2}}  \right\}, \forall \ell \in [r].
\end{equation}
Let $T_{\rm QGSP}$ be the query complexity of oracles $U_A$ and $V_A$ to implement Algorithm~\ref{Basis} once.
Thus, the overall query complexity is 
\begin{equation}\label{4.1_error_runtime_overall_complexity}
T_{\text{basis}} = N T_{\rm QGSP}.
\end{equation}

To begin with, consider the first iteration of Algorithm~\ref{Basis}. The Gram matrix of the sampled basis has the dimension $1\times 1$ with one element $1$. Thus, the condition ${\rm cond}^{(1)}$ always holds. 
We proceed to the general cases inductively. Suppose that a basis with $(\ell-1)$ rows, which satisfies the condition ${\rm cond}^{(\ell-1)}$ in Eq.~(\ref{aim_lower_bound}), has been obtained. 
Next, we move on to the $\ell$-th iteration of Algorithm~\ref{Basis}. We accept the newly sampled row as part of the basis, if the condition ${\rm cond}^{(\ell)}$ holds,  and proceed to the $\ell+1$-th iteration. If the condition is violated, we stop the procedure and repeat Algorithm~\ref{Basis} from the first iteration.
Thus, the conditions in Eq.~(\ref{aim_lower_bound}) would hold during the procedure, with the cost that Algorithm~\ref{Basis} needs to be run $N$ number of times in order to guarantee one basis obtained with high probability.

Now we analyze the complexity of the procedure in detail.
Notice that $T_{\rm QGSP}$ consists of three parts: the cost of oracles $U_A$ and $V_A$ for encoding all rows of the input matrix $A$, the cost of Hadamard Test for calculating coefficients $\{\bm{z}_{\ell}\}_{\ell=1}^{r-1}$, and the cost of implementing gates $\{C(R_\ell)\}_{\ell=1}^{r-1}$. Based on Lemma~\ref{4_1_lemma_error_t_l} and Lemma~\ref{4.1_R_l_main}, the latter two complexities depend on the error in the implementation of $R_\ell$. 
In the following proof, we provide explicit upper bounds of $N$ and $T_{\rm QGSP}$, 
by setting 
\begin{align}\label{4.1_error_runtime_error_C}
\epsilon_C &= \frac{1}{960r^{\frac{23}{2}} \kappa^{6r}} 
\end{align} 
to be the error bound of each element in $C_{r}$.

Firstly we demonstrate that the sampling in each iteration of Algorithm~\ref{Basis} obeys the distribution in Eq.~(\ref{main_distribution_P_noisy}),
i.e., the error of each gate $C(R_j)$ is bounded as
\begin{align}\label{4.1_error_runtime_error_R}
\| \tilde{R}_j - R_j \| \leq {\epsilon_{R} }&= \frac{1}{3r^{5} \kappa^{2r}}, \forall j \in [r-1].
\end{align}
Based on {Lemma~\ref{4_1_lemma_error_t_l}}, the error of $\bm{t}_j$ induced by noisy coefficients is bounded by 
\begin{align}
\| \tilde{\bm{t}}_j -\bm{t}_j\| &\leq \frac{8j^{\frac{5}{2}}}{\sigma_{\min}^2 (C_j)} \epsilon_C \label{4.1_error_runtime_1_1} \\
&\leq 32r^{\frac{13}{2}} \kappa^{4r} \epsilon_C \label{4.1_error_runtime_1_3} \\
&= \frac{1}{30r^5 \kappa^{2r}} \label{4.1_error_runtime_1_4} \\
&= \frac{\epsilon_R}{10}, \label{4.1_error_runtime_1_5}
\end{align}
for $j \in [r-1]$. Eq.~(\ref{4.1_error_runtime_1_1}) follows from Eq.~(\ref{eq_norm_error_t_l}). 
Since the condition ${\rm cond}^{(j)}$ (\ref{aim_lower_bound}) holds, we obtain Eq.~(\ref{4.1_error_runtime_1_3}). Eq.~(\ref{4.1_error_runtime_1_4}) is derived by using Eq.~(\ref{4.1_error_runtime_error_C}).
Eq.~(\ref{4.1_error_runtime_1_5}) is derived by using Eq.~(\ref{4.1_error_runtime_error_R}).

Then, based on Lemma~\ref{4.1_R_l_main},
we could implement the gate $C(R_j)$ with an error {$\epsilon_R$}
by using
\begin{align}
T_{R_j} &= O(j \sigma_{\min}^{-\frac{1}{2}}(C_j) \epsilon_R^{-1}) \label{4.1_error_runtime_2_1} \\
&\leq O(r^2 \kappa^r \epsilon_R^{-1}) \label{4.1_error_runtime_2_3} \\
&\leq O(r^7 \kappa^{3r}) \label{4.1_error_runtime_2_4}
\end{align}
queries to the oracle $U_A$.
Since the condition ${\rm cond}^{(j)}$ (\ref{aim_lower_bound}) holds, we obtain  Eq.~(\ref{4.1_error_runtime_2_3}). 
Eq.~(\ref{4.1_error_runtime_2_4}) follows from the definition of $\epsilon_R$ in Eq.~(\ref{4.1_error_runtime_error_R}).
 
Next we calculate $N$. The number of times to perform  Algorithm 1 is bounded as
\begin{align}
N &= O\left( \frac{1}{{\rm Pr} \big\{ {\rm cond}^{(r)} \big\} } \right),\label{N_Pr}
\end{align}
where the probability follows from the distribution $\tilde{P}$ defined in Eq.~(\ref{main_distribution_P_noisy}). 
Now we proceed to bound ${\rm Pr} \big\{ {\rm cond}^{(r)} \big\}$.
In fact, we have
\begin{align}
&{}\ \ \ \ \left( 1- {\rm Pr} \big\{ {\rm cond}^{(r)} \big\} \right) \frac{1}{2r^2 \kappa^{2r-2}} + {\rm Pr} \big\{ {\rm cond}^{(r)} \big\} \cdot 1 \label{4.1_error_runtime_3_1}\\
&\geq \left( 1- {\rm Pr} \big\{ {\rm cond}^{(r)} \big\} \right) \frac{1}{2r^2 \kappa^{2r-2}} \nonumber \\
&+ {\rm Pr} \big\{ {\rm cond}^{(r)} \big\} \mathbb{E}_{\tilde{P}} \left[ \sigma_{\min}({C}_{r}) \Big|  {\rm cond}^{(r)} \right]\label{4.1_error_runtime_3_2} \\
&\geq \left( 1- {\rm Pr} \big\{ {\rm cond}^{(r)} \big\} \right) \mathbb{E}_{\tilde{P}} \left[ \sigma_{\min}({C}_{r}) \Big|  \text{not cond}^{(r)} \right] \nonumber \\
&+ {\rm Pr} \big\{ {\rm cond}^{(r)} \big\} \mathbb{E}_{\tilde{P}} \left[ \sigma_{\min}({C}_{r}) \Big|  {\rm cond}^{(r)} \right] \label{4.1_error_runtime_3_3} \\
&= \mathbb{E}_{\tilde{P}} [\sigma_{\min}({C}_r)], \label{4.1_error_runtime_3_4}
\end{align}
where Eq.~(\ref{4.1_error_runtime_3_2}) is obtained by noticing 
$$\sigma_{\min}({C}_{r}) \leq \frac{\text{Tr}({C}_r)}{r} = 1$$ 
holds for all choices of basis.
Eq.~(\ref{4.1_error_runtime_3_3}) is derived since $ \frac{1}{2r^2 \kappa^{2r-2}} \geq \sigma_{\min}({C}_{r}) $ when the condition ${\rm cond}^{(r)}$ does not hold.
 
Combining Eq.~(\ref{4.1_error_runtime_error_R}) with Lemma~\ref{bound_sigma_noisy}, we have the following statement:
\begin{align}
\mathbb{E}_{\tilde{P}} [\sigma_{\min}({C}_r)] &\geq \frac{2}{3} \mathop{\mathbb{E}}\limits_{{\text{Pr}}^{(1)}} \cdots \mathop{\mathbb{E}}\limits_{{\text{Pr}}^{(r)}}  [\sigma_{\min} (C_r)] \label{4.1_error_runtime_4_1}\\
&\geq \frac{2}{3r^2 \kappa^{2r-2}}. \label{4.1_error_runtime_4_2}
\end{align}
Thus,  Eq.~(\ref{4.1_error_runtime_4_2}) together with Eq.~(\ref{4.1_error_runtime_3_1}) yields
\begin{equation}
\frac{  1- {\rm Pr} \big\{ {\rm cond}^{(r)} \big\}  }{2r^2 \kappa^{2r-2}} + {\rm Pr} \big\{ {\rm cond}^{(r)} \big\}  \geq \frac{2}{3r^2 \kappa^{2r-2}}.
\end{equation}
We could solve
\begin{equation}
{\rm Pr} \big\{ {\rm cond}^{(r)} \big\} \geq 	\frac{\frac{1}{6r^2 \kappa^{2r-2}}}{1-\frac{1}{2r^2 \kappa^{2r-2}}} \geq \frac{1}{6r^2} \kappa^{2-2r},
\end{equation}
which induces the bound $N\leq O(r^2 \kappa^{2r-2})$ by using Eq.~(\ref{N_Pr}).

Finally we move on to analyze the query complexity $T_{\text{QGSP}}$. 
Based on Lemma~\ref{4_1_lemma_calculate_z_j}, coefficients $\{\z_\ell\}_{\ell=1}^{r-1}$ are calculated using the estimation of $C_r$.
Denote by $T_{C}$ the required query complexity of the oracle $U_A$ to estimate each element in $C_r$ via the Hadamard Test. We have
\begin{align}
T_C &= O (r^2 \epsilon_C^{-2}) \label{4.1_error_runtime_time_C_1} \\
&= O (r^{25} \kappa^{12r} ), \label{4.1_error_runtime_time_C_2}
\end{align}
where Eq.~(\ref{4.1_error_runtime_time_C_1}) is derived by using Eq.~(\ref{4.1_error_runtime_error_C}).
Recall that in each iteration of $\ell = 1,\cdots,r$ in Algorithm \ref{Basis}, we perform operations $U_A, V_A, R_1,R_2,\cdots,R_{\ell-1}$ for $1/P_\ell$ times. Taking the complexity of estimating $C_r$ into account, we have
\begin{align}
T_{\text{QGSP}} &= T_C + \sum_{\ell=1}^{r} \frac{1}{P_\ell} \left(2 + \sum_{m=1}^{\ell-1}T_{R_m} \right) \label{4.1_error_runtime_time_QGSP_1} \\
&= O(r^{25} \kappa^{12r} ) + \sum_{\ell=1}^{r} \frac{1}{P_\ell} \left(2 + \sum_{m=1}^{\ell-1} O(r^7 \kappa^{3r}) \right) \label{4.1_error_runtime_time_QGSP_2} \\
&\leq O(r^{25} \kappa^{12r} ) + O(r^8 \kappa^{3r}) \sum_{\ell=1}^{r} \frac{1}{P_\ell}
\label{4.1_error_runtime_time_QGSP_3} \\
&\leq O(r^{25} \kappa^{12r} ) + O(r^8 \kappa^{3r}) \sum_{\ell=1}^{r} r \kappa^{2} \label{4.1_error_runtime_time_QGSP_4} \\
&\leq O(r^{25} \kappa^{12r} ). \label{4.1_error_runtime_time_QGSP_5}
\end{align}
Eq.~(\ref{4.1_error_runtime_time_QGSP_2}) is obtained by using Eq.~(\ref{4.1_error_runtime_time_C_2}) and Eq.~(\ref{4.1_error_runtime_2_4}).
Eq.~(\ref{4.1_error_runtime_time_QGSP_4}) is derived by using Eq.~(\ref{lower_bound_P_j}).

By considering $N \leq O(r^2 \kappa^{2r-2})$ being the required number of times to run Algorithm~\ref{Basis}, we prove the Theorem~\ref{Basis_theorem_16}.

\end{proof}

\section{Proof of Theorem~\ref{solution_123}}
\label{app_coordinate}
We will first demonstrate that the proposed quantum circuit in Fig.~\ref{sign_circuit_main} is similar to the SWAP test, and provides a $\epsilon$-error estimation to the value ${a}_i' = \<\bm{t}_k|\bm{v}\>\<\bm{v}|\bm{t}_i\>, \forall i \in [r]$, with $O(1/\epsilon^2)$ measurements.

Firstly, after all unitary operations, the state in Fig.~\ref{sign_circuit_main} before the measurements is:
\begin{align*}
& \frac{1}{4} |0\> \Big[  |\bm{v}\> |\bm{t}_k\> + |\bm{v}\> |\bm{t}_i\> + |\bm{t}_k\> |\bm{v}\> + |\bm{t}_i\> |\bm{v}\> \Big] |0\>\\
+& \frac{1}{4} |0\> \Big[|\bm{v}\> |\bm{t}_k\> - |\bm{v}\> |\bm{t}_i\> + |\bm{t}_k\> |\bm{v}\> - |\bm{t}_i\> |\bm{v}\> \Big]|1\>\\
+& \frac{1}{4} |1\> \Big[|\bm{v}\> |\bm{t}_k\> + |\bm{v}\> |\bm{t}_i\> - |\bm{t}_k\> |\bm{v}\> - |\bm{t}_i\>|\bm{v}\> \Big]|0\>\\
+& \frac{1}{4} |1\> \Big[|\bm{v}\> |\bm{t}_k\> - |\bm{v}\> |\bm{t}_i\> - |\bm{t}_k\> |\bm{v}\> + |\bm{t}_i\>|\bm{v}\> \Big]	 |1\>.
\end{align*}
Measuring the first and the last register could result in outcomes $00$ and $11$ with probability:

\begin{align*}
P_{00} &= \frac{2 + |\<\bm{v}|\bm{t}_k\>|^2 + |\<\bm{v}|\bm{t}_i\>|^2 +2 \<\bm{t}_i|\bm{v}\>\<\bm{v}|\bm{t}_k\>}{8} ,\\
P_{11} &= \frac{2 - |\<\bm{v}|\bm{t}_k\>|^2 - |\<\bm{v}|\bm{t}_i\>|^2 +2 \<\bm{t}_i|\bm{v}\>\<\bm{v}|\bm{t}_k\>}{8} .
\end{align*}
We remark that the statistics of outcomes $00$ and $11$ implies the value ${a}_i a_k$.
\begin{equation*}
P_{\textnormal{same}} = P_{00} + P_{11} = \frac{1+ \<\bm{t}_i|\bm{v}\>\<\bm{v}|\bm{t}_k\>}{2} = \frac{1+{a}_i a_k}{2}.
\end{equation*}
The efficiency of the quantum circuit in Fig.~\ref{sign_circuit_main} depends on the efficiency of preparing the input state $(|\bm{t}_k\>|0\> + |\bm{t}_i\>|1\>)/\sqrt{2}$. 
Lemma~\ref{state_t_1_t_i} below proves that it can be prepared with query complexity $O(r \sigma_{\min}^{-1/2}(C_r))$. 
\begin{lemma}
\label{state_t_1_t_i}
Given perturbed coefficients provided in Lemma~\ref{4_1_lemma_error_t_l} for both indices $k$ and $\ell$, the state $\frac{1}{\sqrt{2}} (|0\>|\bm{t}_k\> + |1\>|\bm{t}_\ell\>)$ could be prepared with query complexity $O(r \sigma_{\min}^{-1/2}(C_r))$ with $\ell^2$ norm error bounded by $\epsilon$.
\end{lemma}

\begin{proof}

We sketch the main idea of the proof. Firstly, we generate the superposition state of $|\tilde{\bm{t}}_\ell\>$ and $|\tilde{\bm{t}}_k\>$ using perturbed coefficients, where perturbed vectors are expressed as
\begin{equation}
\tilde{\bm{t}}_\ell = \sum_{i=1}^\ell \tilde{z}_{i\ell} \bm{s}_i/\|\bm{s}_i\|,\ \tilde{\bm{t}}_k = \sum_{i=1}^k \tilde{z}_{ik} \bm{s}_i/\|\bm{s}_i\|.
\end{equation}
Then, we provide the error analysis.
Specifically, Given the coefficients $\tilde{\bm{z}}_k$ and $\tilde{\bm{z}}_\ell$, we prepare the state 
\begin{equation}\label{proof_super_t_k_t_l}
\frac{1}{\sqrt{\|\tilde{\bm{t}}_\ell\|^2 + \|{\tilde{\bm{t}}}_k\|^2 }} (\|\tilde{\bm{t}}_k \| |0\>|\tilde{\bm{t}}_k\> + \|\tilde{\bm{t}}_\ell \| |1\>|\tilde{\bm{t}}_\ell\>)
\end{equation}
by the LCU method as follows. 
Since the notation $k$ and $\ell$ are symmetrical here, we could assume that $\ell \geq k$ for convenience.
Firstly, we initialize the state $ \frac{|0\> + |1\>}{\sqrt{2}} |0\>^{\otimes \log m} |0\>^{\otimes \log n} |0\>$. 
Then, we apply Hadamard operations on the last $\log \ell$ qubits in the second register to create the state:
\begin{equation*}
\frac{|0\> + |1\>}{\sqrt{2\ell}} \sum_{j=1}^{\ell} |j\> |0\>|0\>.
\end{equation*}
Next, we employ the operation $U_{\text{index}}$ defined in (\ref{U_index}) to create the state:
\begin{equation*}
\frac{|0\> + |1\>}{\sqrt{2\ell}} \sum_{j=1}^\ell  |g(j)\>|0\>|0\>.	
\end{equation*}
Then we employ the oracle $U_A$ on the first and the second register, followed by the unitary $U_{\text{index}}^\dag$ to yield:
\begin{align*}
  \frac{|0\> + |1\>}{\sqrt{2l}} \sum_{j=1}^{\ell}  |j\>|A_{g(j)}\> |0\> \equiv  \frac{|0\> + |1\>}{\sqrt{2l}} \sum_{j=1}^{\ell} |j\>|\bm{s}_j\> |0\>. 
\end{align*}
Denote $\tilde{z} \equiv \max(\max_{j}|\tilde{z}_{j \ell}|, \max_{j}|\tilde{z}_{j k}|)$.
Next, we perform the controlled rotation
\begin{equation}
\begin{aligned}
&{}\ \ \ \ |0\>\<0| {}\otimes {}\sum_{j=1}^k |j\>\<j| {}\otimes I {}\otimes e^{-i \sigma_y \arccos(\tilde{z}_{jk}/\tilde{z})} \\
&+ |0\>\<0| {} \otimes {} \sum_{j=k+1}^\ell  |j\>\<j| \otimes I \otimes \sigma_x \\
&+ |0\>\<0| \otimes \sum_{j=\ell+1}^m |j\>\<j| \otimes I \otimes I\\
&+ |1\>\<1| \otimes \sum_{j=1}^\ell |j\>\<j| \otimes I \otimes e^{-i \sigma_y \arccos(\tilde{z}_{j\ell}/\tilde{z})} \\
&+ |1\>\<1| \otimes \sum_{j=\ell+1}^{m} |j\>\<j| \otimes I \otimes I,
\end{aligned}\label{control_rotation}
\end{equation}
to obtain the state
\begin{align*}
&{}\ \ \ \ \frac{1}{\sqrt{2\ell}} |0\> \sum_{j=1}^k |j\> |\bm{s}_j\> \left( \frac{\tilde{z}_{jk}}{\tilde{z}} |0\> + \sqrt{1-\frac{\tilde{z}_{jk}^2}{\tilde{z}^2}} |1\>\right) \\
&+ \frac{1}{\sqrt{2\ell}}  |0\> \sum_{j=k+1}^\ell |j\> |\bm{s}_j\> |1\> \\
&+ \frac{1}{\sqrt{2\ell}}  |1\> \sum_{j=1}^{\ell}  |j\>|\bm{s}_j\> \left(\frac{\tilde{z}_{j\ell}}{\tilde{z}}|0\> + \sqrt{1-\frac{\tilde{z}_{j \ell}^2}{\tilde{z}^2}}|1\>\right) . 
\end{align*}
The unitary (\ref{control_rotation}) could be performed by using $O(\ell)$ quantum operations due to the $O(\ell)$ sparsity.
Finally, we employ Hadamard operations on the last $\log \ell$ qubits in the second register, to obtain the state
\begin{align*}
 &{}\ \ \ \ \frac{1}{\sqrt{2}\ell \tilde{z}} |0\> \sum_{j=1}^k |0\> \tilde{z}_{jk}|\bm{s}_j\>  |0\> + \frac{1}{\sqrt{2}\ell \tilde{z}} |1\> \sum_{j=1}^{\ell} \tilde{z}_{j \ell} |0\>  |\bm{s}_j\> |0\> \\
 &\ \ \ \ +\ orthogonal\ garbage\ state\\
&= \frac{1}{\sqrt{2} \ell \tilde{z}} \left( \|\tilde{\bm{t}}_k\| |0\>|0\>|\tilde{\bm{t}}_k\> |0\> + \|\tilde{\bm{t}}_\ell\| |1\>|0\> |\tilde{\bm{t}}_{\ell}\> |0\> \right) \\
&\ \ \ \ +\ orthogonal\ garbage\ state,
\end{align*}

The measurement on the $2$-nd and the $4$-th registers of the final state could yield state in (\ref{proof_super_t_k_t_l}) with probability 
$$\frac{\|\tilde{\bm{t}}_\ell\|^2 + \|\tilde{\bm{t}}_k\|^2}{2 \ell^2 \tilde{z}^2},$$
so we could prepare this state with 
$${O} \left( \frac{\ell \tilde{z}}{\sqrt{\|\tilde{\bm{t}}_\ell\|^2 + \|\tilde{\bm{t}}_k\|^2}} \right)$$ 
queries to $U_A$ by using the amplitude amplification method~\cite{brassard2002quantum}. By using Eq.~(\ref{bound_z_l}), the complexity is further upper bounded as $O(r \sigma_{\min}^{-1/2}(C_r))$.

Now we analyze the distance between the state $\frac{1}{\sqrt{2}} (|0\>|\bm{t}_k\> + |1\>|\bm{t}_\ell\>)$  and the state in (\ref{proof_super_t_k_t_l}) as follows.
\begin{align}
&\ \ \ \ \left\| \frac{\|\tilde{\bm{t}}_k \| |0\>|\tilde{\bm{t}}_k\> + \|\tilde{\bm{t}}_\ell \| |1\>|\tilde{\bm{t}}_\ell\>}{\sqrt{\|\tilde{\bm{t}}_\ell\|^2 + \|\tilde{\bm{t}}_k\|^2 }}  - \frac{|0\>|\bm{t}_k\> + |1\>|\bm{t}_\ell\>}{\sqrt{2}} \right\| \nonumber \\
&\leq \left\|\frac{\tilde{\bm{t}}_k  }{\sqrt{\|\tilde{\bm{t}}_\ell\|^2 + \|\tilde{\bm{t}}_k\|^2 }}  - \frac{\bm{t}_k}{\sqrt{2}} \right\| + 
\left\|\frac{\tilde{\bm{t}}_\ell }{\sqrt{\|\tilde{\bm{t}}_\ell\|^2 + \|\tilde{\bm{t}}_k\|^2 }}  - \frac{\bm{t}_\ell}{\sqrt{2}} \right\| \label{state_t_1_t_i_1_2} \\
&\leq \left\|\frac{\tilde{\bm{t}}_k }{\sqrt{\|\tilde{\bm{t}}_\ell\|^2 + \|\tilde{\bm{t}}_k\|^2 }}  - \frac{\tilde{\bm{t}}_k}{\sqrt{2}} \right\| + \left\| \frac{\tilde{\bm{t}}_k - \bm{t}_k}{\sqrt{2}} \right\| \nonumber \\
&+ \left\|\frac{\tilde{\bm{t}}_\ell }{\sqrt{\|\tilde{\bm{t}}_\ell\|^2 + \|\tilde{\bm{t}}_k\|^2 }}  - \frac{\tilde{\bm{t}}_\ell}{\sqrt{2}} \right\| + \left\| \frac{\tilde{\bm{t}}_\ell - \bm{t}_\ell}{\sqrt{2}} \right\| \label{state_t_1_t_i_1_3} \\
&\leq \left[ (1+\frac{\epsilon}{4}) \left| \frac{1}{\sqrt{2(1-\frac{\epsilon}{4})^2}} - \frac{1}{\sqrt{2}}\right| + \frac{\epsilon}{4\sqrt{2}} \right] \times 2 \label{state_t_1_t_i_1_4} \\
&\leq \epsilon, \nonumber
\end{align}
where Eq.~(\ref{state_t_1_t_i_1_4}) is derived by using 
$$ 1-\frac{\epsilon}{4} \leq \|\bm{t}_i\| - \|\bm{t}_i - \tilde{\bm{t}}_i\| \leq \|\tilde{\bm{t}}_i\| \leq \|\bm{t}_i\| + \|\bm{t}_i - \tilde{\bm{t}}_i\| \leq  1+\frac{\epsilon}{4},$$
for $i=k/\ell$.

\end{proof}

Now we begin the proof of Theorem~\ref{solution_123} that provides the error analysis of Algorithm~\ref{coordinate_main} for reading out the state $|\bm{v}\>$.

\begin{proof}

We firstly study the error in the read-out procedure and then provide the time analysis.
Specifically, notice that the state $\frac{1}{\sqrt{2}}(|0\>|\bm{t}_k\>+|1\>|\bm{t}_i\>)$ generated by Lemma~\ref{state_t_1_t_i} is perturbed due to the noisy coefficients $\bm{z}_k$ and $\bm{z}_i$. 
Thus, the read-out error consists of two parts: the error on generating $\frac{1}{\sqrt{2}}(|0\>|\bm{t}_k\>+|1\>|\bm{t}_i\>)$, and the error induced by the statistical noise during the measurement in the Fig~\ref{sign_circuit_main}.

Firstly, we analyze the measurement distribution of Fig.~\ref{sign_circuit_main} which uses the perturbed input state $\frac{1}{\sqrt{2}}(|0\>|\bm{t}_k\>+|1\>|\bm{t}_i\>)$.
Denote $\tilde{\bm{z}}_j$ and $\tilde{\bm{t}}_j$ as the perturbed form of $\bm{z}_j$ and $\bm{t}_j$, respectively, $ \forall j \in [r]$. 
In this proof, we assume the $\ell^2$ norm on the error of each $\bm{t}_j$ is bounded by $\epsilon_3 = \frac{1}{14 r^{3/2}} \epsilon$. 
The final state in Fig.~\ref{sign_circuit_main} is:
\begin{align*}
& \frac{1}{4} |0\> \Big[ f_k |\bm{v}\> |\tilde{\bm{t}}_k\> + f_i |\bm{v}\> |\tilde{\bm{t}}_i\> + f_k |\tilde{\bm{t}}_k\> |\bm{v}\> + f_i |\tilde{\bm{t}}_i\> |\bm{v}\> \Big] |0\>\\
+& \frac{1}{4} |0\> \Big[ f_k |\bm{v}\> |\tilde{\bm{t}}_k\> - f_i |\bm{v}\> |\tilde{\bm{t}}_i\> + f_k |\tilde{\bm{t}}_k\> |\bm{v}\> - f_i |\tilde{\bm{t}}_i\> |\bm{v}\> \Big]|1\>\\
+& \frac{1}{4} |1\> \Big[ f_k |\bm{v}\> |\tilde{\bm{t}}_k\> + f_i |\bm{v}\> |\tilde{\bm{t}}_i\> - f_k |\tilde{\bm{t}}_k\> |\bm{v}\> - f_i |\tilde{\bm{t}}_i\>|\bm{v}\> \Big]|0\>\\
+& \frac{1}{4} |1\> \Big[ f_k |\bm{v}\> |\tilde{\bm{t}}_k\> - f_i |\bm{v}\> |\tilde{\bm{t}}_i\> - f_k |\tilde{\bm{t}}_k\> |\bm{v}\> + f_i |\tilde{\bm{t}}_i\>|\bm{v}\> \Big]	 |1\>,
\end{align*}
where we denote 
$$f_{k} = \sqrt{\frac{2\|\tilde{\bm{t}}_{k}\|^2}{\|\tilde{\bm{t}}_k\|^2 + \|\tilde{\bm{t}}_i\|^2}},\ f_{i} = \sqrt{\frac{2\|\tilde{\bm{t}}_{i}\|^2}{\|\tilde{\bm{t}}_k\|^2 + \|\tilde{\bm{t}}_i\|^2}}.$$
Measuring the first and the last register could result in outcomes $00$ and $11$ with probability:
\begin{align*}
\tilde{P}_{00} &= \frac{1 + f_i f_k \<\tilde{\bm{t}}_i|\bm{v}\>\<\bm{v}|\tilde{\bm{t}}_k\>}{4} \\
&{}+ \frac{f_k^2 |\<\bm{v}|\tilde{\bm{t}}_k\>|^2 + f_i^2 |\<\bm{v}|\tilde{\bm{t}}_i\>|^2 + 2 f_i f_k \<\tilde{\bm{t}}_k|\tilde{\bm{t}}_i\>}{8} ,\\
\tilde{P}_{11} &= \frac{1 + f_i f_k \<\tilde{\bm{t}}_i|\bm{v}\>\<\bm{v}|\tilde{\bm{t}}_k\>}{4} \\
&{}- \frac{f_k^2 |\<\bm{v}|\tilde{\bm{t}}_k\>|^2 + f_i^2 |\<\bm{v}|\tilde{\bm{t}}_i\>|^2 + 2 f_i f_k \<\tilde{\bm{t}}_k|\tilde{\bm{t}}_i\>}{8} .
\end{align*}
Thus, the perturbed statistics of outcomes $00$ and $11$ is:
\begin{equation}\label{sample_noisy}
\tilde{P}_{\text{same}} = \tilde{P}_{00} + \tilde{P}_{11} = \frac{1+ f_i f_k \<\tilde{\bm{t}}_i|\bm{v}\>\<\bm{v}|\tilde{\bm{t}}_k\>}{2} = \frac{1+\tilde{a}_i \tilde{a}_k }{2},
\end{equation}
where we denote $\tilde{a}_i = f_i \<\tilde{\bm{t}}_i|\bm{v}\>$.

Next, we analyze the error induced by the statistical noise.
Notice that each $\tilde{a}_i \tilde{a}_k$ in Eq.~(\ref{sample_noisy}) is estimated via the SWAP Test. 
We assume the statistical error of each value in $\{\tilde{a}_i \tilde{a}_k\}_{i=1}^{r}$ is bounded by $\epsilon_2=\frac{1}{14r^{3/2}}\epsilon$, and denote $\widetilde{(\tilde{a}_i \tilde{a}_k)}$ as the approximated value of $\tilde{a}_i \tilde{a}_k$. 
Then, in parallel to the exact form
\begin{equation}\label{express_v_at}
\bm{v} = \sum_{i=1}^r a_i \bm{t}_i,	
\end{equation}
we use the expression 
\begin{equation}\label{express_v_at_noisy}
\tilde{\bm{v}}=\sum_{i=1}^r \tilde{\tilde{a}}_i \tilde{\bm{t}}_i
\end{equation} 
as the perturbed description of the vector $\bm{v}$, where 
\begin{equation}\label{tilde_tilde_a_i}
\tilde{\tilde{a}}_i = \frac{\widetilde{(\tilde{a}_i \tilde{a}_k)}}{\sqrt{\sum_{i=1}^r {\widetilde{(\tilde{a}_i \tilde{a}_k)}}^2}},\ \forall i \in [r].
\end{equation}

Thus, the $\ell^2$ norm of the error on the vector description of the read-out state could be bounded as follows.
\begin{align}
&{}\ \ \ \ \|\tilde{\bm{v}} - \bm{v}\| = \| \sum_{i=1}^r ( \tilde{\tilde{a}}_i \tilde{\bm{t}}_i -  a_i \bm{t}_i ) \| \label{app_readout_error_3_1} \\
& \leq \|\sum_{i=1}^r \tilde{\tilde{a}}_i  (\tilde{\bm{t}}_i - \bm{t}_i) \| + \|\sum_{i=1}^r (\tilde{\tilde{a}}_i - \tilde{a}_i) \bm{t}_i\| \nonumber \\
&+ \| \sum_{i=1}^r (\tilde{a}_i - a_i) \bm{t}_i \| \label{app_readout_error_3_2} \\
&\leq \sum_{i=1}^r |\tilde{\tilde{a}}_i| \epsilon_3 + \sqrt{\sum_{i=1}^r (\tilde{\tilde{a}}_i-\tilde{a}_i)^2} + \sqrt{\sum_{i=1}^r (\tilde{a}_i -a_i)^2} \label{app_readout_error_3_3} \\
&\leq \sqrt{r} \epsilon_3 + \sqrt{\sum_{i=1}^r (\tilde{\tilde{a}}_i-\tilde{a}_i)^2} + \sqrt{\sum_{i=1}^r (\tilde{a}_i -a_i)^2} \label{app_readout_error_3_4} \\
&\leq \frac{\epsilon}{14} + \frac{4\epsilon}{7} + \frac{2\epsilon}{7} \leq \epsilon, \label{app_readout_error_3_5} 
\end{align}
where Eq.~(\ref{app_readout_error_3_1}) is obtained by using Eqs.~(\ref{express_v_at}-\ref{express_v_at_noisy}).
Eq.~(\ref{app_readout_error_3_2}) follows from the triangular inequality.
Eq.~(\ref{app_readout_error_3_3}) holds due to $\| \tilde{\bm{t}}_i - \bm{t}_i \|\leq \epsilon_3$ and $\bm{t}_i^T \bm{t}_j = \delta_{ij}$. 
Eq.~(\ref{app_readout_error_3_4}) is derived by using the definition in Eq.~(\ref{tilde_tilde_a_i}) and 
$$\frac{\sum_{i=1}^{r}|\tilde{\tilde{a}}_i|}{r} \leq \sqrt{\frac{\sum_{i=1}^r |\tilde{\tilde{a}}_i|^2}{r}} = \frac{1}{\sqrt{r}}.$$

Since the term $\epsilon_3 = \frac{1}{14r^{3/2}}\epsilon$ is provided, we notice that Eq.~(\ref{app_readout_error_3_5}) holds if the following statements is true for any $i \in [r]$:
\begin{align}
|\tilde{a}_i - a_i| &\leq \frac{2\epsilon}{7r^{1/2}}, \label{bound_tilde_a_i}\\
|\tilde{\tilde{a}}_i - \tilde{a}_i| &\leq \frac{4\epsilon}{7r^{1/2}}. \label{bound_tilde_tilde_a_i}
\end{align}

So we just need to bound terms $|{\tilde{a}}_i - {a}_i|$ and $|\tilde{\tilde{a}}_i - \tilde{a}_i|$ for deriving the upper bound on $\|\tilde{\bm{v}} - \bm{v}\|$, which can be obtained in Eqs.~(\ref{app_readout_error_1_1}-\ref{app_readout_error_1_5}) and Eqs.~(\ref{app_readout_error_2_1}-\ref{app_readout_error_2_7}), respectively, as follows.
\begin{align}
&{}\ \ \ \ |\tilde{a}_i-a_i| = 	| f_i \<\tilde{\bm{t}}_i|\bm{v}\> - \<\bm{t}_i|\bm{v}\> | \label{app_readout_error_1_1} \\
&\leq \left\|\sqrt{\frac{2}{\|\tilde{\bm{t}}_k\|^2 +\|\tilde{\bm{t}}_i\|^2 }} \tilde{\bm{t}}_i - \bm{t}_i \right\| \cdot \|\bm{v}\|  \label{app_readout_error_1_2}  \\
&\leq \left\|\sqrt{\frac{2}{\|\tilde{\bm{t}}_k\|^2 +\|\tilde{\bm{t}}_i\|^2 }} (\tilde{\bm{t}}_i - \bm{t}_i) \right\| \nonumber\\
&+ \left\| \sqrt{\frac{2}{\|\tilde{\bm{t}}_k\|^2 +\|\tilde{\bm{t}}_i\|^2 }} \bm{t}_i - \bm{t}_i \right\|  \label{app_readout_error_1_6}  \\
&\leq \sqrt{\frac{2}{\|\tilde{\bm{t}}_k\|^2 +\|\tilde{\bm{t}}_i\|^2 }} \epsilon_3 + \left| \sqrt{\frac{2}{\|\tilde{\bm{t}}_k\|^2 +\|\tilde{\bm{t}}_i\|^2 }} -1 \right| \label{app_readout_error_1_3} \\
&\leq \sqrt{\frac{2}{2(1-\epsilon_3)^2}} \epsilon_3 + \left| \sqrt{\frac{2}{2(1-\epsilon_3)^2}} -1 \right| \label{app_readout_error_1_4} \\
&= \frac{2\epsilon_3}{1-\epsilon_3} \leq \frac{2 \epsilon}{13r^{3/2}} \leq \frac{2 \epsilon}{7r^{1/2}} .\label{app_readout_error_1_5}
\end{align}
Eq.~(\ref{app_readout_error_1_1}) follows from the definitions 
$$\tilde{a}_i = f_i \<\tilde{\bm{t}}_i|\bm{v}\>,\ a_i = \<\bm{t}_i|\bm{v}\>.$$
Eq.~(\ref{app_readout_error_1_2}) is obtained by using the definition $f_{i} = \sqrt{\frac{2\|\tilde{\bm{t}}_{i}\|^2}{\|\tilde{\bm{t}}_k\|^2 + \|\tilde{\bm{t}}_i\|^2}}$.
Eq.~(\ref{app_readout_error_1_6}) follows from the triangular inequality and $\|\bm{v}\|=1$.
Eqs.~(\ref{app_readout_error_1_3}) and (\ref{app_readout_error_1_4}) are derived by using $\|\bm{t}_i\| = 1$ and $\|\tilde{\bm{t}}_{k/i}-\bm{t}_{k/i}\| \leq \epsilon_3$.
Eq.~(\ref{app_readout_error_1_5}) is obtained by the assumption $\epsilon_3 = \frac{\epsilon}{14r^{3/2}}$.

On the other hand, the term $\tilde{\tilde{a}}_i$ is bounded around $\tilde{a}_i$ as follows,
\begin{align}
&{}\ \ \ \ |\tilde{\tilde{a}}_i - \tilde{a}_i| =  \left| \frac{\widetilde{(\tilde{a}_i \tilde{a}_k)}}{\sqrt{\sum_{i=1}^r {\widetilde{(\tilde{a}_i \tilde{a}_k)}}^2}} - \tilde{a}_i \right| \label{app_readout_error_2_1} \\
&\leq \left| \frac{|\tilde{a}_i \tilde{a}_k| +\epsilon_2 }{ |\tilde{a}_k| \|\tilde{\bm{a}}\|- \sqrt{r} \epsilon_2} - |\tilde{a}_i| \right| \label{app_readout_error_2_2} \\
&= \frac{ |\tilde{a}_i||\tilde{a}_k| \left(1-\|\tilde{\bm{a}}\|\right)+(\sqrt{r}+|\tilde{a}_i|)\epsilon_2 }{|\tilde{a}_k| \|\tilde{\bm{a}}\| - \sqrt{r} \epsilon_2} \label{app_readout_error_2_3} \\
&\leq \frac{\sqrt{r} \frac{2\epsilon_3}{1-\epsilon_3} + (\sqrt{r}+\frac{1+\epsilon_3}{1-\epsilon_3})\epsilon_2}{(\frac{1}{\sqrt{r}} - \frac{2\epsilon_3}{1-\epsilon_3})(1-\sqrt{r} \frac{2\epsilon_3}{1-\epsilon_3}) - \sqrt{r} \epsilon_2} \label{app_readout_error_2_4} \\
&\leq \frac{(2r \epsilon_3  + 2r \epsilon_2)(1-\epsilon_3)}{(1- \epsilon_3 - 2 \sqrt{r} \epsilon_3)^2 - {r} \epsilon_2(1-\epsilon_3)^2} \label{app_readout_error_2_6} \\
&\leq \frac{2r\epsilon_3 + 2r \epsilon_2}{1-6r\epsilon_3-r\epsilon_2} \label{app_readout_error_2_5}\\
&= \frac{4\epsilon}{7r^{1/2}}. \label{app_readout_error_2_7}
\end{align}
Eq.~(\ref{app_readout_error_2_1}) follows from the definition in Eq.~(\ref{tilde_tilde_a_i}).
Eq.~(\ref{app_readout_error_2_2}) is derived by using $|\widetilde{(\tilde{a}_i \tilde{a}_k)} - \tilde{a}_i \tilde{a}_k| \leq \epsilon_2$. Eq.~(\ref{app_readout_error_2_4}) is derived by using $|\tilde{a}_i \tilde{a}_k| \leq 1$ (Eq.~(\ref{sample_noisy})) and
\begin{align*}
|\tilde{a}_{i}| &\leq |a_{i}| + \frac{2\epsilon_3}{1-\epsilon_3} \leq \frac{1+\epsilon_3}{1-\epsilon_3}, \\
|\tilde{a}_k| &\geq |a_k| - \frac{2\epsilon_3}{1-\epsilon_3} \geq \frac{1}{\sqrt{r}} - \frac{2\epsilon_3}{1-\epsilon_3},\\
\|\tilde{\bm{a}}\| &\geq \|\bm{a}\| - \|\tilde{\bm{a}} - \bm{a}\| \geq 1 - \sqrt{r} \frac{2\epsilon_3}{1-\epsilon_3}.
\end{align*}
Eq.~(\ref{app_readout_error_2_6}) is derived by multiplying $\sqrt{r}(1-\epsilon_3)^2$ on both the numerator and the denominator, and noticing that 
$$ r \epsilon_2 (1-\epsilon_3)^2 + \sqrt{r} \epsilon_2 (1-\epsilon_3^2) \leq 2r\epsilon_2 (1-\epsilon_3).$$
Eq.~(\ref{app_readout_error_2_5}) is obtained since
\begin{align*}
(1-\epsilon_3) &\leq 1, \\
(1-\epsilon_3 - 2\sqrt{r} \epsilon_3)^2 &\geq 1-6r\epsilon_3.
\end{align*}
We further derive Eq.~(\ref{app_readout_error_2_7}) by inserting 
$$\epsilon_2 = \epsilon_3 = \frac{1}{14r^{3/2}} \epsilon.$$

Finally, we analyze the time complexity of the protocol.
Notice that the error $ \|\tilde{\bm{t}}_i - \bm{t}_i\| \leq \epsilon_3$ could be achieved for all $i \in [r]$ by using 
$$r^2 \cdot O(r^5 \sigma_{\min}^{-4}(C) \epsilon_3^{-2})=O(r^{10} \sigma_{\min}^{-4}(C) \epsilon^{-2})$$ 
queries to input oracles due to Lemma~\ref{4_1_lemma_error_t_l}.  
Besides, the error $\epsilon_2$ induced as the statistical noise during the measurement in Fig.~\ref{sign_circuit_main} could be achieved by using $ \epsilon_2^{-2}=O(r^{3}\epsilon^{-2})$ copies of states $|\bm{v}\>$ and $\frac{1}{\sqrt{2}} (|0\>|\bm{t}_k\> + |1\>|\bm{t}_i\>)$ for $i \in [r]$, where the latter state could be prepared by using $O(r\sigma_{\min}^{-1/2}(C_r))$ queries to input oracles. By summing for each $i \in [r]$, we need $O(r^{4}\epsilon^{-2})$ copies of the state $|\bm{v}\>$ and $O(r^{5} \sigma_{\min}^{-1/2}(C) \epsilon^{-2})$ queries to input oracles in the measurement stage.
By counting the required resources in two stages, we have proved Theorem~\ref{solution_123}.

\end{proof}


\begin{thebibliography}{0}%
\makeatletter
\providecommand \@ifxundefined [1]{%
 \@ifx{#1\undefined}
}%
\providecommand \@ifnum [1]{%
 \ifnum #1\expandafter \@firstoftwo
 \else \expandafter \@secondoftwo
 \fi
}%
\providecommand \@ifx [1]{%
 \ifx #1\expandafter \@firstoftwo
 \else \expandafter \@secondoftwo
 \fi
}%
\providecommand \natexlab [1]{#1}%
\providecommand \enquote  [1]{``#1''}%
\providecommand \bibnamefont  [1]{#1}%
\providecommand \bibfnamefont [1]{#1}%
\providecommand \citenamefont [1]{#1}%
\providecommand \href@noop [0]{\@secondoftwo}%
\providecommand \href [0]{\begingroup \@sanitize@url \@href}%
\providecommand \@href[1]{\@@startlink{#1}\@@href}%
\providecommand \@@href[1]{\endgroup#1\@@endlink}%
\providecommand \@sanitize@url [0]{\catcode `\\12\catcode `\$12\catcode
  `\&12\catcode `\#12\catcode `\^12\catcode `\_12\catcode `\%12\relax}%
\providecommand \@@startlink[1]{}%
\providecommand \@@endlink[0]{}%
\providecommand \url  [0]{\begingroup\@sanitize@url \@url }%
\providecommand \@url [1]{\endgroup\@href {#1}{\urlprefix }}%
\providecommand \urlprefix  [0]{URL }%
\providecommand \Eprint [0]{\href }%
\providecommand \doibase [0]{https://doi.org/}%
\providecommand \selectlanguage [0]{\@gobble}%
\providecommand \bibinfo  [0]{\@secondoftwo}%
\providecommand \bibfield  [0]{\@secondoftwo}%
\providecommand \translation [1]{[#1]}%
\providecommand \BibitemOpen [0]{}%
\providecommand \bibitemStop [0]{}%
\providecommand \bibitemNoStop [0]{.\EOS\space}%
\providecommand \EOS [0]{\spacefactor3000\relax}%
\providecommand \BibitemShut  [1]{\csname bibitem#1\endcsname}%
\let\auto@bib@innerbib\@empty
\end{thebibliography}%


\begin{thebibliography}{67}%
\makeatletter
\providecommand \@ifxundefined [1]{%
 \@ifx{#1\undefined}
}%
\providecommand \@ifnum [1]{%
 \ifnum #1\expandafter \@firstoftwo
 \else \expandafter \@secondoftwo
 \fi
}%
\providecommand \@ifx [1]{%
 \ifx #1\expandafter \@firstoftwo
 \else \expandafter \@secondoftwo
 \fi
}%
\providecommand \natexlab [1]{#1}%
\providecommand \enquote  [1]{``#1''}%
\providecommand \bibnamefont  [1]{#1}%
\providecommand \bibfnamefont [1]{#1}%
\providecommand \citenamefont [1]{#1}%
\providecommand \href@noop [0]{\@secondoftwo}%
\providecommand \href [0]{\begingroup \@sanitize@url \@href}%
\providecommand \@href[1]{\@@startlink{#1}\@@href}%
\providecommand \@@href[1]{\endgroup#1\@@endlink}%
\providecommand \@sanitize@url [0]{\catcode `\\12\catcode `\$12\catcode
  `\&12\catcode `\#12\catcode `\^12\catcode `\_12\catcode `\%12\relax}%
\providecommand \@@startlink[1]{}%
\providecommand \@@endlink[0]{}%
\providecommand \url  [0]{\begingroup\@sanitize@url \@url }%
\providecommand \@url [1]{\endgroup\@href {#1}{\urlprefix }}%
\providecommand \urlprefix  [0]{URL }%
\providecommand \Eprint [0]{\href }%
\providecommand \doibase [0]{https://doi.org/}%
\providecommand \selectlanguage [0]{\@gobble}%
\providecommand \bibinfo  [0]{\@secondoftwo}%
\providecommand \bibfield  [0]{\@secondoftwo}%
\providecommand \translation [1]{[#1]}%
\providecommand \BibitemOpen [0]{}%
\providecommand \bibitemStop [0]{}%
\providecommand \bibitemNoStop [0]{.\EOS\space}%
\providecommand \EOS [0]{\spacefactor3000\relax}%
\providecommand \BibitemShut  [1]{\csname bibitem#1\endcsname}%
\let\auto@bib@innerbib\@empty
\bibitem [{\citenamefont {Hempel}\ \emph {et~al.}(2018)\citenamefont {Hempel},
  \citenamefont {Maier}, \citenamefont {Romero}, \citenamefont {McClean},
  \citenamefont {Monz}, \citenamefont {Shen}, \citenamefont {Jurcevic},
  \citenamefont {Lanyon}, \citenamefont {Love}, \citenamefont {Babbush},
  \citenamefont {Aspuru-Guzik}, \citenamefont {Blatt},\ and\ \citenamefont
  {Roos}}]{PhysRevX.8.031022}%
  \BibitemOpen
  \bibfield  {author} {\bibinfo {author} {\bibfnamefont {C.}~\bibnamefont
  {Hempel}}, \bibinfo {author} {\bibfnamefont {C.}~\bibnamefont {Maier}},
  \bibinfo {author} {\bibfnamefont {J.}~\bibnamefont {Romero}}, \bibinfo
  {author} {\bibfnamefont {J.}~\bibnamefont {McClean}}, \bibinfo {author}
  {\bibfnamefont {T.}~\bibnamefont {Monz}}, \bibinfo {author} {\bibfnamefont
  {H.}~\bibnamefont {Shen}}, \bibinfo {author} {\bibfnamefont {P.}~\bibnamefont
  {Jurcevic}}, \bibinfo {author} {\bibfnamefont {B.~P.}\ \bibnamefont
  {Lanyon}}, \bibinfo {author} {\bibfnamefont {P.}~\bibnamefont {Love}},
  \bibinfo {author} {\bibfnamefont {R.}~\bibnamefont {Babbush}}, \bibinfo
  {author} {\bibfnamefont {A.}~\bibnamefont {Aspuru-Guzik}}, \bibinfo {author}
  {\bibfnamefont {R.}~\bibnamefont {Blatt}},\ and\ \bibinfo {author}
  {\bibfnamefont {C.~F.}\ \bibnamefont {Roos}},\ }\bibfield  {title} {\bibinfo
  {title} {Quantum chemistry calculations on a trapped-ion quantum simulator},\
  }\href {https://doi.org/10.1103/PhysRevX.8.031022} {\bibfield  {journal}
  {\bibinfo  {journal} {Phys. Rev. X}\ }\textbf {\bibinfo {volume} {8}},\
  \bibinfo {pages} {031022} (\bibinfo {year} {2018})}\BibitemShut {NoStop}%
\bibitem [{\citenamefont {McArdle}\ \emph {et~al.}(2020)\citenamefont
  {McArdle}, \citenamefont {Endo}, \citenamefont {Aspuru-Guzik}, \citenamefont
  {Benjamin},\ and\ \citenamefont {Yuan}}]{RevModPhys.92.015003}%
  \BibitemOpen
  \bibfield  {author} {\bibinfo {author} {\bibfnamefont {S.}~\bibnamefont
  {McArdle}}, \bibinfo {author} {\bibfnamefont {S.}~\bibnamefont {Endo}},
  \bibinfo {author} {\bibfnamefont {A.}~\bibnamefont {Aspuru-Guzik}}, \bibinfo
  {author} {\bibfnamefont {S.~C.}\ \bibnamefont {Benjamin}},\ and\ \bibinfo
  {author} {\bibfnamefont {X.}~\bibnamefont {Yuan}},\ }\bibfield  {title}
  {\bibinfo {title} {Quantum computational chemistry},\ }\href
  {https://doi.org/10.1103/RevModPhys.92.015003} {\bibfield  {journal}
  {\bibinfo  {journal} {Rev. Mod. Phys.}\ }\textbf {\bibinfo {volume} {92}},\
  \bibinfo {pages} {015003} (\bibinfo {year} {2020})}\BibitemShut {NoStop}%
\bibitem [{\citenamefont {Nachman}\ \emph {et~al.}(2021)\citenamefont
  {Nachman}, \citenamefont {Provasoli}, \citenamefont {de~Jong},\ and\
  \citenamefont {Bauer}}]{PhysRevLett.126.062001}%
  \BibitemOpen
  \bibfield  {author} {\bibinfo {author} {\bibfnamefont {B.}~\bibnamefont
  {Nachman}}, \bibinfo {author} {\bibfnamefont {D.}~\bibnamefont {Provasoli}},
  \bibinfo {author} {\bibfnamefont {W.~A.}\ \bibnamefont {de~Jong}},\ and\
  \bibinfo {author} {\bibfnamefont {C.~W.}\ \bibnamefont {Bauer}},\ }\bibfield
  {title} {\bibinfo {title} {Quantum algorithm for high energy physics
  simulations},\ }\href {https://doi.org/10.1103/PhysRevLett.126.062001}
  {\bibfield  {journal} {\bibinfo  {journal} {Phys. Rev. Lett.}\ }\textbf
  {\bibinfo {volume} {126}},\ \bibinfo {pages} {062001} (\bibinfo {year}
  {2021})}\BibitemShut {NoStop}%
\bibitem [{\citenamefont {Guerreschi}\ and\ \citenamefont
  {Matsuura}(2019)}]{Guerreschi_2019}%
  \BibitemOpen
  \bibfield  {author} {\bibinfo {author} {\bibfnamefont {G.~G.}\ \bibnamefont
  {Guerreschi}}\ and\ \bibinfo {author} {\bibfnamefont {A.~Y.}\ \bibnamefont
  {Matsuura}},\ }\bibfield  {title} {\bibinfo {title} {Qaoa for max-cut
  requires hundreds of qubits for quantum speed-up},\ }\href@noop {} {\bibfield
   {journal} {\bibinfo  {journal} {Sci. Rep.}\ }\textbf {\bibinfo {volume}
  {9}},\ \bibinfo {pages} {1} (\bibinfo {year} {2019})}\BibitemShut {NoStop}%
\bibitem [{\citenamefont {Sanders}\ \emph {et~al.}(2020)\citenamefont
  {Sanders}, \citenamefont {Berry}, \citenamefont {Costa}, \citenamefont
  {Tessler}, \citenamefont {Wiebe}, \citenamefont {Gidney}, \citenamefont
  {Neven},\ and\ \citenamefont {Babbush}}]{PRXQuantum.1.020312}%
  \BibitemOpen
  \bibfield  {author} {\bibinfo {author} {\bibfnamefont {Y.~R.}\ \bibnamefont
  {Sanders}}, \bibinfo {author} {\bibfnamefont {D.~W.}\ \bibnamefont {Berry}},
  \bibinfo {author} {\bibfnamefont {P.~C.}\ \bibnamefont {Costa}}, \bibinfo
  {author} {\bibfnamefont {L.~W.}\ \bibnamefont {Tessler}}, \bibinfo {author}
  {\bibfnamefont {N.}~\bibnamefont {Wiebe}}, \bibinfo {author} {\bibfnamefont
  {C.}~\bibnamefont {Gidney}}, \bibinfo {author} {\bibfnamefont
  {H.}~\bibnamefont {Neven}},\ and\ \bibinfo {author} {\bibfnamefont
  {R.}~\bibnamefont {Babbush}},\ }\bibfield  {title} {\bibinfo {title}
  {Compilation of fault-tolerant quantum heuristics for combinatorial
  optimization},\ }\href {https://doi.org/10.1103/PRXQuantum.1.020312}
  {\bibfield  {journal} {\bibinfo  {journal} {PRX Quantum}\ }\textbf {\bibinfo
  {volume} {1}},\ \bibinfo {pages} {020312} (\bibinfo {year}
  {2020})}\BibitemShut {NoStop}%
\bibitem [{\citenamefont {Prakash}(2014)}]{prakash2014quantum}%
  \BibitemOpen
  \bibfield  {author} {\bibinfo {author} {\bibfnamefont {A.}~\bibnamefont
  {Prakash}},\ }\emph {\bibinfo {title} {Quantum algorithms for linear algebra
  and machine learning}},\ \href@noop {} {Ph.D. thesis},\ \bibinfo  {school}
  {UC Berkeley} (\bibinfo {year} {2014})\BibitemShut {NoStop}%
\bibitem [{\citenamefont {Biamonte}\ \emph {et~al.}(2017)\citenamefont
  {Biamonte}, \citenamefont {Wittek}, \citenamefont {Pancotti}, \citenamefont
  {Rebentrost}, \citenamefont {Wiebe},\ and\ \citenamefont
  {Lloyd}}]{biamonte2017quantum}%
  \BibitemOpen
  \bibfield  {author} {\bibinfo {author} {\bibfnamefont {J.}~\bibnamefont
  {Biamonte}}, \bibinfo {author} {\bibfnamefont {P.}~\bibnamefont {Wittek}},
  \bibinfo {author} {\bibfnamefont {N.}~\bibnamefont {Pancotti}}, \bibinfo
  {author} {\bibfnamefont {P.}~\bibnamefont {Rebentrost}}, \bibinfo {author}
  {\bibfnamefont {N.}~\bibnamefont {Wiebe}},\ and\ \bibinfo {author}
  {\bibfnamefont {S.}~\bibnamefont {Lloyd}},\ }\bibfield  {title} {\bibinfo
  {title} {Quantum machine learning},\ }\href@noop {} {\bibfield  {journal}
  {\bibinfo  {journal} {Nature}\ }\textbf {\bibinfo {volume} {549}},\ \bibinfo
  {pages} {195} (\bibinfo {year} {2017})}\BibitemShut {NoStop}%
\bibitem [{\citenamefont {Havlicek}\ \emph {et~al.}(2019)\citenamefont
  {Havlicek}, \citenamefont {Corcoles}, \citenamefont {Temme}, \citenamefont
  {Harrow}, \citenamefont {Kandala}, \citenamefont {Chow},\ and\ \citenamefont
  {Gambetta}}]{havlivcek2019supervised}%
  \BibitemOpen
  \bibfield  {author} {\bibinfo {author} {\bibfnamefont {V.}~\bibnamefont
  {Havlicek}}, \bibinfo {author} {\bibfnamefont {A.~D.}\ \bibnamefont
  {Corcoles}}, \bibinfo {author} {\bibfnamefont {K.}~\bibnamefont {Temme}},
  \bibinfo {author} {\bibfnamefont {A.~W.}\ \bibnamefont {Harrow}}, \bibinfo
  {author} {\bibfnamefont {A.}~\bibnamefont {Kandala}}, \bibinfo {author}
  {\bibfnamefont {J.~M.}\ \bibnamefont {Chow}},\ and\ \bibinfo {author}
  {\bibfnamefont {J.~M.}\ \bibnamefont {Gambetta}},\ }\bibfield  {title}
  {\bibinfo {title} {Supervised learning with quantum-enhanced feature
  spaces},\ }\href@noop {} {\bibfield  {journal} {\bibinfo  {journal} {Nature}\
  }\textbf {\bibinfo {volume} {567}},\ \bibinfo {pages} {209} (\bibinfo {year}
  {2019})}\BibitemShut {NoStop}%
\bibitem [{\citenamefont {Harrow}\ \emph {et~al.}(2009)\citenamefont {Harrow},
  \citenamefont {Hassidim},\ and\ \citenamefont {Lloyd}}]{harrow2009quantum}%
  \BibitemOpen
  \bibfield  {author} {\bibinfo {author} {\bibfnamefont {A.~W.}\ \bibnamefont
  {Harrow}}, \bibinfo {author} {\bibfnamefont {A.}~\bibnamefont {Hassidim}},\
  and\ \bibinfo {author} {\bibfnamefont {S.}~\bibnamefont {Lloyd}},\ }\bibfield
   {title} {\bibinfo {title} {Quantum algorithm for linear systems of
  equations},\ }\href@noop {} {\bibfield  {journal} {\bibinfo  {journal} {Phys.
  Rev. Lett.}\ }\textbf {\bibinfo {volume} {103}},\ \bibinfo {pages} {150502}
  (\bibinfo {year} {2009})}\BibitemShut {NoStop}%
\bibitem [{\citenamefont {Rebentrost}\ \emph {et~al.}(2018)\citenamefont
  {Rebentrost}, \citenamefont {Steffens}, \citenamefont {Marvian},\ and\
  \citenamefont {Lloyd}}]{PhysRevA.97.012327}%
  \BibitemOpen
  \bibfield  {author} {\bibinfo {author} {\bibfnamefont {P.}~\bibnamefont
  {Rebentrost}}, \bibinfo {author} {\bibfnamefont {A.}~\bibnamefont
  {Steffens}}, \bibinfo {author} {\bibfnamefont {I.}~\bibnamefont {Marvian}},\
  and\ \bibinfo {author} {\bibfnamefont {S.}~\bibnamefont {Lloyd}},\ }\bibfield
   {title} {\bibinfo {title} {Quantum singular-value decomposition of nonsparse
  low-rank matrices},\ }\href {https://doi.org/10.1103/PhysRevA.97.012327}
  {\bibfield  {journal} {\bibinfo  {journal} {Phys. Rev. A}\ }\textbf {\bibinfo
  {volume} {97}},\ \bibinfo {pages} {012327} (\bibinfo {year}
  {2018})}\BibitemShut {NoStop}%
\bibitem [{\citenamefont {Somma}\ and\ \citenamefont
  {Subasi}(2021)}]{PRXQuantum.2.010315}%
  \BibitemOpen
  \bibfield  {author} {\bibinfo {author} {\bibfnamefont {R.~D.}\ \bibnamefont
  {Somma}}\ and\ \bibinfo {author} {\bibfnamefont {Y.}~\bibnamefont {Subasi}},\
  }\bibfield  {title} {\bibinfo {title} {Complexity of quantum state
  verification in the quantum linear systems problem},\ }\href
  {https://doi.org/10.1103/PRXQuantum.2.010315} {\bibfield  {journal} {\bibinfo
   {journal} {PRX Quantum}\ }\textbf {\bibinfo {volume} {2}},\ \bibinfo {pages}
  {010315} (\bibinfo {year} {2021})}\BibitemShut {NoStop}%
\bibitem [{\citenamefont {Rebentrost}\ \emph {et~al.}(2014)\citenamefont
  {Rebentrost}, \citenamefont {Mohseni},\ and\ \citenamefont
  {Lloyd}}]{PhysRevLett.113.130503}%
  \BibitemOpen
  \bibfield  {author} {\bibinfo {author} {\bibfnamefont {P.}~\bibnamefont
  {Rebentrost}}, \bibinfo {author} {\bibfnamefont {M.}~\bibnamefont
  {Mohseni}},\ and\ \bibinfo {author} {\bibfnamefont {S.}~\bibnamefont
  {Lloyd}},\ }\bibfield  {title} {\bibinfo {title} {Quantum support vector
  machine for big data classification},\ }\href
  {https://doi.org/10.1103/PhysRevLett.113.130503} {\bibfield  {journal}
  {\bibinfo  {journal} {Phys. Rev. Lett.}\ }\textbf {\bibinfo {volume} {113}},\
  \bibinfo {pages} {130503} (\bibinfo {year} {2014})}\BibitemShut {NoStop}%
\bibitem [{\citenamefont {Lloyd}\ \emph {et~al.}(2013)\citenamefont {Lloyd},
  \citenamefont {Mohseni},\ and\ \citenamefont
  {Rebentrost}}]{lloyd2013quantum}%
  \BibitemOpen
  \bibfield  {author} {\bibinfo {author} {\bibfnamefont {S.}~\bibnamefont
  {Lloyd}}, \bibinfo {author} {\bibfnamefont {M.}~\bibnamefont {Mohseni}},\
  and\ \bibinfo {author} {\bibfnamefont {P.}~\bibnamefont {Rebentrost}},\
  }\bibfield  {title} {\bibinfo {title} {Quantum algorithms for supervised and
  unsupervised machine learning},\ }\href@noop {} {\bibfield  {journal}
  {\bibinfo  {journal} {arXiv:1307.0411}\ } (\bibinfo {year}
  {2013})}\BibitemShut {NoStop}%
\bibitem [{\citenamefont {Arunachalam}\ and\ \citenamefont
  {de~Wolf}(2017)}]{10.1145/3106700.3106710}%
  \BibitemOpen
  \bibfield  {author} {\bibinfo {author} {\bibfnamefont {S.}~\bibnamefont
  {Arunachalam}}\ and\ \bibinfo {author} {\bibfnamefont {R.}~\bibnamefont
  {de~Wolf}},\ }\bibfield  {title} {\bibinfo {title} {Guest column: A survey of
  quantum learning theory},\ }\href {https://doi.org/10.1145/3106700.3106710}
  {\bibfield  {journal} {\bibinfo  {journal} {SIGACT News}\ }\textbf {\bibinfo
  {volume} {48}},\ \bibinfo {pages} {41–67} (\bibinfo {year}
  {2017})}\BibitemShut {NoStop}%
\bibitem [{\citenamefont {Kerenidis}\ \emph {et~al.}(2019)\citenamefont
  {Kerenidis}, \citenamefont {Landman}, \citenamefont {Luongo},\ and\
  \citenamefont {Prakash}}]{kerenidis2019q}%
  \BibitemOpen
  \bibfield  {author} {\bibinfo {author} {\bibfnamefont {I.}~\bibnamefont
  {Kerenidis}}, \bibinfo {author} {\bibfnamefont {J.}~\bibnamefont {Landman}},
  \bibinfo {author} {\bibfnamefont {A.}~\bibnamefont {Luongo}},\ and\ \bibinfo
  {author} {\bibfnamefont {A.}~\bibnamefont {Prakash}},\ }\bibfield  {title}
  {\bibinfo {title} {q-means: A quantum algorithm for unsupervised machine
  learning},\ }in\ \href@noop {} {\emph {\bibinfo {booktitle} {Advances in
  Neural Information Processing Systems}}}\ (\bibinfo {year} {2019})\ pp.\
  \bibinfo {pages} {4136--4146}\BibitemShut {NoStop}%
\bibitem [{\citenamefont {Kapoor}\ \emph {et~al.}(2016)\citenamefont {Kapoor},
  \citenamefont {Wiebe},\ and\ \citenamefont {Svore}}]{kapoor2016quantum}%
  \BibitemOpen
  \bibfield  {author} {\bibinfo {author} {\bibfnamefont {A.}~\bibnamefont
  {Kapoor}}, \bibinfo {author} {\bibfnamefont {N.}~\bibnamefont {Wiebe}},\ and\
  \bibinfo {author} {\bibfnamefont {K.}~\bibnamefont {Svore}},\ }\bibfield
  {title} {\bibinfo {title} {Quantum perceptron models},\ }in\ \href@noop {}
  {\emph {\bibinfo {booktitle} {Advances in Neural Information Processing
  Systems}}}\ (\bibinfo {year} {2016})\ pp.\ \bibinfo {pages}
  {3999--4007}\BibitemShut {NoStop}%
\bibitem [{\citenamefont {Bausch}(2020)}]{bausch2020recurrent}%
  \BibitemOpen
  \bibfield  {author} {\bibinfo {author} {\bibfnamefont {J.}~\bibnamefont
  {Bausch}},\ }\bibfield  {title} {\bibinfo {title} {Recurrent quantum neural
  networks},\ }\bibfield  {booktitle} {\emph {\bibinfo {booktitle} {Advances in
  Neural Information Processing Systems}},\ }\href@noop {} {\ \textbf {\bibinfo
  {volume} {33}} (\bibinfo {year} {2020})}\BibitemShut {NoStop}%
\bibitem [{\citenamefont {Giovannetti}\ \emph {et~al.}(2008)\citenamefont
  {Giovannetti}, \citenamefont {Lloyd},\ and\ \citenamefont
  {Maccone}}]{PhysRevLett.100.160501}%
  \BibitemOpen
  \bibfield  {author} {\bibinfo {author} {\bibfnamefont {V.}~\bibnamefont
  {Giovannetti}}, \bibinfo {author} {\bibfnamefont {S.}~\bibnamefont {Lloyd}},\
  and\ \bibinfo {author} {\bibfnamefont {L.}~\bibnamefont {Maccone}},\
  }\bibfield  {title} {\bibinfo {title} {Quantum random access memory},\ }\href
  {https://doi.org/10.1103/PhysRevLett.100.160501} {\bibfield  {journal}
  {\bibinfo  {journal} {Phys. Rev. Lett.}\ }\textbf {\bibinfo {volume} {100}},\
  \bibinfo {pages} {160501} (\bibinfo {year} {2008})}\BibitemShut {NoStop}%
\bibitem [{\citenamefont {Gily{\'e}n}\ and\ \citenamefont
  {Li}(2020)}]{gilyen2019distributional}%
  \BibitemOpen
  \bibfield  {author} {\bibinfo {author} {\bibfnamefont {A.}~\bibnamefont
  {Gily{\'e}n}}\ and\ \bibinfo {author} {\bibfnamefont {T.}~\bibnamefont
  {Li}},\ }\bibfield  {title} {\bibinfo {title} {Distributional property
  testing in a quantum world},\ }in\ \href@noop {} {\emph {\bibinfo {booktitle}
  {11th Innovations in Theoretical Computer Science Conference}}}\ (\bibinfo
  {year} {2020})\BibitemShut {NoStop}%
\bibitem [{\citenamefont {Wossnig}\ \emph {et~al.}(2018)\citenamefont
  {Wossnig}, \citenamefont {Zhao},\ and\ \citenamefont
  {Prakash}}]{wossnig2018quantum}%
  \BibitemOpen
  \bibfield  {author} {\bibinfo {author} {\bibfnamefont {L.}~\bibnamefont
  {Wossnig}}, \bibinfo {author} {\bibfnamefont {Z.}~\bibnamefont {Zhao}},\ and\
  \bibinfo {author} {\bibfnamefont {A.}~\bibnamefont {Prakash}},\ }\bibfield
  {title} {\bibinfo {title} {Quantum linear system algorithm for dense
  matrices},\ }\href@noop {} {\bibfield  {journal} {\bibinfo  {journal} {Phys.
  Rev. Lett.}\ }\textbf {\bibinfo {volume} {120}},\ \bibinfo {pages} {050502}
  (\bibinfo {year} {2018})}\BibitemShut {NoStop}%
\bibitem [{\citenamefont {Kerenidis}\ and\ \citenamefont
  {Prakash}(2020{\natexlab{a}})}]{PhysRevA.101.022316}%
  \BibitemOpen
  \bibfield  {author} {\bibinfo {author} {\bibfnamefont {I.}~\bibnamefont
  {Kerenidis}}\ and\ \bibinfo {author} {\bibfnamefont {A.}~\bibnamefont
  {Prakash}},\ }\bibfield  {title} {\bibinfo {title} {Quantum gradient descent
  for linear systems and least squares},\ }\href
  {https://doi.org/10.1103/PhysRevA.101.022316} {\bibfield  {journal} {\bibinfo
   {journal} {Phys. Rev. A}\ }\textbf {\bibinfo {volume} {101}},\ \bibinfo
  {pages} {022316} (\bibinfo {year} {2020}{\natexlab{a}})}\BibitemShut
  {NoStop}%
\bibitem [{\citenamefont {Li}\ \emph {et~al.}(2015)\citenamefont {Li},
  \citenamefont {Liu}, \citenamefont {Xu},\ and\ \citenamefont
  {Du}}]{li2015experimental}%
  \BibitemOpen
  \bibfield  {author} {\bibinfo {author} {\bibfnamefont {Z.}~\bibnamefont
  {Li}}, \bibinfo {author} {\bibfnamefont {X.}~\bibnamefont {Liu}}, \bibinfo
  {author} {\bibfnamefont {N.}~\bibnamefont {Xu}},\ and\ \bibinfo {author}
  {\bibfnamefont {J.}~\bibnamefont {Du}},\ }\bibfield  {title} {\bibinfo
  {title} {Experimental realization of a quantum support vector machine},\
  }\href {https://doi.org/10.1103/PhysRevLett.114.140504} {\bibfield  {journal}
  {\bibinfo  {journal} {Phys. Rev. Lett.}\ }\textbf {\bibinfo {volume} {114}},\
  \bibinfo {pages} {140504} (\bibinfo {year} {2015})}\BibitemShut {NoStop}%
\bibitem [{\citenamefont {Kerenidis}\ \emph {et~al.}(2021)\citenamefont
  {Kerenidis}, \citenamefont {Prakash},\ and\ \citenamefont
  {Szil{\'a}gyi}}]{kerenidis2019quantum}%
  \BibitemOpen
  \bibfield  {author} {\bibinfo {author} {\bibfnamefont {I.}~\bibnamefont
  {Kerenidis}}, \bibinfo {author} {\bibfnamefont {A.}~\bibnamefont {Prakash}},\
  and\ \bibinfo {author} {\bibfnamefont {D.}~\bibnamefont {Szil{\'a}gyi}},\
  }\bibfield  {title} {\bibinfo {title} {Quantum algorithms for second-order
  cone programming and support vector machines},\ }\href@noop {} {\bibfield
  {journal} {\bibinfo  {journal} {Quantum}\ }\textbf {\bibinfo {volume} {5}},\
  \bibinfo {pages} {427} (\bibinfo {year} {2021})}\BibitemShut {NoStop}%
\bibitem [{\citenamefont {Allcock}\ \emph {et~al.}(2020)\citenamefont
  {Allcock}, \citenamefont {Hsieh}, \citenamefont {Kerenidis},\ and\
  \citenamefont {Zhang}}]{allcock2018quantum}%
  \BibitemOpen
  \bibfield  {author} {\bibinfo {author} {\bibfnamefont {J.}~\bibnamefont
  {Allcock}}, \bibinfo {author} {\bibfnamefont {C.-Y.}\ \bibnamefont {Hsieh}},
  \bibinfo {author} {\bibfnamefont {I.}~\bibnamefont {Kerenidis}},\ and\
  \bibinfo {author} {\bibfnamefont {S.}~\bibnamefont {Zhang}},\ }\bibfield
  {title} {\bibinfo {title} {Quantum algorithms for feedforward neural
  networks},\ }\href@noop {} {\bibfield  {journal} {\bibinfo  {journal} {ACM
  Transactions on Quantum Computing}\ }\textbf {\bibinfo {volume} {1}},\
  \bibinfo {pages} {1} (\bibinfo {year} {2020})}\BibitemShut {NoStop}%
\bibitem [{\citenamefont {Kerenidis}\ \emph
  {et~al.}(2020{\natexlab{a}})\citenamefont {Kerenidis}, \citenamefont
  {Landman},\ and\ \citenamefont {Prakash}}]{kerenidis2019quantum_dcnn}%
  \BibitemOpen
  \bibfield  {author} {\bibinfo {author} {\bibfnamefont {I.}~\bibnamefont
  {Kerenidis}}, \bibinfo {author} {\bibfnamefont {J.}~\bibnamefont {Landman}},\
  and\ \bibinfo {author} {\bibfnamefont {A.}~\bibnamefont {Prakash}},\
  }\bibfield  {title} {\bibinfo {title} {Quantum algorithms for deep
  convolutional neural networks},\ }in\ \href
  {https://openreview.net/forum?id=Hygab1rKDS} {\emph {\bibinfo {booktitle}
  {International Conference on Learning Representations}}}\ (\bibinfo {year}
  {2020})\BibitemShut {NoStop}%
\bibitem [{\citenamefont {Kerenidis}\ \emph
  {et~al.}(2020{\natexlab{b}})\citenamefont {Kerenidis}, \citenamefont
  {Luongo},\ and\ \citenamefont {Prakash}}]{kerenidis2019quantum_gauss}%
  \BibitemOpen
  \bibfield  {author} {\bibinfo {author} {\bibfnamefont {I.}~\bibnamefont
  {Kerenidis}}, \bibinfo {author} {\bibfnamefont {A.}~\bibnamefont {Luongo}},\
  and\ \bibinfo {author} {\bibfnamefont {A.}~\bibnamefont {Prakash}},\
  }\bibfield  {title} {\bibinfo {title} {Quantum expectation-maximization for
  gaussian mixture models},\ }in\ \href@noop {} {\emph {\bibinfo {booktitle}
  {International Conference on Machine Learning}}}\ (\bibinfo {organization}
  {PMLR},\ \bibinfo {year} {2020})\ pp.\ \bibinfo {pages}
  {5187--5197}\BibitemShut {NoStop}%
\bibitem [{\citenamefont {Kerenidis}\ and\ \citenamefont
  {Prakash}(2017)}]{kerenidis2016quantum}%
  \BibitemOpen
  \bibfield  {author} {\bibinfo {author} {\bibfnamefont {I.}~\bibnamefont
  {Kerenidis}}\ and\ \bibinfo {author} {\bibfnamefont {A.}~\bibnamefont
  {Prakash}},\ }\bibfield  {title} {\bibinfo {title} {Quantum recommendation
  systems},\ }in\ \href@noop {} {\emph {\bibinfo {booktitle} {8th Innovations
  in Theoretical Computer Science Conference}}}\ (\bibinfo {organization}
  {Schloss Dagstuhl-Leibniz-Zentrum fuer Informatik},\ \bibinfo {year}
  {2017})\BibitemShut {NoStop}%
\bibitem [{\citenamefont {Kerenidis}\ and\ \citenamefont
  {Prakash}(2020{\natexlab{b}})}]{kerenidis2018quantum}%
  \BibitemOpen
  \bibfield  {author} {\bibinfo {author} {\bibfnamefont {I.}~\bibnamefont
  {Kerenidis}}\ and\ \bibinfo {author} {\bibfnamefont {A.}~\bibnamefont
  {Prakash}},\ }\bibfield  {title} {\bibinfo {title} {A quantum interior point
  method for lps and sdps},\ }\href@noop {} {\bibfield  {journal} {\bibinfo
  {journal} {ACM Transactions on Quantum Computing}\ }\textbf {\bibinfo
  {volume} {1}},\ \bibinfo {pages} {1} (\bibinfo {year}
  {2020}{\natexlab{b}})}\BibitemShut {NoStop}%
\bibitem [{Note1()}]{Note1}%
  \BibitemOpen
  \bibinfo {note} {Another implementation of the QRAM oracle is proposed in
  Ref. \cite {kerenidis2016quantum}, which requires $O(k {\protect \rm
  polylog}(md))$ quantum operations and physical resources for $m \times d$
  matrix with $k$ non-zero elements}\BibitemShut {NoStop}%
\bibitem [{\citenamefont {Hann}\ \emph {et~al.}(2019)\citenamefont {Hann},
  \citenamefont {Zou}, \citenamefont {Zhang}, \citenamefont {Chu},
  \citenamefont {Schoelkopf}, \citenamefont {Girvin},\ and\ \citenamefont
  {Jiang}}]{PhysRevLett.123.250501}%
  \BibitemOpen
  \bibfield  {author} {\bibinfo {author} {\bibfnamefont {C.~T.}\ \bibnamefont
  {Hann}}, \bibinfo {author} {\bibfnamefont {C.-L.}\ \bibnamefont {Zou}},
  \bibinfo {author} {\bibfnamefont {Y.}~\bibnamefont {Zhang}}, \bibinfo
  {author} {\bibfnamefont {Y.}~\bibnamefont {Chu}}, \bibinfo {author}
  {\bibfnamefont {R.~J.}\ \bibnamefont {Schoelkopf}}, \bibinfo {author}
  {\bibfnamefont {S.~M.}\ \bibnamefont {Girvin}},\ and\ \bibinfo {author}
  {\bibfnamefont {L.}~\bibnamefont {Jiang}},\ }\bibfield  {title} {\bibinfo
  {title} {Hardware-efficient quantum random access memory with hybrid quantum
  acoustic systems},\ }\href {https://doi.org/10.1103/PhysRevLett.123.250501}
  {\bibfield  {journal} {\bibinfo  {journal} {Phys. Rev. Lett.}\ }\textbf
  {\bibinfo {volume} {123}},\ \bibinfo {pages} {250501} (\bibinfo {year}
  {2019})}\BibitemShut {NoStop}%
\bibitem [{\citenamefont {Hann}\ \emph {et~al.}(2021)\citenamefont {Hann},
  \citenamefont {Lee}, \citenamefont {Girvin},\ and\ \citenamefont
  {Jiang}}]{PRXQuantum.2.020311}%
  \BibitemOpen
  \bibfield  {author} {\bibinfo {author} {\bibfnamefont {C.~T.}\ \bibnamefont
  {Hann}}, \bibinfo {author} {\bibfnamefont {G.}~\bibnamefont {Lee}}, \bibinfo
  {author} {\bibfnamefont {S.}~\bibnamefont {Girvin}},\ and\ \bibinfo {author}
  {\bibfnamefont {L.}~\bibnamefont {Jiang}},\ }\bibfield  {title} {\bibinfo
  {title} {Resilience of quantum random access memory to generic noise},\
  }\href {https://doi.org/10.1103/PRXQuantum.2.020311} {\bibfield  {journal}
  {\bibinfo  {journal} {PRX Quantum}\ }\textbf {\bibinfo {volume} {2}},\
  \bibinfo {pages} {020311} (\bibinfo {year} {2021})}\BibitemShut {NoStop}%
\bibitem [{\citenamefont {Aaronson}(2015)}]{aaronson2015read}%
  \BibitemOpen
  \bibfield  {author} {\bibinfo {author} {\bibfnamefont {S.}~\bibnamefont
  {Aaronson}},\ }\bibfield  {title} {\bibinfo {title} {Read the fine print},\
  }\href@noop {} {\bibfield  {journal} {\bibinfo  {journal} {Nature Physics}\
  }\textbf {\bibinfo {volume} {11}},\ \bibinfo {pages} {291} (\bibinfo {year}
  {2015})}\BibitemShut {NoStop}%
\bibitem [{\citenamefont {Gross}\ \emph {et~al.}(2010)\citenamefont {Gross},
  \citenamefont {Liu}, \citenamefont {Flammia}, \citenamefont {Becker},\ and\
  \citenamefont {Jens}}]{PhysRevLett.105.150401}%
  \BibitemOpen
  \bibfield  {author} {\bibinfo {author} {\bibfnamefont {D.}~\bibnamefont
  {Gross}}, \bibinfo {author} {\bibfnamefont {Y.-K.}\ \bibnamefont {Liu}},
  \bibinfo {author} {\bibfnamefont {S.~T.}\ \bibnamefont {Flammia}}, \bibinfo
  {author} {\bibfnamefont {S.}~\bibnamefont {Becker}},\ and\ \bibinfo {author}
  {\bibfnamefont {E.}~\bibnamefont {Jens}},\ }\bibfield  {title} {\bibinfo
  {title} {Quantum state tomography via compressed sensing},\ }\href
  {https://doi.org/10.1103/PhysRevLett.105.150401} {\bibfield  {journal}
  {\bibinfo  {journal} {Phys. Rev. Lett.}\ }\textbf {\bibinfo {volume} {105}},\
  \bibinfo {pages} {150401} (\bibinfo {year} {2010})}\BibitemShut {NoStop}%
\bibitem [{\citenamefont {Kyrillidis}\ \emph {et~al.}(2018)\citenamefont
  {Kyrillidis}, \citenamefont {Kalev}, \citenamefont {Park}, \citenamefont
  {Bhojanapalli}, \citenamefont {Caramanis},\ and\ \citenamefont
  {Sanghavi}}]{kyrillidis2018provable}%
  \BibitemOpen
  \bibfield  {author} {\bibinfo {author} {\bibfnamefont {A.}~\bibnamefont
  {Kyrillidis}}, \bibinfo {author} {\bibfnamefont {A.}~\bibnamefont {Kalev}},
  \bibinfo {author} {\bibfnamefont {D.}~\bibnamefont {Park}}, \bibinfo {author}
  {\bibfnamefont {S.}~\bibnamefont {Bhojanapalli}}, \bibinfo {author}
  {\bibfnamefont {C.}~\bibnamefont {Caramanis}},\ and\ \bibinfo {author}
  {\bibfnamefont {S.}~\bibnamefont {Sanghavi}},\ }\bibfield  {title} {\bibinfo
  {title} {Provable compressed sensing quantum state tomography via non-convex
  methods},\ }\href@noop {} {\bibfield  {journal} {\bibinfo  {journal} {npj
  Quantum Information}\ }\textbf {\bibinfo {volume} {4}},\ \bibinfo {pages}
  {36} (\bibinfo {year} {2018})}\BibitemShut {NoStop}%
\bibitem [{\citenamefont {Haah}\ \emph {et~al.}(2017)\citenamefont {Haah},
  \citenamefont {Harrow}, \citenamefont {Ji}, \citenamefont {Wu},\ and\
  \citenamefont {Yu}}]{haah2017sample}%
  \BibitemOpen
  \bibfield  {author} {\bibinfo {author} {\bibfnamefont {J.}~\bibnamefont
  {Haah}}, \bibinfo {author} {\bibfnamefont {A.~W.}\ \bibnamefont {Harrow}},
  \bibinfo {author} {\bibfnamefont {Z.}~\bibnamefont {Ji}}, \bibinfo {author}
  {\bibfnamefont {X.}~\bibnamefont {Wu}},\ and\ \bibinfo {author}
  {\bibfnamefont {N.}~\bibnamefont {Yu}},\ }\bibfield  {title} {\bibinfo
  {title} {Sample-optimal tomography of quantum states},\ }\href@noop {}
  {\bibfield  {journal} {\bibinfo  {journal} {IEEE Transactions on Information
  Theory}\ }\textbf {\bibinfo {volume} {63}},\ \bibinfo {pages} {5628}
  (\bibinfo {year} {2017})}\BibitemShut {NoStop}%
\bibitem [{\citenamefont {O'Donnell}\ and\ \citenamefont
  {Wright}(2016)}]{o2016efficient}%
  \BibitemOpen
  \bibfield  {author} {\bibinfo {author} {\bibfnamefont {R.}~\bibnamefont
  {O'Donnell}}\ and\ \bibinfo {author} {\bibfnamefont {J.}~\bibnamefont
  {Wright}},\ }\bibfield  {title} {\bibinfo {title} {Efficient quantum
  tomography},\ }in\ \href@noop {} {\emph {\bibinfo {booktitle} {Proceedings of
  the 48th annual ACM symposium on Theory of Computing}}}\ (\bibinfo
  {organization} {ACM},\ \bibinfo {year} {2016})\ pp.\ \bibinfo {pages}
  {899--912}\BibitemShut {NoStop}%
\bibitem [{\citenamefont {Cramer}\ \emph {et~al.}(2010)\citenamefont {Cramer},
  \citenamefont {Plenio}, \citenamefont {Flammia}, \citenamefont {Somma},
  \citenamefont {Gross}, \citenamefont {Bartlett}, \citenamefont
  {Landon-Cardinal}, \citenamefont {Poulin},\ and\ \citenamefont
  {Liu}}]{cramer2010efficient}%
  \BibitemOpen
  \bibfield  {author} {\bibinfo {author} {\bibfnamefont {M.}~\bibnamefont
  {Cramer}}, \bibinfo {author} {\bibfnamefont {M.~B.}\ \bibnamefont {Plenio}},
  \bibinfo {author} {\bibfnamefont {S.~T.}\ \bibnamefont {Flammia}}, \bibinfo
  {author} {\bibfnamefont {R.}~\bibnamefont {Somma}}, \bibinfo {author}
  {\bibfnamefont {D.}~\bibnamefont {Gross}}, \bibinfo {author} {\bibfnamefont
  {S.~D.}\ \bibnamefont {Bartlett}}, \bibinfo {author} {\bibfnamefont
  {O.}~\bibnamefont {Landon-Cardinal}}, \bibinfo {author} {\bibfnamefont
  {D.}~\bibnamefont {Poulin}},\ and\ \bibinfo {author} {\bibfnamefont {Y.-K.}\
  \bibnamefont {Liu}},\ }\bibfield  {title} {\bibinfo {title} {Efficient
  quantum state tomography},\ }\href@noop {} {\bibfield  {journal} {\bibinfo
  {journal} {Nature communications}\ }\textbf {\bibinfo {volume} {1}},\
  \bibinfo {pages} {149} (\bibinfo {year} {2010})}\BibitemShut {NoStop}%
\bibitem [{\citenamefont {Xin}\ \emph {et~al.}(2017)\citenamefont {Xin},
  \citenamefont {Lu}, \citenamefont {Klassen}, \citenamefont {Yu},
  \citenamefont {Ji}, \citenamefont {Chen}, \citenamefont {Ma}, \citenamefont
  {Long}, \citenamefont {Zeng},\ and\ \citenamefont
  {Laflamme}}]{xin2017quantum}%
  \BibitemOpen
  \bibfield  {author} {\bibinfo {author} {\bibfnamefont {T.}~\bibnamefont
  {Xin}}, \bibinfo {author} {\bibfnamefont {D.}~\bibnamefont {Lu}}, \bibinfo
  {author} {\bibfnamefont {J.}~\bibnamefont {Klassen}}, \bibinfo {author}
  {\bibfnamefont {N.}~\bibnamefont {Yu}}, \bibinfo {author} {\bibfnamefont
  {Z.}~\bibnamefont {Ji}}, \bibinfo {author} {\bibfnamefont {J.}~\bibnamefont
  {Chen}}, \bibinfo {author} {\bibfnamefont {X.}~\bibnamefont {Ma}}, \bibinfo
  {author} {\bibfnamefont {G.}~\bibnamefont {Long}}, \bibinfo {author}
  {\bibfnamefont {B.}~\bibnamefont {Zeng}},\ and\ \bibinfo {author}
  {\bibfnamefont {R.}~\bibnamefont {Laflamme}},\ }\bibfield  {title} {\bibinfo
  {title} {Quantum state tomography via reduced density matrices},\ }\href@noop
  {} {\bibfield  {journal} {\bibinfo  {journal} {Phys. Rev. Lett.}\ }\textbf
  {\bibinfo {volume} {118}},\ \bibinfo {pages} {020401} (\bibinfo {year}
  {2017})}\BibitemShut {NoStop}%
\bibitem [{\citenamefont {Haffner}\ \emph {et~al.}(2005)\citenamefont
  {Haffner}, \citenamefont {Hansel}, \citenamefont {Roos} \emph
  {et~al.}}]{haffner2005scalable}%
  \BibitemOpen
  \bibfield  {author} {\bibinfo {author} {\bibfnamefont {H.}~\bibnamefont
  {Haffner}}, \bibinfo {author} {\bibfnamefont {W.}~\bibnamefont {Hansel}},
  \bibinfo {author} {\bibfnamefont {C.}~\bibnamefont {Roos}}, \emph {et~al.},\
  }\bibfield  {title} {\bibinfo {title} {Scalable multiparticle entanglement of
  trapped ions},\ }\href@noop {} {\bibfield  {journal} {\bibinfo  {journal}
  {Nature}\ }\textbf {\bibinfo {volume} {438}},\ \bibinfo {pages} {643}
  (\bibinfo {year} {2005})}\BibitemShut {NoStop}%
\bibitem [{\citenamefont {Riebe}\ \emph {et~al.}(2006)\citenamefont {Riebe},
  \citenamefont {Kim}, \citenamefont {Schindler}, \citenamefont {Monz},
  \citenamefont {Schmidt}, \citenamefont {Korber}, \citenamefont {Hansel},
  \citenamefont {Haffner}, \citenamefont {Roos},\ and\ \citenamefont
  {Blatt}}]{riebe2006process}%
  \BibitemOpen
  \bibfield  {author} {\bibinfo {author} {\bibfnamefont {M.}~\bibnamefont
  {Riebe}}, \bibinfo {author} {\bibfnamefont {K.}~\bibnamefont {Kim}}, \bibinfo
  {author} {\bibfnamefont {P.}~\bibnamefont {Schindler}}, \bibinfo {author}
  {\bibfnamefont {T.}~\bibnamefont {Monz}}, \bibinfo {author} {\bibfnamefont
  {P.~O.}\ \bibnamefont {Schmidt}}, \bibinfo {author} {\bibfnamefont {T.~K.}\
  \bibnamefont {Korber}}, \bibinfo {author} {\bibfnamefont {W.}~\bibnamefont
  {Hansel}}, \bibinfo {author} {\bibfnamefont {H.}~\bibnamefont {Haffner}},
  \bibinfo {author} {\bibfnamefont {C.~F.}\ \bibnamefont {Roos}},\ and\
  \bibinfo {author} {\bibfnamefont {R.}~\bibnamefont {Blatt}},\ }\bibfield
  {title} {\bibinfo {title} {Process tomography of ion trap quantum gates},\
  }\href@noop {} {\bibfield  {journal} {\bibinfo  {journal} {Phys. Rev. Lett.}\
  }\textbf {\bibinfo {volume} {97}},\ \bibinfo {pages} {220407} (\bibinfo
  {year} {2006})}\BibitemShut {NoStop}%
\bibitem [{\citenamefont {Lvovsky}\ and\ \citenamefont
  {Raymer}(2009)}]{RevModPhys.81.299}%
  \BibitemOpen
  \bibfield  {author} {\bibinfo {author} {\bibfnamefont {A.~I.}\ \bibnamefont
  {Lvovsky}}\ and\ \bibinfo {author} {\bibfnamefont {M.~G.}\ \bibnamefont
  {Raymer}},\ }\bibfield  {title} {\bibinfo {title} {Continuous-variable
  optical quantum-state tomography},\ }\href
  {https://doi.org/10.1103/RevModPhys.81.299} {\bibfield  {journal} {\bibinfo
  {journal} {Rev. Mod. Phys.}\ }\textbf {\bibinfo {volume} {81}},\ \bibinfo
  {pages} {299} (\bibinfo {year} {2009})}\BibitemShut {NoStop}%
\bibitem [{\citenamefont {Gupta}\ \emph {et~al.}(2021)\citenamefont {Gupta},
  \citenamefont {Xia}, \citenamefont {Levine},\ and\ \citenamefont
  {Kais}}]{PRXQuantum.2.010318}%
  \BibitemOpen
  \bibfield  {author} {\bibinfo {author} {\bibfnamefont {R.}~\bibnamefont
  {Gupta}}, \bibinfo {author} {\bibfnamefont {R.}~\bibnamefont {Xia}}, \bibinfo
  {author} {\bibfnamefont {R.~D.}\ \bibnamefont {Levine}},\ and\ \bibinfo
  {author} {\bibfnamefont {S.}~\bibnamefont {Kais}},\ }\bibfield  {title}
  {\bibinfo {title} {Maximal entropy approach for quantum state tomography},\
  }\href {https://doi.org/10.1103/PRXQuantum.2.010318} {\bibfield  {journal}
  {\bibinfo  {journal} {PRX Quantum}\ }\textbf {\bibinfo {volume} {2}},\
  \bibinfo {pages} {010318} (\bibinfo {year} {2021})}\BibitemShut {NoStop}%
\bibitem [{\citenamefont {Bonet-Monroig}\ \emph {et~al.}(2020)\citenamefont
  {Bonet-Monroig}, \citenamefont {Babbush},\ and\ \citenamefont
  {O'Brien}}]{PhysRevX.10.031064}%
  \BibitemOpen
  \bibfield  {author} {\bibinfo {author} {\bibfnamefont {X.}~\bibnamefont
  {Bonet-Monroig}}, \bibinfo {author} {\bibfnamefont {R.}~\bibnamefont
  {Babbush}},\ and\ \bibinfo {author} {\bibfnamefont {T.~E.}\ \bibnamefont
  {O'Brien}},\ }\bibfield  {title} {\bibinfo {title} {Nearly optimal
  measurement scheduling for partial tomography of quantum states},\ }\href
  {https://doi.org/10.1103/PhysRevX.10.031064} {\bibfield  {journal} {\bibinfo
  {journal} {Phys. Rev. X}\ }\textbf {\bibinfo {volume} {10}},\ \bibinfo
  {pages} {031064} (\bibinfo {year} {2020})}\BibitemShut {NoStop}%
\bibitem [{\citenamefont {Bent}\ \emph {et~al.}(2015)\citenamefont {Bent},
  \citenamefont {Qassim}, \citenamefont {Tahir}, \citenamefont {Sych},
  \citenamefont {Leuchs}, \citenamefont {S\'anchez-Soto}, \citenamefont
  {Karimi},\ and\ \citenamefont {Boyd}}]{PhysRevX.5.041006}%
  \BibitemOpen
  \bibfield  {author} {\bibinfo {author} {\bibfnamefont {N.}~\bibnamefont
  {Bent}}, \bibinfo {author} {\bibfnamefont {H.}~\bibnamefont {Qassim}},
  \bibinfo {author} {\bibfnamefont {A.~A.}\ \bibnamefont {Tahir}}, \bibinfo
  {author} {\bibfnamefont {D.}~\bibnamefont {Sych}}, \bibinfo {author}
  {\bibfnamefont {G.}~\bibnamefont {Leuchs}}, \bibinfo {author} {\bibfnamefont
  {L.~L.}\ \bibnamefont {S\'anchez-Soto}}, \bibinfo {author} {\bibfnamefont
  {E.}~\bibnamefont {Karimi}},\ and\ \bibinfo {author} {\bibfnamefont {R.~W.}\
  \bibnamefont {Boyd}},\ }\bibfield  {title} {\bibinfo {title} {Experimental
  realization of quantum tomography of photonic qudits via symmetric
  informationally complete positive operator-valued measures},\ }\href
  {https://doi.org/10.1103/PhysRevX.5.041006} {\bibfield  {journal} {\bibinfo
  {journal} {Phys. Rev. X}\ }\textbf {\bibinfo {volume} {5}},\ \bibinfo {pages}
  {041006} (\bibinfo {year} {2015})}\BibitemShut {NoStop}%
\bibitem [{\citenamefont {Struchalin}\ \emph {et~al.}(2021)\citenamefont
  {Struchalin}, \citenamefont {Zagorovskii}, \citenamefont {Kovlakov},
  \citenamefont {Straupe},\ and\ \citenamefont {Kulik}}]{PRXQuantum.2.010307}%
  \BibitemOpen
  \bibfield  {author} {\bibinfo {author} {\bibfnamefont {G.}~\bibnamefont
  {Struchalin}}, \bibinfo {author} {\bibfnamefont {Y.~A.}\ \bibnamefont
  {Zagorovskii}}, \bibinfo {author} {\bibfnamefont {E.}~\bibnamefont
  {Kovlakov}}, \bibinfo {author} {\bibfnamefont {S.}~\bibnamefont {Straupe}},\
  and\ \bibinfo {author} {\bibfnamefont {S.}~\bibnamefont {Kulik}},\ }\bibfield
   {title} {\bibinfo {title} {Experimental estimation of quantum state
  properties from classical shadows},\ }\href
  {https://doi.org/10.1103/PRXQuantum.2.010307} {\bibfield  {journal} {\bibinfo
   {journal} {PRX Quantum}\ }\textbf {\bibinfo {volume} {2}},\ \bibinfo {pages}
  {010307} (\bibinfo {year} {2021})}\BibitemShut {NoStop}%
\bibitem [{\citenamefont {Kulis}\ \emph {et~al.}(2006)\citenamefont {Kulis},
  \citenamefont {Sustik},\ and\ \citenamefont {Dhillon}}]{kulis2006learning}%
  \BibitemOpen
  \bibfield  {author} {\bibinfo {author} {\bibfnamefont {B.}~\bibnamefont
  {Kulis}}, \bibinfo {author} {\bibfnamefont {M.}~\bibnamefont {Sustik}},\ and\
  \bibinfo {author} {\bibfnamefont {I.}~\bibnamefont {Dhillon}},\ }\bibfield
  {title} {\bibinfo {title} {Learning low-rank kernel matrices},\ }in\
  \href@noop {} {\emph {\bibinfo {booktitle} {Proceedings of the 23rd
  international conference on Machine learning}}}\ (\bibinfo {year} {2006})\
  pp.\ \bibinfo {pages} {505--512}\BibitemShut {NoStop}%
\bibitem [{\citenamefont {Yao}\ \emph {et~al.}(2018)\citenamefont {Yao},
  \citenamefont {Kwok}, \citenamefont {Wang},\ and\ \citenamefont
  {Liu}}]{yao2018large}%
  \BibitemOpen
  \bibfield  {author} {\bibinfo {author} {\bibfnamefont {Q.}~\bibnamefont
  {Yao}}, \bibinfo {author} {\bibfnamefont {J.~T.}\ \bibnamefont {Kwok}},
  \bibinfo {author} {\bibfnamefont {T.}~\bibnamefont {Wang}},\ and\ \bibinfo
  {author} {\bibfnamefont {T.-Y.}\ \bibnamefont {Liu}},\ }\bibfield  {title}
  {\bibinfo {title} {Large-scale low-rank matrix learning with nonconvex
  regularizers},\ }\href@noop {} {\bibfield  {journal} {\bibinfo  {journal}
  {IEEE transactions on pattern analysis and machine intelligence}\ }\textbf
  {\bibinfo {volume} {41}},\ \bibinfo {pages} {2628} (\bibinfo {year}
  {2018})}\BibitemShut {NoStop}%
\bibitem [{\citenamefont {Udell}\ and\ \citenamefont
  {Townsend}(2019)}]{udell2019big}%
  \BibitemOpen
  \bibfield  {author} {\bibinfo {author} {\bibfnamefont {M.}~\bibnamefont
  {Udell}}\ and\ \bibinfo {author} {\bibfnamefont {A.}~\bibnamefont
  {Townsend}},\ }\bibfield  {title} {\bibinfo {title} {Why are big data
  matrices approximately low rank?},\ }\href@noop {} {\bibfield  {journal}
  {\bibinfo  {journal} {SIAM Journal on Mathematics of Data Science}\ }\textbf
  {\bibinfo {volume} {1}},\ \bibinfo {pages} {144} (\bibinfo {year}
  {2019})}\BibitemShut {NoStop}%
\bibitem [{\citenamefont {Wang}\ \emph {et~al.}(2016)\citenamefont {Wang},
  \citenamefont {Zhang}, \citenamefont {Liu}, \citenamefont {Liu},\ and\
  \citenamefont {Wang}}]{wang2016unsupervised}%
  \BibitemOpen
  \bibfield  {author} {\bibinfo {author} {\bibfnamefont {D.}~\bibnamefont
  {Wang}}, \bibinfo {author} {\bibfnamefont {H.}~\bibnamefont {Zhang}},
  \bibinfo {author} {\bibfnamefont {R.}~\bibnamefont {Liu}}, \bibinfo {author}
  {\bibfnamefont {X.}~\bibnamefont {Liu}},\ and\ \bibinfo {author}
  {\bibfnamefont {J.}~\bibnamefont {Wang}},\ }\bibfield  {title} {\bibinfo
  {title} {Unsupervised feature selection through gram--schmidt
  orthogonalization—a word co-occurrence perspective},\ }\href@noop {}
  {\bibfield  {journal} {\bibinfo  {journal} {Neurocomputing}\ }\textbf
  {\bibinfo {volume} {173}},\ \bibinfo {pages} {845} (\bibinfo {year}
  {2016})}\BibitemShut {NoStop}%
\bibitem [{\citenamefont {Zhao}\ \emph {et~al.}(2017)\citenamefont {Zhao},
  \citenamefont {Li}, \citenamefont {Xi}, \citenamefont {Liang}, \citenamefont
  {Sun},\ and\ \citenamefont {Chen}}]{zhao2017gram}%
  \BibitemOpen
  \bibfield  {author} {\bibinfo {author} {\bibfnamefont {Y.-P.}\ \bibnamefont
  {Zhao}}, \bibinfo {author} {\bibfnamefont {Z.-Q.}\ \bibnamefont {Li}},
  \bibinfo {author} {\bibfnamefont {P.-P.}\ \bibnamefont {Xi}}, \bibinfo
  {author} {\bibfnamefont {D.}~\bibnamefont {Liang}}, \bibinfo {author}
  {\bibfnamefont {L.}~\bibnamefont {Sun}},\ and\ \bibinfo {author}
  {\bibfnamefont {T.-H.}\ \bibnamefont {Chen}},\ }\bibfield  {title} {\bibinfo
  {title} {Gram--schmidt process based incremental extreme learning machine},\
  }\href@noop {} {\bibfield  {journal} {\bibinfo  {journal} {Neurocomputing}\
  }\textbf {\bibinfo {volume} {241}},\ \bibinfo {pages} {1} (\bibinfo {year}
  {2017})}\BibitemShut {NoStop}%
\bibitem [{\citenamefont {Bansal}\ \emph {et~al.}(2018)\citenamefont {Bansal},
  \citenamefont {Chen},\ and\ \citenamefont {Wang}}]{bansal2018can}%
  \BibitemOpen
  \bibfield  {author} {\bibinfo {author} {\bibfnamefont {N.}~\bibnamefont
  {Bansal}}, \bibinfo {author} {\bibfnamefont {X.}~\bibnamefont {Chen}},\ and\
  \bibinfo {author} {\bibfnamefont {Z.}~\bibnamefont {Wang}},\ }\bibfield
  {title} {\bibinfo {title} {Can we gain more from orthogonality
  regularizations in training deep networks?},\ }in\ \href@noop {} {\emph
  {\bibinfo {booktitle} {Advances in Neural Information Processing Systems}}}\
  (\bibinfo {year} {2018})\ pp.\ \bibinfo {pages} {4261--4271}\BibitemShut
  {NoStop}%
\bibitem [{\citenamefont {Vanner}\ \emph {et~al.}(2013)\citenamefont {Vanner},
  \citenamefont {Aspelmeyer},\ and\ \citenamefont {Kim}}]{vanner2013quantum}%
  \BibitemOpen
  \bibfield  {author} {\bibinfo {author} {\bibfnamefont {M.}~\bibnamefont
  {Vanner}}, \bibinfo {author} {\bibfnamefont {M.}~\bibnamefont {Aspelmeyer}},\
  and\ \bibinfo {author} {\bibfnamefont {M.}~\bibnamefont {Kim}},\ }\bibfield
  {title} {\bibinfo {title} {Quantum state orthogonalization and a toolset for
  quantum optomechanical phonon control},\ }\href@noop {} {\bibfield  {journal}
  {\bibinfo  {journal} {Phys. Rev. Lett.}\ }\textbf {\bibinfo {volume} {110}},\
  \bibinfo {pages} {010504} (\bibinfo {year} {2013})}\BibitemShut {NoStop}%
\bibitem [{\citenamefont {Je{\v{z}}ek}\ \emph {et~al.}(2014)\citenamefont
  {Je{\v{z}}ek}, \citenamefont {Mi{\v{c}}uda}, \citenamefont {Straka},
  \citenamefont {Mikova}, \citenamefont {Du{\v{s}}ek},\ and\ \citenamefont
  {Fiur{\'a}{\v{s}}ek}}]{jevzek2014orthogonalization}%
  \BibitemOpen
  \bibfield  {author} {\bibinfo {author} {\bibfnamefont {M.}~\bibnamefont
  {Je{\v{z}}ek}}, \bibinfo {author} {\bibfnamefont {M.}~\bibnamefont
  {Mi{\v{c}}uda}}, \bibinfo {author} {\bibfnamefont {I.}~\bibnamefont
  {Straka}}, \bibinfo {author} {\bibfnamefont {M.}~\bibnamefont {Mikova}},
  \bibinfo {author} {\bibfnamefont {M.}~\bibnamefont {Du{\v{s}}ek}},\ and\
  \bibinfo {author} {\bibfnamefont {J.}~\bibnamefont {Fiur{\'a}{\v{s}}ek}},\
  }\bibfield  {title} {\bibinfo {title} {Orthogonalization of partly unknown
  quantum states},\ }\href@noop {} {\bibfield  {journal} {\bibinfo  {journal}
  {Phys. Rev. A}\ }\textbf {\bibinfo {volume} {89}},\ \bibinfo {pages} {042316}
  (\bibinfo {year} {2014})}\BibitemShut {NoStop}%
\bibitem [{\citenamefont {Coelho}\ \emph {et~al.}(2016)\citenamefont {Coelho},
  \citenamefont {Costanzo}, \citenamefont {Zavatta}, \citenamefont {Hughes},
  \citenamefont {Kim},\ and\ \citenamefont {Bellini}}]{Coelho_2016}%
  \BibitemOpen
  \bibfield  {author} {\bibinfo {author} {\bibfnamefont {A.~S.}\ \bibnamefont
  {Coelho}}, \bibinfo {author} {\bibfnamefont {L.~S.}\ \bibnamefont
  {Costanzo}}, \bibinfo {author} {\bibfnamefont {A.}~\bibnamefont {Zavatta}},
  \bibinfo {author} {\bibfnamefont {C.}~\bibnamefont {Hughes}}, \bibinfo
  {author} {\bibfnamefont {M.~S.}\ \bibnamefont {Kim}},\ and\ \bibinfo {author}
  {\bibfnamefont {M.}~\bibnamefont {Bellini}},\ }\bibfield  {title} {\bibinfo
  {title} {Universal continuous-variable state orthogonalizer and qubit
  generator},\ }\href {https://doi.org/10.1103/PhysRevLett.116.110501}
  {\bibfield  {journal} {\bibinfo  {journal} {Phys. Rev. Lett.}\ }\textbf
  {\bibinfo {volume} {116}},\ \bibinfo {pages} {110501} (\bibinfo {year}
  {2016})}\BibitemShut {NoStop}%
\bibitem [{\citenamefont {Havlicek}\ and\ \citenamefont
  {Svozil}(2018)}]{Havlicek_2018}%
  \BibitemOpen
  \bibfield  {author} {\bibinfo {author} {\bibfnamefont {H.}~\bibnamefont
  {Havlicek}}\ and\ \bibinfo {author} {\bibfnamefont {K.}~\bibnamefont
  {Svozil}},\ }\bibfield  {title} {\bibinfo {title} {Dimensional lifting
  through the generalized gram–schmidt process},\ }\href
  {https://doi.org/10.3390/e20040284} {\bibfield  {journal} {\bibinfo
  {journal} {Entropy}\ }\textbf {\bibinfo {volume} {20}},\ \bibinfo {pages}
  {284} (\bibinfo {year} {2018})}\BibitemShut {NoStop}%
\bibitem [{\citenamefont {Childs}\ and\ \citenamefont
  {Wiebe}(2012)}]{childs2012hamiltonian}%
  \BibitemOpen
  \bibfield  {author} {\bibinfo {author} {\bibfnamefont {A.~M.}\ \bibnamefont
  {Childs}}\ and\ \bibinfo {author} {\bibfnamefont {N.}~\bibnamefont {Wiebe}},\
  }\bibfield  {title} {\bibinfo {title} {Hamiltonian simulation using linear
  combinations of unitary operations},\ }\href@noop {} {\bibfield  {journal}
  {\bibinfo  {journal} {Quantum Information \& Computation}\ }\textbf {\bibinfo
  {volume} {12}},\ \bibinfo {pages} {901} (\bibinfo {year} {2012})}\BibitemShut
  {NoStop}%
\bibitem [{\citenamefont {Lloyd}\ \emph {et~al.}(2014)\citenamefont {Lloyd},
  \citenamefont {Mohseni},\ and\ \citenamefont {Rebentrost}}]{Lloyd_2014}%
  \BibitemOpen
  \bibfield  {author} {\bibinfo {author} {\bibfnamefont {S.}~\bibnamefont
  {Lloyd}}, \bibinfo {author} {\bibfnamefont {M.}~\bibnamefont {Mohseni}},\
  and\ \bibinfo {author} {\bibfnamefont {P.}~\bibnamefont {Rebentrost}},\
  }\bibfield  {title} {\bibinfo {title} {Quantum principal component
  analysis},\ }\href {https://doi.org/10.1038/nphys3029} {\bibfield  {journal}
  {\bibinfo  {journal} {Nature Physics}\ }\textbf {\bibinfo {volume} {10}},\
  \bibinfo {pages} {631–633} (\bibinfo {year} {2014})}\BibitemShut {NoStop}%
\bibitem [{\citenamefont {Aharonov}\ \emph {et~al.}(2009)\citenamefont
  {Aharonov}, \citenamefont {Jones},\ and\ \citenamefont
  {Landau}}]{aharonov2009polynomial}%
  \BibitemOpen
  \bibfield  {author} {\bibinfo {author} {\bibfnamefont {D.}~\bibnamefont
  {Aharonov}}, \bibinfo {author} {\bibfnamefont {V.}~\bibnamefont {Jones}},\
  and\ \bibinfo {author} {\bibfnamefont {Z.}~\bibnamefont {Landau}},\
  }\bibfield  {title} {\bibinfo {title} {A polynomial quantum algorithm for
  approximating the jones polynomial},\ }\href@noop {} {\bibfield  {journal}
  {\bibinfo  {journal} {Algorithmica}\ }\textbf {\bibinfo {volume} {55}},\
  \bibinfo {pages} {395} (\bibinfo {year} {2009})}\BibitemShut {NoStop}%
\bibitem [{\citenamefont {Buhrman}\ \emph {et~al.}(2001)\citenamefont
  {Buhrman}, \citenamefont {Cleve}, \citenamefont {Watrous},\ and\
  \citenamefont {de~Wolf}}]{PhysRevLett.87.167902}%
  \BibitemOpen
  \bibfield  {author} {\bibinfo {author} {\bibfnamefont {H.}~\bibnamefont
  {Buhrman}}, \bibinfo {author} {\bibfnamefont {R.}~\bibnamefont {Cleve}},
  \bibinfo {author} {\bibfnamefont {J.}~\bibnamefont {Watrous}},\ and\ \bibinfo
  {author} {\bibfnamefont {R.}~\bibnamefont {de~Wolf}},\ }\bibfield  {title}
  {\bibinfo {title} {Quantum fingerprinting},\ }\href
  {https://doi.org/10.1103/PhysRevLett.87.167902} {\bibfield  {journal}
  {\bibinfo  {journal} {Phys. Rev. Lett.}\ }\textbf {\bibinfo {volume} {87}},\
  \bibinfo {pages} {167902} (\bibinfo {year} {2001})}\BibitemShut {NoStop}%
\bibitem [{\citenamefont {Wiebe}\ \emph {et~al.}(2012)\citenamefont {Wiebe},
  \citenamefont {Braun},\ and\ \citenamefont {Lloyd}}]{PhysRevLett.109.050505}%
  \BibitemOpen
  \bibfield  {author} {\bibinfo {author} {\bibfnamefont {N.}~\bibnamefont
  {Wiebe}}, \bibinfo {author} {\bibfnamefont {D.}~\bibnamefont {Braun}},\ and\
  \bibinfo {author} {\bibfnamefont {S.}~\bibnamefont {Lloyd}},\ }\bibfield
  {title} {\bibinfo {title} {Quantum algorithm for data fitting},\ }\href
  {https://doi.org/10.1103/PhysRevLett.109.050505} {\bibfield  {journal}
  {\bibinfo  {journal} {Phys. Rev. Lett.}\ }\textbf {\bibinfo {volume} {109}},\
  \bibinfo {pages} {050505} (\bibinfo {year} {2012})}\BibitemShut {NoStop}%
\bibitem [{\citenamefont {Childs}\ \emph {et~al.}(2017)\citenamefont {Childs},
  \citenamefont {Kothari},\ and\ \citenamefont {Somma}}]{childs2017quantum}%
  \BibitemOpen
  \bibfield  {author} {\bibinfo {author} {\bibfnamefont {A.~M.}\ \bibnamefont
  {Childs}}, \bibinfo {author} {\bibfnamefont {R.}~\bibnamefont {Kothari}},\
  and\ \bibinfo {author} {\bibfnamefont {R.~D.}\ \bibnamefont {Somma}},\
  }\bibfield  {title} {\bibinfo {title} {Quantum algorithm for systems of
  linear equations with exponentially improved dependence on precision},\
  }\href@noop {} {\bibfield  {journal} {\bibinfo  {journal} {SIAM Journal on
  Computing}\ }\textbf {\bibinfo {volume} {46}},\ \bibinfo {pages} {1920}
  (\bibinfo {year} {2017})}\BibitemShut {NoStop}%
\bibitem [{\citenamefont {Tang}(2019)}]{3313276.3316310}%
  \BibitemOpen
  \bibfield  {author} {\bibinfo {author} {\bibfnamefont {E.}~\bibnamefont
  {Tang}},\ }\bibfield  {title} {\bibinfo {title} {A quantum-inspired classical
  algorithm for recommendation systems},\ }in\ \href
  {https://doi.org/10.1145/3313276.3316310} {\emph {\bibinfo {booktitle}
  {Proceedings of the 51st Annual ACM SIGACT Symposium on Theory of
  Computing}}}\ (\bibinfo  {publisher} {Association for Computing Machinery},\
  \bibinfo {address} {New York, NY, USA},\ \bibinfo {year} {2019})\ p.\
  \bibinfo {pages} {217–228}\BibitemShut {NoStop}%
\bibitem [{\citenamefont {Chia}\ \emph {et~al.}(2018)\citenamefont {Chia},
  \citenamefont {Lin},\ and\ \citenamefont {Wang}}]{chia2018quantuminspired}%
  \BibitemOpen
  \bibfield  {author} {\bibinfo {author} {\bibfnamefont {N.-H.}\ \bibnamefont
  {Chia}}, \bibinfo {author} {\bibfnamefont {H.-H.}\ \bibnamefont {Lin}},\ and\
  \bibinfo {author} {\bibfnamefont {C.}~\bibnamefont {Wang}},\ }\href@noop {}
  {\bibinfo {title} {Quantum-inspired sublinear classical algorithms for
  solving low-rank linear systems}} (\bibinfo {year} {2018}),\ \Eprint
  {https://arxiv.org/abs/1811.04852} {arXiv:1811.04852} \BibitemShut {NoStop}%
\bibitem [{\citenamefont {Jethwani}\ \emph {et~al.}(2020)\citenamefont
  {Jethwani}, \citenamefont {Le~Gall},\ and\ \citenamefont
  {Singh}}]{jethwani2019quantuminspired}%
  \BibitemOpen
  \bibfield  {author} {\bibinfo {author} {\bibfnamefont {D.}~\bibnamefont
  {Jethwani}}, \bibinfo {author} {\bibfnamefont {F.}~\bibnamefont {Le~Gall}},\
  and\ \bibinfo {author} {\bibfnamefont {S.~K.}\ \bibnamefont {Singh}},\
  }\bibfield  {title} {\bibinfo {title} {Quantum-inspired classical algorithms
  for singular value transformation},\ }in\ \href@noop {} {\emph {\bibinfo
  {booktitle} {45th International Symposium on Mathematical Foundations of
  Computer Science}}}\ (\bibinfo {year} {2020})\BibitemShut {NoStop}%
\bibitem [{\citenamefont {Du}\ \emph {et~al.}(2020)\citenamefont {Du},
  \citenamefont {Hsieh}, \citenamefont {Liu},\ and\ \citenamefont
  {Tao}}]{du2019quantuminspired}%
  \BibitemOpen
  \bibfield  {author} {\bibinfo {author} {\bibfnamefont {Y.}~\bibnamefont
  {Du}}, \bibinfo {author} {\bibfnamefont {M.-H.}\ \bibnamefont {Hsieh}},
  \bibinfo {author} {\bibfnamefont {T.}~\bibnamefont {Liu}},\ and\ \bibinfo
  {author} {\bibfnamefont {D.}~\bibnamefont {Tao}},\ }\bibfield  {title}
  {\bibinfo {title} {Quantum-inspired algorithm for general minimum conical
  hull problems},\ }\href {https://doi.org/10.1103/PhysRevResearch.2.033199}
  {\bibfield  {journal} {\bibinfo  {journal} {Phys. Rev. Research}\ }\textbf
  {\bibinfo {volume} {2}},\ \bibinfo {pages} {033199} (\bibinfo {year}
  {2020})}\BibitemShut {NoStop}%
\bibitem [{\citenamefont {Godunov}()}]{godunov2013guaranteed}%
  \BibitemOpen
  \bibfield  {author} {\bibinfo {author} {\bibfnamefont {S.}~\bibnamefont
  {Godunov}},\ }\href@noop {} {\emph {\bibinfo {title} {Guaranteed accuracy in
  numerical linear algebra}}}\BibitemShut {NoStop}%
\bibitem [{\citenamefont {Brassard}\ \emph {et~al.}(2002)\citenamefont
  {Brassard}, \citenamefont {Hoyer}, \citenamefont {Mosca},\ and\ \citenamefont
  {Tapp}}]{brassard2002quantum}%
  \BibitemOpen
  \bibfield  {author} {\bibinfo {author} {\bibfnamefont {G.}~\bibnamefont
  {Brassard}}, \bibinfo {author} {\bibfnamefont {P.}~\bibnamefont {Hoyer}},
  \bibinfo {author} {\bibfnamefont {M.}~\bibnamefont {Mosca}},\ and\ \bibinfo
  {author} {\bibfnamefont {A.}~\bibnamefont {Tapp}},\ }\bibfield  {title}
  {\bibinfo {title} {Quantum amplitude amplification and estimation},\
  }\href@noop {} {\bibfield  {journal} {\bibinfo  {journal} {Contemporary
  Mathematics}\ }\textbf {\bibinfo {volume} {305}},\ \bibinfo {pages} {53}
  (\bibinfo {year} {2002})}\BibitemShut {NoStop}%
\end{thebibliography}


%

\end{document}